%% file: bell.tex
\documentclass[twocolumn,prd,aps,showpacs,nofootinbib]{revtex4}

\usepackage{bm}
\usepackage{amssymb, amsmath}
\usepackage{graphicx}
\usepackage{hyperref}
\usepackage{xspace}

\makeatletter
\gdef\@fpheader{}
\g@addto@macro\bfseries{\boldmath}
\makeatother

\hypersetup{
    bookmarks=true,         
    unicode=false,          
    pdftoolbar=true,        
    pdfmenubar=true,        
    pdffitwindow=false,     
    pdfstartview={FitH},    
    pdfnewwindow=true,      
    colorlinks=true,        
    linkcolor=red,          
    citecolor=cyan,         
    filecolor=magenta,      
    urlcolor=blue,         
    linktocpage=true
}

\input{newcommands}

\begin{document}

\title{Bell's Inequalities for Continuous-Variable Systems in Generic
  Squeezed States}

\author{J\'er\^ome Martin} \email{jmartin@iap.fr}
\affiliation{Institut d'Astrophysique de Paris, UMR 7095-CNRS,
Universit\'e Pierre et Marie Curie, 98 bis boulevard Arago, 75014
Paris, France}

\author{Vincent Vennin} \email{vincent.vennin@port.ac.uk}
\affiliation{Institute of Cosmology and Gravitation, University of
  Portsmouth, Dennis Sciama Building, Burnaby Road, Portsmouth, PO1
  3FX, United Kingdom}

\date{\today}

\begin{abstract}
  Bell's inequality for continuous-variable bipartite systems is
  studied. The inequality is expressed in terms of pseudo-spin operators 
  and quantum expectation values are
  calculated for generic two-mode squeezed states characterized by a
  squeezing parameter $r$ and a squeezing angle $\varphi$. 
  Allowing for generic values of the squeezing angle
  is especially relevant when $\varphi$ is not under experimental control, 
  such as in cosmic inflation, where small quantum fluctuations in the early
  Universe are responsible for structures formation.  
  Compared to previous studies restricted to $\varphi=0$ and to a
  fixed orientation of the
  pseudo-spin operators, allowing for $\varphi\neq 0$
  and optimizing the angular configuration leads to a completely new and
  rich phenomenology. Two dual schemes
  of approximation are designed that allow for comprehensive exploration of the
  squeezing parameters space. In particular, it is found that Bell's inequality can be violated
  when the squeezing parameter $r$ is large enough, $r\gtrsim 1.12$,
  and the squeezing angle $\varphi$ is small enough, $\varphi\lesssim
  0.34\,\ee^{-r}$. 
\end{abstract}

\maketitle

\section{Introduction}
\label{sec:intro}

Bell's inequalities~\cite{Bell:1964kc} play a major role in
Physics. Their experimental
violation~\cite{Aspect:1981nv,Aspect:1982fx,Weihs:1998gy,Hensen:2015ccp}
demonstrates that Nature cannot be described by a strongly local (no
causal influence travels faster than light) and deterministic theory.
Instead, it obeys the laws of Quantum Mechanics, where a violation can
occur when the system is placed in an entangled state. Historically,
this type of quantum states was considered for the first time by
Einstein, Podolsky and Rosen (EPR) in \Ref{Einstein:1935rr}. In that
paper, they studied a system made of two particles with a quantum
state entangled in position space. Subsequent works have rather
formulated the problem in terms of discrete variables, typically spin
variables for which the Bell inequalities usually takes the
Clauser-Horne-Shimony-Holt (CHSH)
form~\cite{Clauser:1969ny}. Recently, however, in
Ref.~\cite{2004PhRvA..70b2102L}, it was shown how (discrete)
pseudo-spin operators can be constructed out of (continuous) position
operators, thus opening the possibility to test the Bell inequalities in
its CHSH form for continuous variable systems.

\par

Testing Bell's inequalities for continuous variable
systems~\cite{PhysRevLett.88.040406,PhysRevLett.99.170408,2003PhRvA..67b2108L,2003PhRvA..68f2105B,2009JPhA...42B5309D}
is interesting not only for investigating the EPR argument in its
original formulation but also because it allows us to treat the case
of quantum fields where the role of the continuous variable is played
by the (Fourier) amplitude of the field. This is especially relevant
because when a quantum field interacts with a classical source,
particle creation occurs and, typically, the corresponding system is
in a squeezed state, an example of entangled state. Such states
contains genuine quantum correlations as can also be checked by
computing their quantum
discord~\cite{Henderson:2001,Zurek:2001,Martin:2015qta}. The above
mentioned situation arises, for instance, in the Schwinger
effect~\cite{Schwinger:1951nm} but also in the cosmic inflationary
mechanism~\cite{Starobinsky:1979ty,Starobinsky:1980te,Sato:1980yn,Guth:1980zm,Linde:1981mu,Mukhanov:1981xt,Albrecht:1982wi,Hawking:1982cz,Starobinsky:1982ee,Guth:1982ec,Bardeen:1983qw,Linde:1983gd}
(for reviews, see
\Refs{Martin:2004yj,Martin:2004um,Martin:2003bt,Martin:2013gra,Vennin:2015eaa,Martin:2015dha})
for large scale structures growth and Cosmic Microwave Background
Radiation (CMBR) anisotropies~\cite{Grishchuk:1990bj,Martin:2007bw}.

\par

In this paper, we reconsider the work of
Ref.~\cite{2004PhRvA..70b2102L} and extend it in various directions,
in order to be able to treat the case of quantum field theory and
cosmology, topics that we plan to address in subsequent
papers~\cite{MV}. Compared to Ref.~\cite{2004PhRvA..70b2102L}, we have
obtained several new and important results that we now briefly
describe.  Firstly, we have treated the general case of a two-mode
squeezed state with non-vanishing squeezing angle. In
\Ref{2004PhRvA..70b2102L}, the squeezing angle was set to zero since
it is a controllable parameter in the laboratory and, hence, might be
tuned to zero by working with specific experimental set-ups. However,
in the case of \eg cosmic inflation, the squeezing angle is
``God-given'' and, crucially, is non-vanishing and
dynamical~\cite{Martin:2012pea,Martin:2012ua}. This leads to a new and
rich phenomenology. Secondly, we have numerically computed the
pseudo-spin correlation functions and checked, when possible, our
results with those of Ref.~\cite{2004PhRvA..70b2102L}. Global
agreement is usually found even if we have also detected some slight
differences. Thirdly, we have derived the optimal configuration
leading to Bell's inequality violation and have shown that very
relevant differences can happen compared to the standard configuration
used in Ref.~\cite{2004PhRvA..70b2102L}. For instance, we have
exhibited squeezing parameters and angles such that a violation occurs
for the optimal configuration but not for the standard one. Fourthly,
we have obtained several approximated formulas regarding the
pseudo-spin correlation functions and Bell's operator expectation values
which allowed us to better understand their dependence on the
squeezing parameter and angle, and to interpret the numerical
calculations.  This also made possible studying violation of Bell's
inequality in regimes that would be impossible to reach
numerically. Finally, for the first time to our knowledge, we have
produced a map in the two-dimensional squeezing space (squeezing
parameter and angle) of Bell's inequality violation, see
\Fig{fig:map}. This can serve as a useful guide to find the optimal
squeezing parameter and angle given a specific experimental design.

\par

The paper is organized as follows. In the next section,
\Sec{sec:spin}, we introduce the pseudo-spin operators and study their
properties. In \Sec{sec:BellInequality}, we define the Bell operator
and compute its expectation value for a two-mode squeezed state with
arbitrary squeezing parameter and angle. We also pay special attention
to the four angles involved in the definition of the Bell operator and
derive the corresponding optimal configuration. In
\Sec{sec:BellViolation}, we then investigate for which value of the
squeezing parameter and angle Bell's inequality is violated. Finally,
in \Sec{sec:conclusion}, we present our conclusions. The technical
aspects of the work are summarized in six appendices. In
\Sec{sec:spinPheno}, we numerically calculate the spin correlation
functions. In \Secs{sec:Appr} and~\ref{sec:appr:scheme}, we design
generic approximation schemes allowing us to interpret the numerical
computations and explore regions that cannot be accessed
numerically. In \Secs{sec:largesqueezing} and~\ref{sec:varphi=pi/2},
we work out the large squeezing limit in the two dual cases where the
squeezing angle is close to $0$ and $\pi/2$ respectively. In
\Sec{sec:PhaseSpaceRotation} finally, we show how different
orientations of the pseudo-spin operators in phase space can be dealt
with.

\section{Spin Operators for Continuous Variable Systems}
\label{sec:spin}
The standard formulation of the Bell-CHSH inequality is written in
term of spin variables. In this section, following
\Ref{2004PhRvA..70b2102L}, we explain how, in the case of a continuous
variable system, one can define such quantities.

\par

Let $Q$ be some continuous variable taking values in $\setR$. It can
be the position of a particle but also the (Fourier) amplitude of some
quantum field. We divide the real axis in an infinite number of
intervals $\left[n\ell, (n+1)\ell\right]$ of length $\ell$, where $n$
is an integer number running from $-\infty $ to $+\infty$. Then, we define
the operator $\hat{P}(n,\ell)$ by
\begin{equation}
\hat{P}(n,\ell)\equiv \int_{n\ell}^{(n+1)\ell}{\rm d}Q\left\vert Q\rangle 
\langle Q\right\vert .
\label{eq:projectoru:def}
\end{equation}
Clearly, $\hat{P}(n,\ell)$ is a projector since we have
$\hat{P}(n,\ell)\hat{P}(m,\ell)=\delta _{nm}\hat{P}(n,\ell)$. Its
eigenvectors can be written as (up to normalization)
$\left[\hat{\mathbb{I}}-\hat{P}(n,\ell)\right]\vert \Psi\rangle$ and
$\hat{P}(n,\ell)\vert \Psi\rangle$, where $\vert \Psi\rangle $ is any
wavefunction, with corresponding eigenvalues $0$ and $1$
respectively. If one starts from a state $\vert \Psi\rangle $ which
has support everywhere on $\setR$, then the state
$\hat{P}(n,\ell)\vert \Psi\rangle$ has support only in the interval
$\left[n\ell, (n+1)\ell\right]$ and vanishes elsewhere. In some sense,
it only retains the part of $\vert \Psi \rangle $ present in that
interval. Moreover, the mean value of $\hat{P}(n,\ell)$ in the state
$\vert \Psi\rangle $ gives the probability to find the system in the
interval $\left[n\ell, (n+1)\ell\right]$.

\par

The next step consists in introducing the following operator
\begin{align}
\label{eq:defsz}
\hat{S}_z(\ell)&=\sum_{n=-\infty}^{+\infty}(-1)^n\hat{P}(n,\ell)
\\
\label{eq:defsz2}
& =\sum_{n=-\infty}^{\infty}(-1)^n\int _{n\ell}
^{(n+1)\ell}{\rm d}Q \vert Q\rangle 
\langle Q\vert \, .
\end{align}
This defines a spin variable because the eigenvalues of this operator
are $\pm 1$. This can be proven by noticing that
$\hat{S}_z^2(\ell)=\hat{\mathbb{I}}$. Indeed, one has
\begin{align}
\hat{S}_z^2(\ell)& =\sum_{n=-\infty}^{+\infty}\sum_{m=-\infty}^{+\infty}(-1)^{n+m}
\nonumber \\ & \times
\int_{n\ell}^{(n+1)\ell}\int_{m\ell}^{(m+1)\ell}
{\rm d}Q{\rm d}\overline{Q}
\left\vert Q\rangle 
\langle Q\right\vert \overline{Q}\rangle \langle \overline{Q}\vert.
\label{eq:szsquared}
\end{align}
The scalar product $\langle Q\vert \overline{Q}\rangle $ gives a Dirac
function $\delta(Q-\overline{Q})$ but only if $Q$ and $\overline{Q}$
belong to the same interval (otherwise they cannot be equal). This
means that one must have $n=m$. As a consequence,
\begin{align}
\hat{S}_z^2(\ell)& =\sum_{n,m}(-1)^{n+m}
\delta_{nm}\int_{n\ell}^{(n+1)\ell}
{\rm d}Q
\left\vert Q\rangle 
\langle Q\right\vert
\\
& =\sum_{n=-\infty}^{+\infty} \int_{n\ell}^{(n+1)\ell}
{\rm d}Q
\left\vert Q\rangle 
\langle Q\right\vert
=\hat{\mathbb{I}},
\end{align}
since the probability to find the particle somewhere on the real axis
is always one.  The eigenvectors of $\hat{S}_z(\ell)$ are the
wavefunctions having support within $\cup_{n} [2n\ell,(2n+1)\ell]$
(eigenvalue $+1$) and within $\cup_{n} [(2n-1)\ell,2n\ell]$
(eigenvalue $-1$). In particular, $\hat{P}(n,\ell)\vert \Psi\rangle$
is an eigenvector of $\hat{S}_z(\ell)$
with eigenvalue $(-1)^n$.

\par

Having defined the spin operator along the $z$-axis, we now need to
introduce the operators $\hat{S}_x(\ell)$ and $\hat{S}_y(\ell)$. To do
this, we define an operator $\hat{T}(n,\ell)$ by the following
expression
\begin{align}
\label{eq:defT}
\hat{T}(n,\ell)
=\int_{n\ell}^{(n+1)\ell}{\rm d}Q\left\vert Q\rangle 
\langle Q+\ell\right\vert .
\end{align}
Given the interval $\left[n\ell, (n+1)\ell\right]$, this operator
takes the ``translated'' part (by $\ell$) of $\vert \Psi \rangle$ and
restricts it to $\left[n\ell, (n+1)\ell\right]$. The fact that
$\hat{T}(n,\ell)\vert \Psi\rangle $ has support only in this interval
can be checked from the relation
$\hat{P}(n,\ell)\hat{T}(n,\ell)=\hat{T}(n,\ell)$. In the same manner,
one can show that the adjoint of $\hat{T}(n,\ell)$,
\begin{align}
\label{eq:defT}
\hat{T}^{\dagger}(n,\ell)
=\int_{n\ell+\ell}^{(n+2)\ell}{\rm d}Q\left\vert Q\rangle 
\langle Q-\ell\right\vert ,
\end{align}
takes the translated part (by $-\ell$) of $\vert \Psi\rangle $ and
restricts it to the interval $\left[(n+1)\ell, (n+2)\ell\right]$, as
confirmed by the fact that
$\hat{P}(n+1,\ell)\hat{T}^{\dagger}(n,\ell)=\hat{T}^{\dagger}(n,\ell)$.

\par

One can then define the ``spin step'' operators $\hat{S}_+$ and
$\hat{S}_-$ through the relations
\begin{align}
\hat{S}_+(\ell)=\sum_{n=-\infty}^{\infty}\hat{T}(2n,\ell)
=\sum_{n=-\infty}^{\infty}\int_{2n\ell}^{(2n+1)\ell}{\rm
    d}Q\left\vert Q\rangle \langle Q+\ell\right\vert
\label{eq:defsplus}
\end{align}
and $\hat{S}_-(\ell)=\hat{S}_+^{\dagger}(\ell)$. The operator
$\hat{S}_+(\ell)$ takes an eigenstate of $\hat{S}_z(\ell)$ with
eigenvalue $-1$ and transforms it into another eigenstate of
$\hat{S}_z(\ell)$ but, this time, with eigenvalue $+1$. The proof goes
as follows: an eigenstate of $\hat{S}_z(\ell)$ with eigenvalue $-1$
can be written as $\hat{P}(2n+1,\ell)\vert \Psi\rangle$. In order to
check that $\hat{S}_+(\ell)\hat{P}(2n+1,\ell)\vert \Psi\rangle$ is
also an eigenstate of $\hat{S}_z(\ell)$, one has to evaluate
$\hat{S}_z(\ell)\hat{S}_+(\ell)\hat{P}(2n+1,\ell)\vert
\Psi\rangle$. Using the relations
$\hat{S}_+(\ell)\hat{P}(2n+1,\ell)=\hat{T}(2n,\ell)$ and
$\hat{S}_z(\ell)\hat{T}(n,\ell)=(-1)^n\hat{T}(n,\ell)$, one obtains
that $\hat{S}_z(\ell)\hat{S}_+(\ell)\hat{P}(2n+1,\ell)\vert
\Psi\rangle =\hat{S}_z(\ell)\hat{T}(2n,\ell)\vert \Psi\rangle
=(-1)^{2n}\hat{T}(2n,\ell)\vert \Psi\rangle
=\hat{S}_+(\ell)\hat{P}(2n+1,\ell)\vert \Psi\rangle $, which
demonstrates that, indeed, $\hat{S}_+(\ell)\hat{P}(2n+1,\ell)\vert
\Psi\rangle$ is an eigenstate of $\hat{S}_z(\ell)$ with eigenvalue
$+1$. A similar proof showing that $\hat{S}_-(\ell)$ takes an
eigenstate of $\hat{S}_z(\ell)$ with eigenvalue $+1$ and transforms it
into another eigenstate but with eigenvalue $-1$ can easily be
constructed along the same lines.

\par

One is now in a position to introduce the $x$ and $y$ components of the
pseudo-spin system. They are defined in terms of the spin step
operators by the usual expressions, namely
\begin{align}
\label{eq:defsx}
\hat{S}_x(\ell)& =\hat{S}_+(\ell)+\hat{S}_-(\ell), \\
\label{eq:defsy}
\hat{S}_y(\ell) & =-i\left[\hat{S}_+(\ell)-\hat{S}_-(\ell)\right].
\end{align}
Using the result that
$[\hat{S}_z(\ell),\hat{T}(n,\ell)]=2(-1)^{n}\hat{T}(n,\ell)$, one can
verify that $[\hat{S}_z(\ell),\hat{S}_x(\ell)]=2i\hat{S}_y(\ell)$ and
$[\hat{S}_y(\ell),\hat{S}_z(\ell)]=2i\hat{S}_x(\ell)$. Since one also
has
$[\hat{T}(2n,\ell),\hat{T}^{\dagger}(2m,\ell)]=\delta_{n,m}[\hat{P}(2n,\ell)-\hat{P}(2n+1,\ell)]$,
one obtains
$[\hat{S}_x(\ell),\hat{S}_y(\ell)]=2i[\hat{S}_+(\ell),\hat{S}_-(\ell)]=2i\hat{S}_z(\ell)$.
Finally, one can check that
$\hat{S}_x^2(\ell)=\hat{S}_y^2(\ell)=\hat{\mathbb{I}}$, which
completes the construction of our spin operators. Notice that here,
the choice of $\ell$ is entirely controllable by the observer.
\section{Bell's Inequality}
\label{sec:BellInequality}
Having defined pseudo-spin operators in the last section, one can now
proceed and introduce a Bell operator. Let us consider the case of a bipartite system $(1)\otimes (2)$. Since we want to study the Bell's 
inequalities in their CHSH form, we define the following operator
\begin{align}
\hat{B}(\ell) &\equiv\left[{\bm n}\cdot \hat{{\bm S}}^{(1)}(\ell)\right]
\otimes \left[{\bm m}\cdot \hat{{\bm S}}^{(2)}(\ell)\right]\\
& +
\left[{\bm n}\cdot \hat{{\bm S}}^{(1)}(\ell)\right]
\otimes \left[{\bm m}'\cdot \hat{{\bm S}}^{(2)}(\ell)\right]\\
& +
\left[{\bm n}'\cdot \hat{{\bm S}}^{(1)}(\ell)\right]
\otimes \left[{\bm m}\cdot \hat{{\bm S}}^{(2)}(\ell)\right]\\
& -
\left[{\bm n}'\cdot \hat{{\bm S}}^{(1)}(\ell)\right]
\otimes \left[{\bm m}'\cdot \hat{{\bm S}}^{(2)}(\ell)\right],
\end{align}
where ${\bm n}$, ${\bm n}'$, ${\bm m}$ and ${\bm m}'$ are four
arbitrary unit vectors that can be expanded in terms of their polar
and azimuthal angles, ${\bm n}=\left(\sin \theta _{\bm n}\cos
  \phi_{\bm n}, \sin \theta_{\bm n}\sin \phi_{\bm n},\cos \theta_{\bm
    n}\right)$ (and similar expressions for the three other vectors).
From now on, without loss of generality, we set all azimuthal angles
to zero. Let us then introduce the correlation function
\begin{align}
E\left(\ell,{\bm n},{\bm m}\right) 
 \equiv & \left \langle \left[{\bm n}
\cdot \hat{{\bm S}}^{(1)}(\ell)\right]
\otimes \left[{\bm m}\cdot \hat{{\bm S}}^{(2)}(\ell)\right]
\right \rangle\\
 =& \sin \theta _{\bm n}\sin \theta_{\bm m} 
\langle \hat{S}^{(1)}_x(\ell)
\hat{S}^{(2)}_x(\ell)\rangle 
\nonumber \\ &
+\cos \theta _{\bm n}\cos \theta_{\bm m} 
\langle \hat{S}^{(1)}_z(\ell)
\hat{S}^{(2)}_z(\ell)\rangle \, ,
\label{eq:defcorrelationE}
\end{align}
where the expectation value is with respect to the quantum state of
the system under scrutiny. In general, terms proportional to $\langle
\hat{S}^{(1)}_z(\ell) \hat{S}^{(2)}_x(\ell)\rangle$ and $\langle
\hat{S}^{(1)}_x(\ell) \hat{S}^{(2)}_z(\ell)\rangle$ are also present,
but for the two-mode squeezed state we consider below, it is shown in
\Sec{sec:spinPheno:xz} that these correlators vanish. This expression
allows us to calculate the expectation value of the Bell operator,
namely
\begin{align}
\label{eq:meanbell}
  \langle \hat{B}(\ell)\rangle = & E\left(\ell, \theta_{\bm
      n},\theta_{\bm m}\right) +E\left(\ell, \theta_{\bm
      n},\theta_{\bm m}'\right)+E\left(\ell, \theta_{\bm
      n}',\theta_{\bm m}\right) \nonumber \\ 
&
- E\left(\ell, \theta_{\bm
      n}',\theta_{\bm m}'\right)\, .
\end{align}
Several remarks are in order here. Firstly, the correlation function
does not involve the $y$-axis component of the spin and this is of
course a consequence of the fact that we have taken vanishing
azimuthal angles. Secondly, we notice that the expectation value of
the Bell operator is entirely calculable in terms of the two-point
correlation functions of the spin operators. This is the reason why we
study them in detail in \Sec{sec:spinPheno}. Thirdly, as is
well-known, local realistic theories imply $\vert \langle
\hat{B}(\ell)\rangle \vert \le 2$ while quantum mechanics only imposes
the so-called Cirel'son bound~\cite{1980LMaPh...4...93C}, namely
$\vert \langle \hat{B}(\ell)\rangle \vert \le 2\sqrt{2}$. Therefore,
any situation such that $2<\vert \langle \hat{B}(\ell)\rangle \vert
\le 2\sqrt{2}$ will be referred to as violation of Bell's
inequalities.
\begin{figure}[t]
\begin{center}
\includegraphics[width=0.45\textwidth,clip=true]{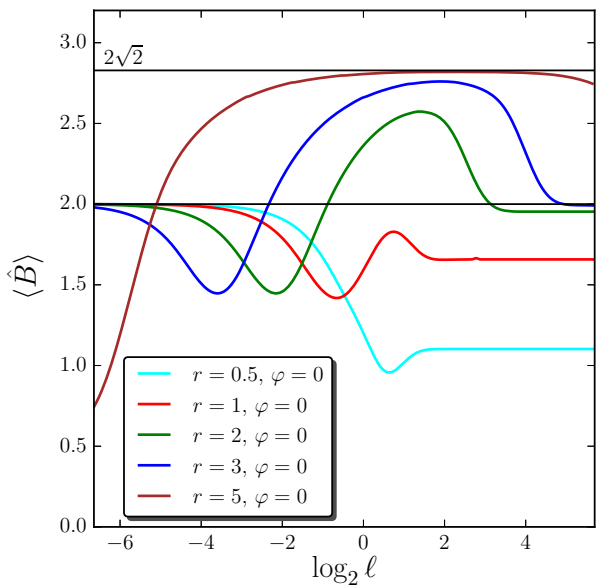}
\caption{Expectation value of the Bell operator $\hat{B}$ as a
  function of $\ell$, for different squeezing parameters $r$ and
  squeezing angle $\varphi=0$. Situations where
  $\langle\hat{B}\rangle>2$ indicate violation of Bell's inequalities,
  and one can check that the Cirel'son bound
  $\langle\hat{B}\rangle<2\sqrt{2}$ is always satisfied.}
\label{fig:exactBell_phi=0}
\end{center}
\end{figure}
\begin{figure*}[t]
\begin{center}
\includegraphics[width=0.45\textwidth,clip=true]{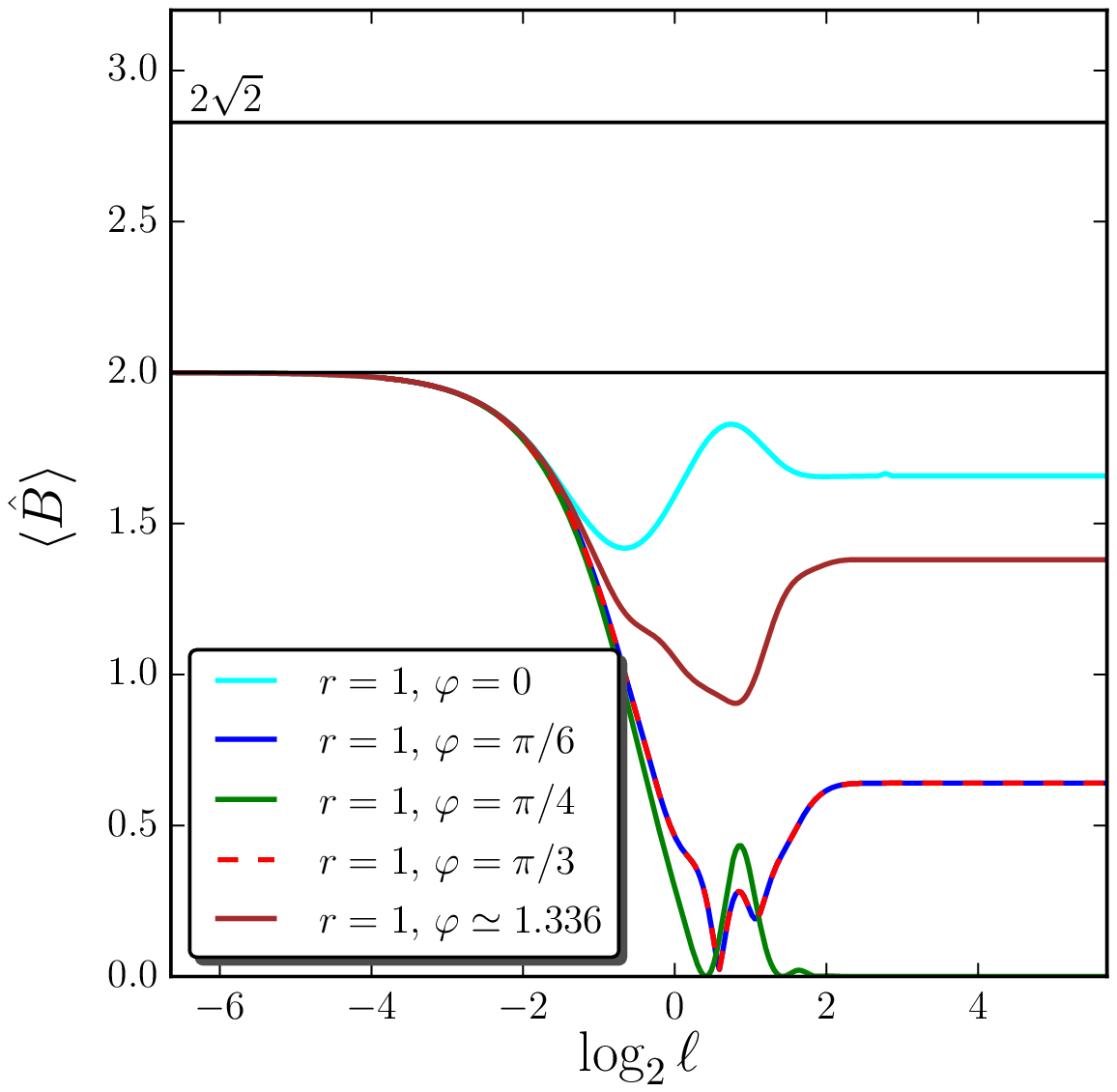}
\includegraphics[width=0.465\textwidth,clip=true]{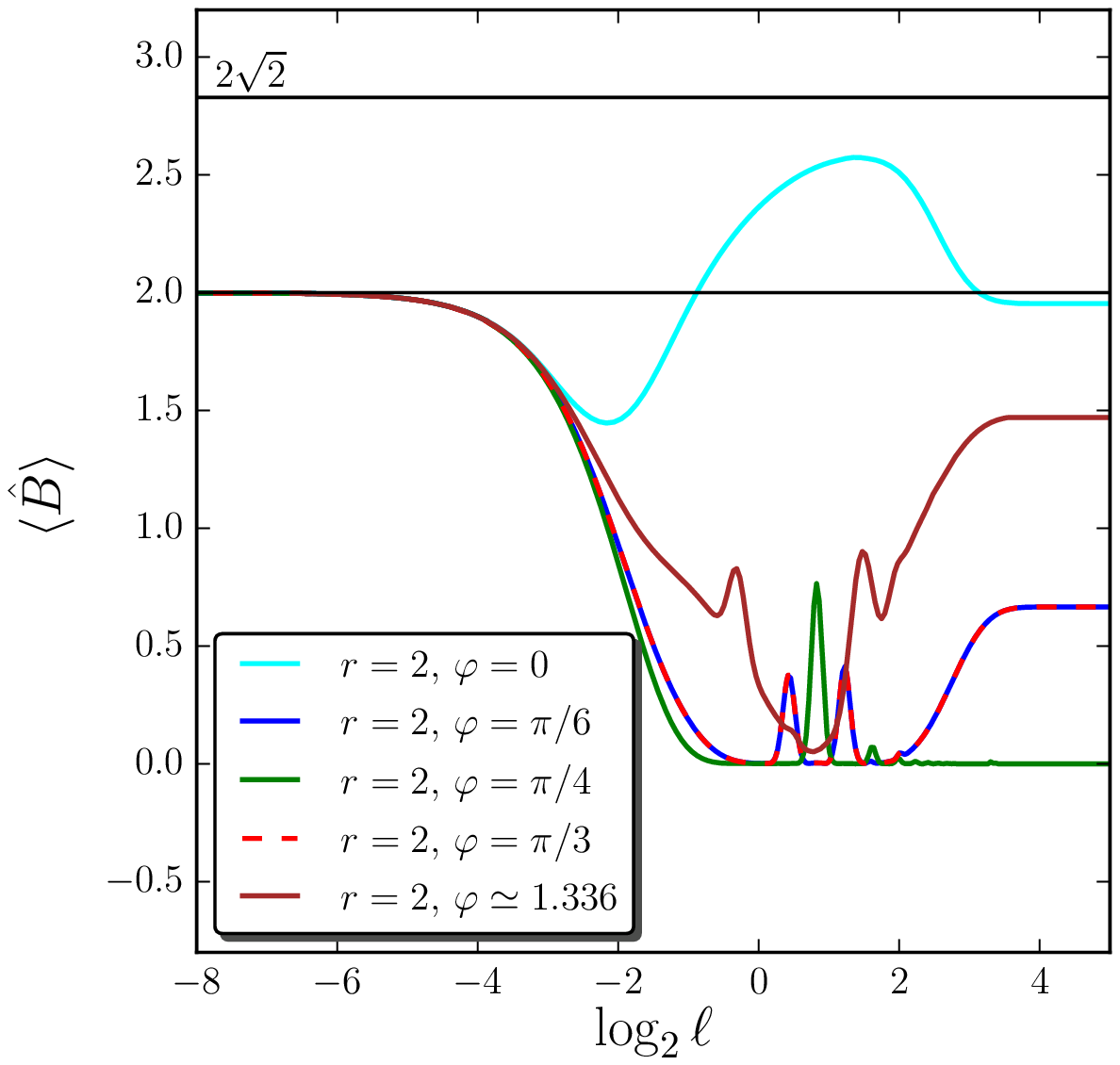}
\caption{Expectation value of the Bell operator $\hat{B}$ as a
  function of $\ell$, for $r=1$ (left panel) and $r=2$ (right panel),
  and different squeezing angles $\varphi$. As in
  \Fig{fig:exactBell_phi=0}, Bell's inequality violation threshold $\langle
  \hat{B} \rangle =2$ and the Cirel'son bound $\langle \hat{B} \rangle
  <2\sqrt{2}$ are displayed, see the two horizontal black lines. One
  can see that, in general, increasing $\varphi$ from $0$ tends to
  make Bell's inequalities violation more difficult to achieve. The situations where
  $\varphi=\pi/6$ and $\varphi=\pi/3$ give rise to exactly the same
  result (as a consequence of the invariance of $\langle \hat{B}
  \rangle$ under $\varphi\rightarrow \pi/2-\varphi$), which is why we
  have used a dashed line for $\varphi=\pi/3$.}
\label{fig:exactBell_differentphi}
\end{center}
\end{figure*}
Fourthly, the polar angles $\theta_{\bm n}$, $\theta_{\bm m}$,
$\theta_{\bm n}'$ and $\theta_{\bm m}'$ need to be carefully
chosen. The standard choice, made in \Ref{2004PhRvA..70b2102L}, is to
take $\theta_{\bm n}=0$, $\theta_{\bm m}=\pi/4$, $\theta_{\bm
  n}'=\pi/2$ and $\theta_{\bm m}'=-\pi/4$. However, this does not
correspond to the optimal configuration. By varying \Eq{eq:meanbell}
with respect to the four polar angles, one can show that the later is
given by $\theta_{\bm n}=0$, $\theta_{\bm n}'=\pi/2$ and $\theta_{\bm
  m}'=-\theta_{\bm m}$ with
\begin{align}
\label{eq:theta:optimal}
\theta_{\bm m}=\arctan \left[\frac{\langle
\hat{S}^{(1)}_x(\ell) \hat{S}^{(2)}_x(\ell)\rangle}
{\langle
\hat{S}^{(1)}_z(\ell) \hat{S}^{(2)}_z(\ell)\rangle}\right].
\end{align}
In the following, we always work with this choice unless
explicitly specified otherwise. With these angles, one has
\begin{align}
\langle \hat{B}\rangle  = 2\sqrt{\langle  
\hat{S}^{(1)}_z(\ell) \hat{S}^{(2)}_z(\ell)\rangle^2 
+ \langle  \hat{S}^{(1)}_x(\ell) \hat{S}^{(2)}_x(\ell) \rangle^2}\, .
\label{eq:B:optimizedAngles}
\end{align}
From this expression, the Cirel'son bound $\langle \hat{B}\rangle <
2\sqrt{2}$ is easily obtained since the two-point correlators of
the spin operators must be comprised between $-1$ and $1$ (having
eigenvalues $-1$ and $1$).

\par

In order to calculate concretely the expectation value of the Bell
operator, see \Eq{eq:meanbell}, we need to specify the quantum state
in which the system is placed. In this paper, we consider the two-mode
squeezed state
\begin{align}
  \vert\Psi_{2\, \mathrm{sq}}\rangle & = \frac{1}{\cosh r} \sum
  _{n=0}^{\infty} \ee^{-2in\varphi}\tanh ^n r \vert n_1,n_2\rangle ,
\label{eq:qstate}
\end{align}
where $r>0$ is the squeezing parameter and $\varphi$ is the squeezing
angle. So far only the case $\varphi=0$ was studied but, here, we
treat the most general situation. The ket $\vert n_1,n_2\rangle$
represents the state of the bipartite system such that the sub-systems
$(1)$ and $(2)$ have the same number of quantas, namely $n$ (not to be
confused with a situation where there would be $n_1$ quantas in the
first system and $n_2$ quantas in the second one; in our notations,
``$1$'' and ``$2$'' are labels for the two sub-systems).

The quantum state~(\ref{eq:qstate}) is exactly the state in which the
cosmological fluctuations are placed at the end of inflation, see for
instance \Refs{Grishchuk:1990bj,Martin:2007bw}. It is also worth
noticing that, in this context, $r\simeq 100$, a much larger number
than what can be achieved in the laboratory (typically a few). Since
we have in mind applications to cosmology, this reinforces the
motivation for deriving a large-squeezing limit as done in
\Sec{sec:largesqueezing}. It is well-known that the CMBR is the best
black body known in Nature~\cite{Fixsen:1996nj} since it is not
possible to reproduce a thermal spectrum at this level of accuracy in
the laboratory. In some sense, cosmological perturbations have a
similar property, since they are placed in a quantum state the
squeezing parameter of which is much larger than what can be obtained
in the laboratory. As a final comment, let us also notice that, when
$r$ goes to infinity, the state~(\ref{eq:qstate}) exactly tends
towards the EPR state.

It is also interesting to recall that the state~(\ref{eq:qstate}) has
a positive definite Wigner function~\cite{Martin:2015qta}. Bell
suggested~\cite{Bell:1987hh} that the non-negativity of the Wigner
function, which can therefore be interpreted as a stochastic
distribution, would prevent Bell's inequality violation. However, in
\Refs{2004PhLA..324..415G,2005PhRvA..71b2103R,2006FoPh...36..546R}, it
was shown that Bell's inequalities violation can occur, in particular
if the operators used are non-analytical in the dynamical variables,
which is precisely the case of the pseudo-spin operators used in this
work.

As mentioned before, calculating the expectation value of the Bell
operator implies to evaluate the two-point correlation functions of
the spin operators. Unfortunately, for the state~(\ref{eq:qstate}),
this cannot be done analytically and we have to rely on numerical
calculations. The details of those computations are explained in
\Sec{sec:spinPheno} and, here, we just quote the results.

Let us first describe the case where the squeezing angle vanishes, see
\Fig{fig:exactBell_phi=0}, where $\langle \hat{B}(\ell)\rangle $ is
represented versus $\log_2 \ell$ for different values of $r$. In the
small $\ell$ limit, we have shown in \Sec{sec:smallelllimit} that
$\langle \hat{S}^{(1)}_x(\ell) \hat{S}^{(2)}_x(\ell)\rangle
\rightarrow 1$ and $\langle \hat{S}^{(1)}_y(\ell)
\hat{S}^{(2)}_y(\ell)\rangle =\langle \hat{S}^{(1)}_z(\ell)
\hat{S}^{(2)}_z(\ell)\rangle \rightarrow 0$, hence
$\langle\hat{B}(\ell)\rangle\rightarrow 2$ from
\Eq{eq:B:optimizedAngles} (in \Fig{fig:exactBell_phi=0}, this result
is also valid for $r=5$ but, in that case, just happens for values of
$\ell$ smaller than those plotted in the figure).
In the large $\ell$ limit, we notice that
we also have a plateau, the value of which depends on the squeezing
parameter. It is shown in \Sec{sec:largeelllimit} that, in this limit,
$\langle \hat{S}^{(1)}_x(\ell) \hat{S}^{(2)}_x(\ell)\rangle =\langle
\hat{S}^{(1)}_y(\ell) \hat{S}^{(2)}_y(\ell)\rangle \rightarrow 0$ and
$\langle \hat{S}^{(1)}_z(\ell) \hat{S}^{(2)}_z(\ell)\rangle
\rightarrow 2/\pi \arctan[\sinh(2r)]$, this last formula having
already been obtained in \Ref{2004PhRvA..70b2102L}. From
\Eq{eq:B:optimizedAngles}, one then has
\begin{align}
\label{eq:largeell:Bell:varphiEq0}
\langle \hat{B}(\ell)\rangle \rightarrow \frac{4}{\pi}
\arctan\left[\sinh(2r)\right]
\end{align}
in this limit and one can check that this value fits
very well the ones observed in the plot.
Between the two plateaus, we observe a more complicated structure with
a bump. In \Sec{sec:appr:scheme}, we develop an approximation scheme
which is able to reproduce with very high accuracy the shape of this
bump. Of course the most striking feature is that, for values such
that $r \gtrsim 1.12$, one has violation of Bell's inequality. One
notices that, when $r$ becomes large, the Cirel'son bound is quickly
saturated but never crossed, which further checks the consistency of
our numerical computation. Let us also remark that the larger $r$, the
wider the region where Bell's inequality is violated.

\par

Let us now study the case of a non-vanishing squeezing angle. As
already mentioned, this is the first time that this is done. The
results are presented in \Figs{fig:exactBell_differentphi} for $r=1$
and $\varphi=0$, $\pi/6$, $\pi/4$, $\pi/3$ and $1.366$ (left panel) and $r=2$
(right panel) for the same values of $\varphi$. As explained in
\Sec{sec:spinPheno}, the cases where $\varphi $ does not belong to
$[0,\pi/4]$ can be easily deduced from the situations where $\varphi
\in [0,\pi/4]$ applying straightforward transformation rules. In
particular, it is shown that $\varphi$ and $\pi/2-\varphi$ give rise
to the same expectation value of the Bell operator, and one can check
that the curves for $\varphi=\pi/6$ and $\pi/3$ are indeed the same in
\Figs{fig:exactBell_differentphi}.
%
\begin{figure*}[t]
\begin{center}
\includegraphics[width=0.465\textwidth,clip=true]{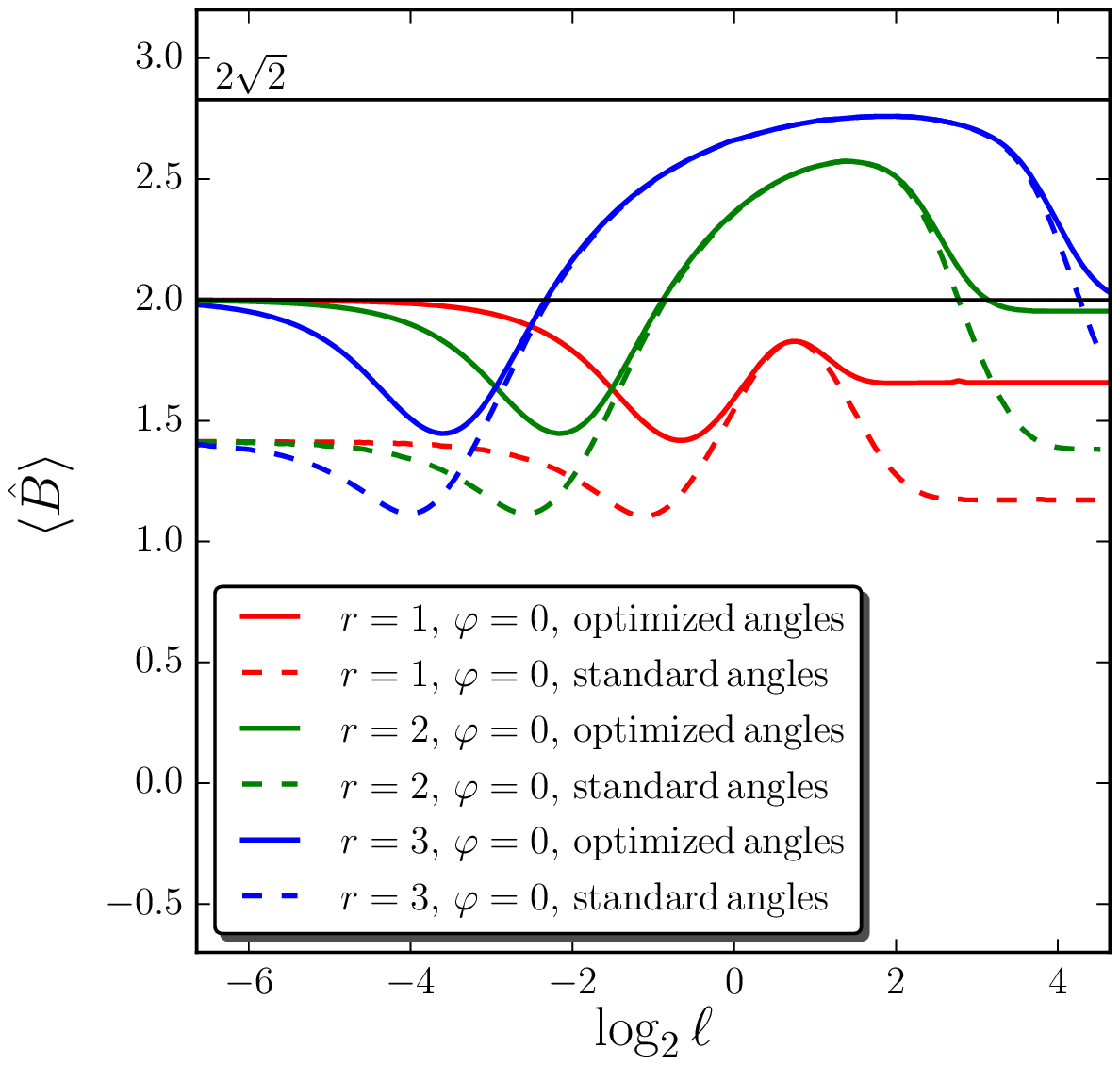}
\includegraphics[width=0.45\textwidth,clip=true]{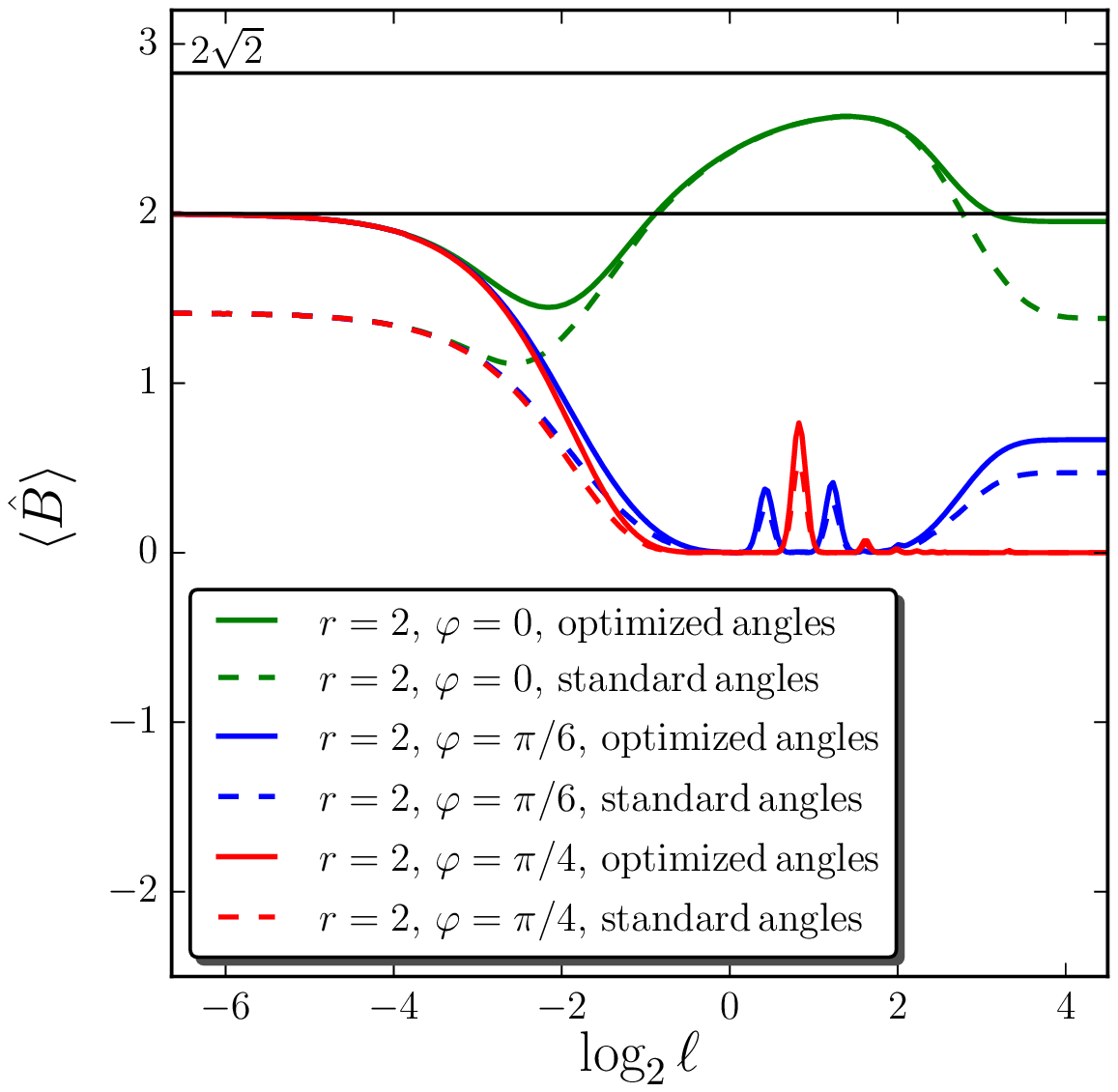}
\caption{Expectation value of the Bell operator $\hat{B}$ as a
  function of $\ell$, with the standard polar angles $\theta_{\bm
    n}=0$, $\theta_{\bm m}=\pi/4$, $\theta_{\bm n}'=\pi/2$ and
  $\theta_{\bm m}'=-\pi/4$ used in \Ref{2004PhRvA..70b2102L} notably
  (dashed lines), and in the optimal configuration where $\theta_{\bm
    n}=0$, $\theta_{\bm n}'=\pi/2$ and $\theta_{\bm m}'=-\theta_{\bm
    m}$ is given by \Eq{eq:theta:optimal} (solid lines). The left
  panel presents the result for different values of $r$ and
  $\varphi=0$, while in the right panel, $r=2$ and different values of
  $\varphi$ are used.}
\label{fig:angleBell}
\end{center}
\end{figure*}
In the small $\ell $ limit, one still has the plateau at $ \langle
\hat{B}(\ell)\rangle \simeq 2 $ in agreement with
\Sec{sec:smallelllimit} since this limit is independent of
$\varphi$. In the large $\ell$ limit, a plateau is also present, the
value of which depends this time both on $r$ and $\varphi$. In
\Sec{sec:largeelllimit}, we shown that $\langle \hat{S}^{(1)}_x(\ell)
\hat{S}^{(2)}_x(\ell)\rangle =\langle \hat{S}^{(1)}_y(\ell)
\hat{S}^{(2)}_y(\ell)\rangle \rightarrow 0$ and that the limit of
$\langle \hat{S}^{(1)}_z(\ell) \hat{S}^{(2)}_z(\ell)\rangle $ is given
by \Eq{eq:largeell:SzSz}. From \Eq{eq:B:optimizedAngles}, this gives
rise to
\begin{align}
\langle
\hat{B}(\ell)\rangle & 
{\rightarrow} 
\frac{4}{\pi}\arctan\left[\frac{\cos(2\varphi)\sinh(2r)}
{\sqrt{\cosh^2(2r)-\cos^2(2\varphi)\sinh^2(2r)}}\right], 
\label{eq:largeell:Bell}
\end{align}
which directly generalizes \Eq{eq:largeell:Bell:varphiEq0}.  Between
the two plateaus, one notices oscillatory patterns with various peaks
and dips. This behavior is more complicated than for the case
$\varphi=0$ where one just has a simple bump. But the most important
difference between these configurations is of course that it seems
more difficult to violate Bell's inequalities when $\varphi \neq
0$. For instance, for $r=2$ and $\varphi=0$, there is a regime where $
\langle \hat{B}(\ell)\rangle \ge 2$ while, for the other values of
$\varphi \neq 0$ considered, this is not the case.

\par

\begin{figure}[t]
\begin{center}
\includegraphics[width=0.45\textwidth,clip=true]{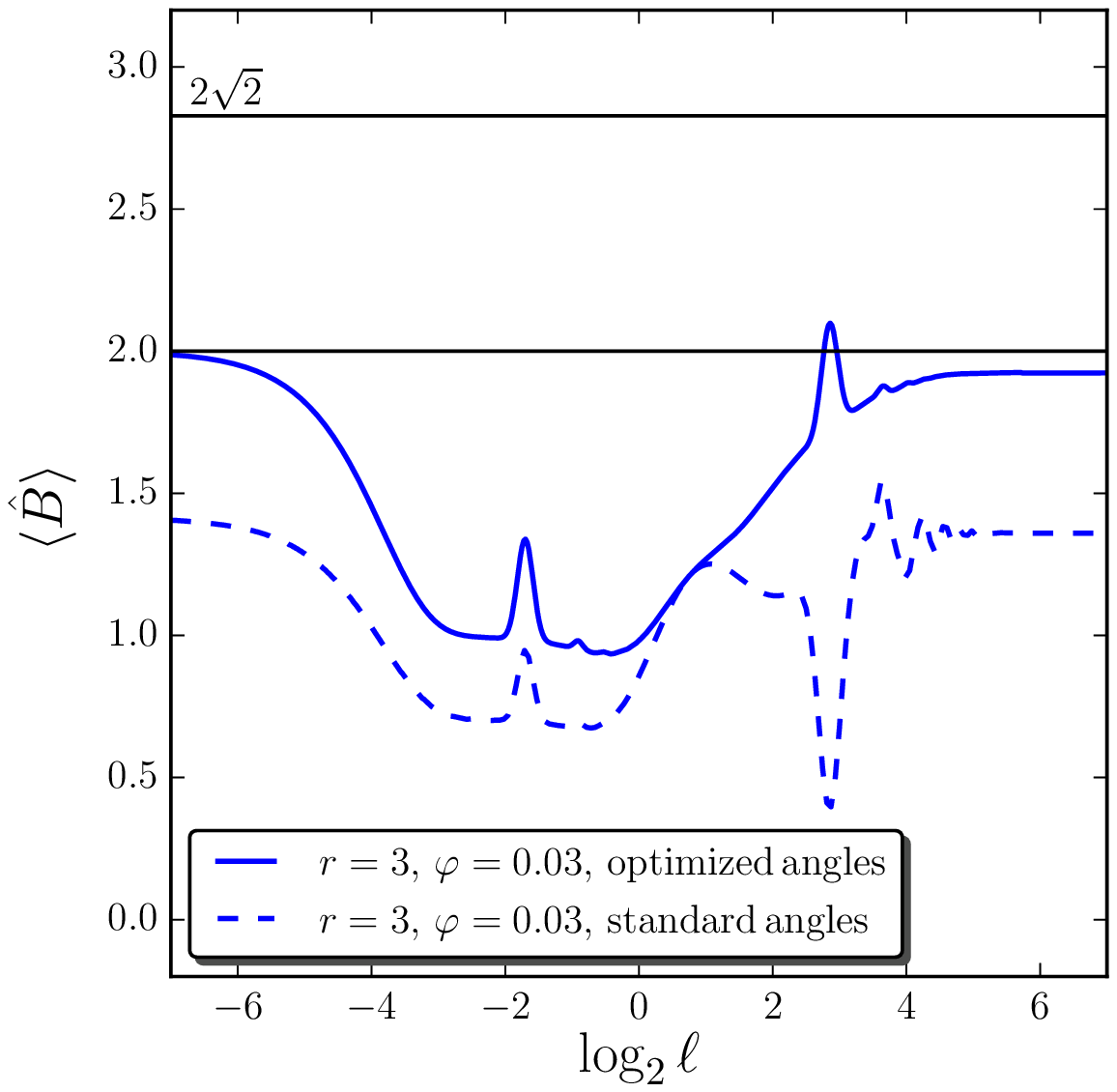}
\caption{Expectation value of the Bell operator $\hat{B}$ as a
  function of $\ell$ for $r=3$ and $\varphi=0.03$, with the standard
  polar angles $\theta_{\bm n}=0$, $\theta_{\bm m}=\pi/4$,
  $\theta_{\bm n}'=\pi/2$ and $\theta_{\bm m}'=-\pi/4$ used in
  \Ref{2004PhRvA..70b2102L} notably (dashed line), and in the optimal
  configuration where $\theta_{\bm n}=0$, $\theta_{\bm n}'=\pi/2$ and
  $\theta_{\bm m}'=-\theta_{\bm m}$ is given by \Eq{eq:theta:optimal}
  (solid line). Bell's inequalities violation can be obtained in this
  state, but is detected only when working with the optimal angles.}
\label{fig:featureBell_opt_standard}
\end{center}
\end{figure}

Finally, it is interesting to study the impact of working with
optimized angles rather than with the standard choice, $\theta_{\bm
  n}=0$, $\theta_{\bm m}=\pi/4$, $\theta_{\bm n}'=\pi/2$ and
$\theta_{\bm m}'=-\pi/4$. In \Fig{fig:angleBell}, we have represented
$ \langle \hat{B}(\ell)\rangle $ for different values of $r$ and
$\varphi=0$ (left panel) and $r=2$ and different values of $\varphi$
(right panel) in the cases of standard and optimized angles. One
notices that, of course, the optimized result is always above the
standard one. One also remarks that, although the two cases are
similar in the vicinity of the bump, they strongly differ for the
small and large $\ell$ plateaus. However, one could argue that working
with optimized angles is, after all, not that important since around
the region where Bell's inequality is violated the standard angles
approximately lead to the same result. But this is not always the
case as revealed, for instance, in \Fig{fig:featureBell_opt_standard}
where we have plotted $\langle \hat{B}(\ell)\rangle $ for $r=3$ and
$\varphi=0.03$. We see that for the standard choice, no violation
occurs while, for the optimized angles, the presence of a ``feature''
enables to cross the $ \langle \hat{B}(\ell)\rangle =2$
threshold. Of course, this is only a specific case but, as a matter of
fact, we have checked that it happens in many other situations.


%
\section{Bell's Inequality Violation in Squeezing  Parameters Space}
\label{sec:BellViolation}
\begin{figure*}[t]
\begin{center}
\includegraphics[width=0.65\textwidth,clip=true]{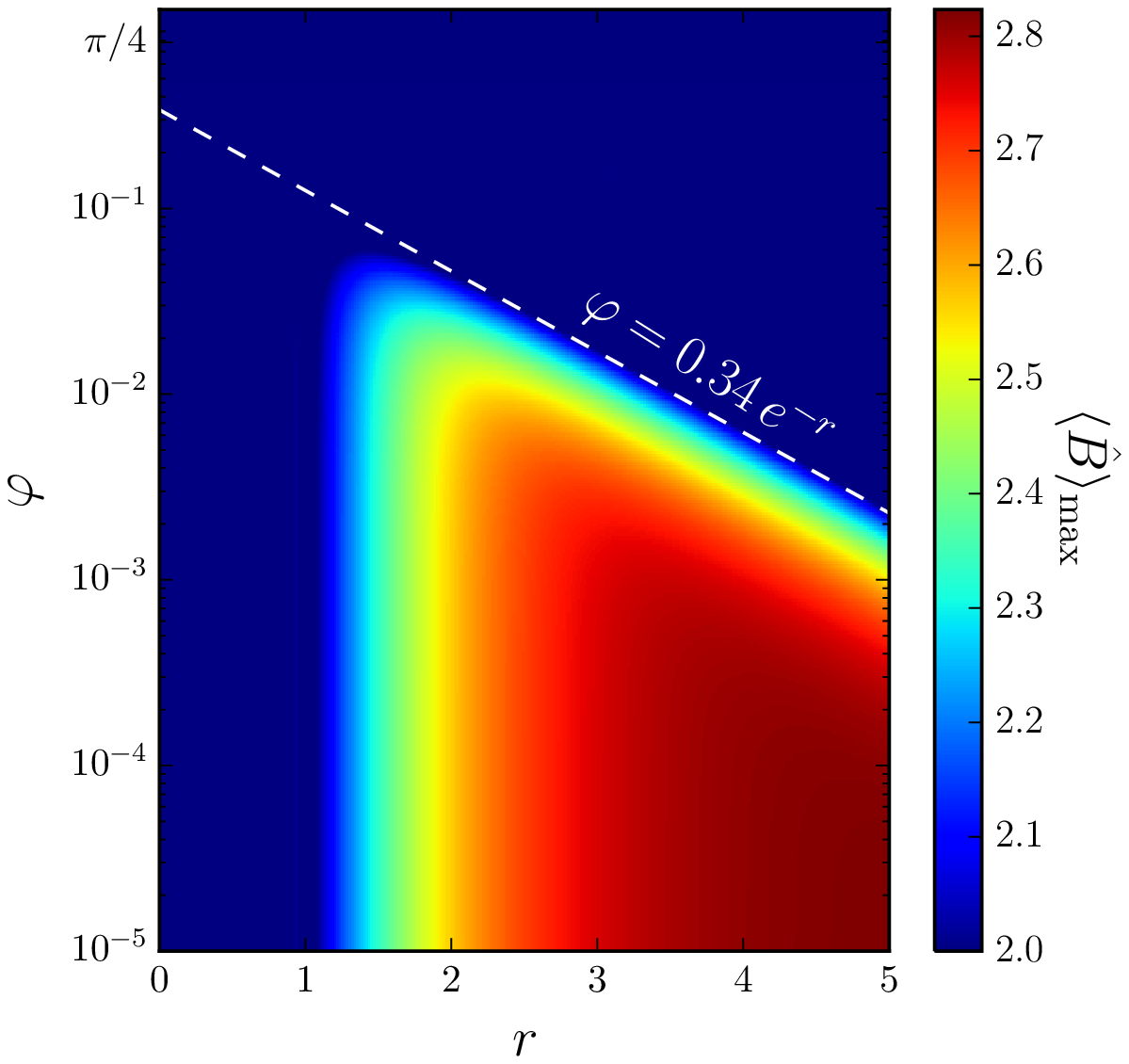}
\caption{Maximum Bell's operator expectation value $\langle \hat{B}
  \rangle_\mathrm{max}$ (where extremization has been performed over
  $\ell$ at the bump) as a function of the squeezing parameters $r$
  and $\varphi$. The dashed white line stands for $\varphi\propto
  0.34\ee^{-r}$, see \Eq{eq:violationCriterion:largesqueezing}, which
  delimits Bell's inequality violation domain $\langle \hat{B} \rangle_\umax
  >2$ in the large squeezing limit. The map is displayed for $\varphi
  \in [0,\pi/4]$ only, since any other value of $\varphi$ can be
  inferred from such configurations making use of
  \Eqs{eq:dual:corr:SzSz}-(\ref{eq:dual:corr:SySy}), and $r\in[0,5]$,
  since for larger values of $r$ the result is very accurately
  provided by \Fig{fig:largesqueezing}.}
\label{fig:map}
\end{center}
\end{figure*}

In the previous section, we have explicitly demonstrated that, for
some values of $r$ and $\varphi$, the spin operators can lead to
violation of Bell's inequalities. The next obvious question is for
which values of $r$ and $\varphi$ such a violation can be obtained. To
answer it, in \Fig{fig:map}, we present a map of $ \langle
\hat{B}\rangle_{\umax} $ in the $(r,\varphi)$ space. This map was
obtained by constructing a grid of $300\times 300$ points in the
$(r,\varphi)$ space and, for each value of $r$ and $\varphi$,
determining the value of $\langle \hat{B}\rangle $ at the top of the
bump (when this value was found to be smaller than $2$, we have put $
\langle \hat{B}\rangle_{\umax}=2$ since we know that, in the small
$\ell$ limit, $ \langle \hat{B}\rangle\rightarrow 2$).

\begin{figure*}[t]
\begin{center}
\includegraphics[width=0.45\textwidth,clip=true]{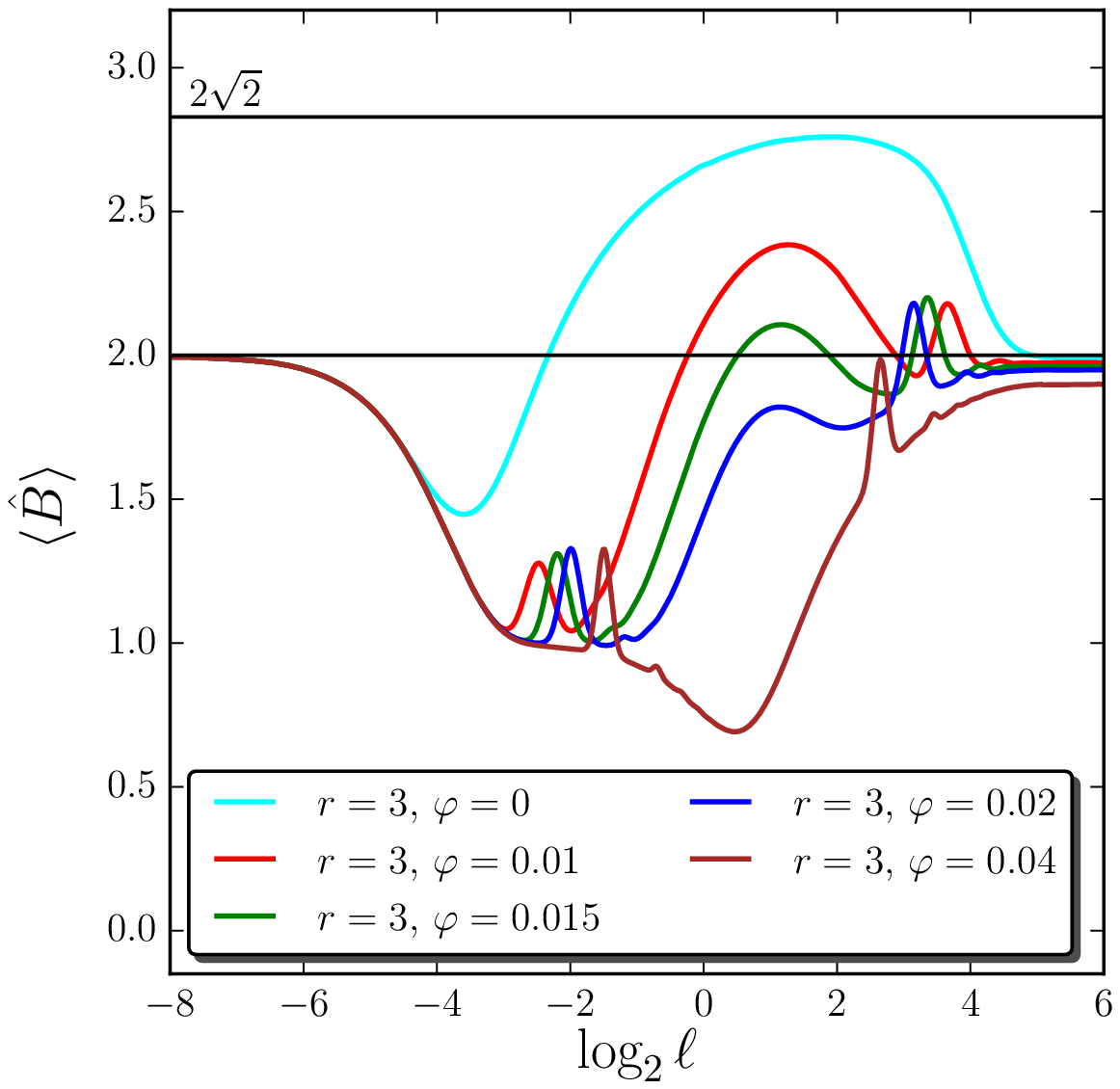}
\includegraphics[width=0.45\textwidth,clip=true]{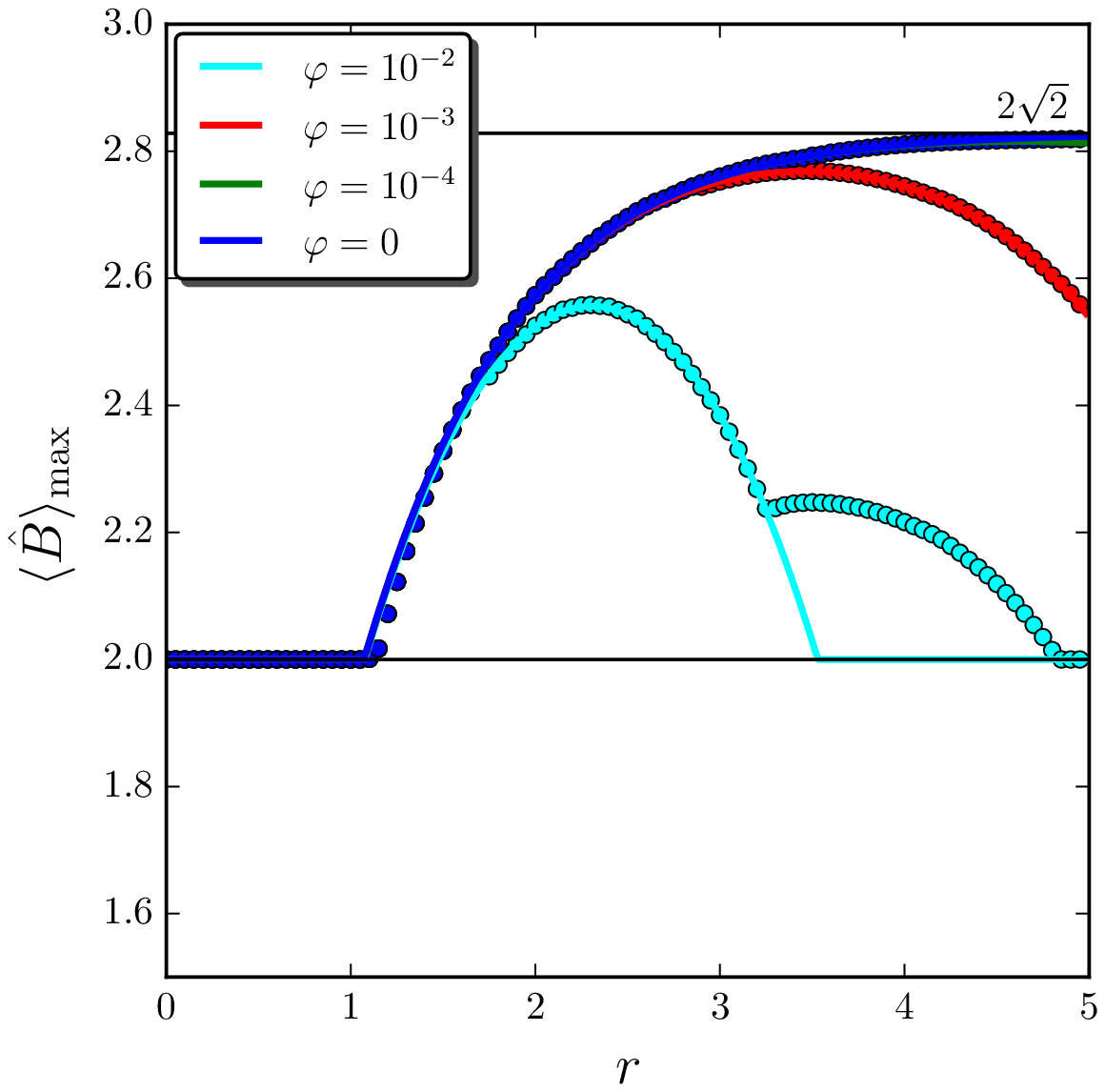}
\caption{Left panel: Expectation value of the Bell operator as a
  function of $\ell$ for $r=3$ and different values of $\varphi$. When
  $\varphi\neq 0$, on top of the usual bump, several features are
  present, that can lead to stronger Bell's inequalities violation than in the bump
  only (see \eg the case $r=3$ and $\varphi=0.015$ or
  $\varphi=0.02$). Right panel: Maximal expectation value of the Bell
  operator, where extremization has been performed over the bump
  making use of the approximation scheme developed in
  \Sec{sec:appr:scheme} (solid lines), and over the entire profile
  with the numerical results presented in \Sec{sec:spinPheno}
  (circles). When a feature leads to stronger violation than the bump,
  as in the left panel for $r=3$ and $\varphi=0.015$ for instance, the
  two results are obviously different, otherwise the agreement is
  excellent.}
\label{fig:checkBell}
\end{center}
\end{figure*}

Exploring the squeezing parameter space can be, for some values of $r$
and $\varphi$, numerically very demanding. This is the reason why we
have in fact determined the value of the Bell operator at the bump by
means of the approximation developed in \Sec{sec:Appr}. In that
section, we have indeed shown that it always reproduces the bump very
accurately. We have further checked that this method is efficient by
comparing it with numerical results in the right panel of
\Fig{fig:checkBell}. In this plot, the solid lines correspond to the
analytical approximation of \Sec{sec:Appr} while the circles stand for
the numerical results presented in \Sec{sec:spinPheno}, where the
maximum of $ \langle \hat{B}\rangle$ is found in a given, sufficiently
wide, range of $\log_2 \ell$. For $\varphi=0$ for instance, they match
very well and this validates our approach.

We notice, however, that for the case $\varphi=10^{-2}$ and $r\gtrsim
3.2$, the numerical method seems to predict results that strongly
deviate from those obtained by means of the analytical
approximation. The former clearly predicts $ \langle
\hat{B}\rangle_{\umax} >2$ while the latter indicates that $ \langle
\hat{B}\rangle_{\umax} =2$. This can be understood studying the left
panel of \Fig{fig:checkBell} where we have represented $ \langle
\hat{B}\rangle $ for $r=3$ and $\varphi=0$, $0.01$, $0.015$, $0.02$
and $0.04$. For $\varphi=0$ and $\varphi=0.01$, we see that the
maximum of $ \langle \hat{B}\rangle $ is located at the bump. However,
as soon as $\varphi\neq 0$, we notice the presence of features, \ie
secondary, smaller, bumps, located at smaller or larger values of
$\log_2 \ell$. For $\varphi\ge 0.015$, the first right feature
actually corresponds to the maximum of $ \langle \hat{B}\rangle$ which
is, therefore, no longer given by the bump. For $\varphi=0.015$, the
bump and this feature both correspond to situations where Bell's
inequality is violated but violation is stronger in the feature than
in the bump. For $\varphi\ge 0.02$, Bell's inequality is only violated
at the feature and no longer at the bump. Clearly, when this happens,
our analytical method breaks down and can no longer identify the
maximum of $ \langle \hat{B}\rangle$ over the full range of $\ell$.
In this sense, the map in \Fig{fig:map} only provides a sufficient,
but not necessary, condition for violating Bell's inequalities in the
squeezing parameter space since only the maximal value of $ \langle
\hat{B}\rangle$ over the bump is displayed.

\par

Let us now discuss the physical implications of the map given in
\Fig{fig:map}. Firstly, in order to obtain violation of Bell's
inequalities, we see that there is a threshold in $r$, namely
$r\gtrsim 1.12$. Of course, the larger $r$, the larger the
violation. This is consistent with the fact~\cite{Martin:2015qta} that
the squeezing parameter $r$ measures the entanglement level of the
state. Secondly, we notice that $ \langle \hat{B}\rangle_{\umax}$
decreases rapidly with $\varphi$. In other words, for a value of $r$
such that Bell's inequality violation is obtained for $\varphi=0$,
only very small non-vanishing squeezing angles still lead to
violation. Moreover, the violation is always maximal for $\varphi=0$
and can only be less important for $\varphi\neq 0$. Thirdly, we
clearly notice in \Fig{fig:map} that the width of the $\varphi$
interval for which one has violation decreases with $r$. In fact, as
shown in \Sec{sec:largesqueezing}, one can demonstrate that for
sufficiently large $r$, $ \langle \hat{B}\rangle_{\umax} $ depends on
$\varphi \ee^{r}$ only, and that Bell's inequality violation occurs
provided
\begin{align}
\varphi < 0.34\ee^{-r}.
\end{align}
This law is very important since it provides a simple criterion for
Bell's inequality violation in a regime (large squeezing) that cannot
be reached numerically.

\par

To summarize, Bell's inequality violation is obtained if two
conditions are met: firstly, $r$ must be large, quite an obvious
conclusion indeed (and one notices that when $r\gtrsim 4$, the
Cirel'son bound is completely saturated and there would be no point in
going much further); secondly, $\varphi$ must be sufficiently small,
and its fine-tuning close to $0$ increases with $r$ (in the most
favorable case, namely close to the threshold $r\simeq 1.12$, one
still must have $\varphi \lesssim 0.05$).
\section{Discussion and Conclusion}
\label{sec:conclusion}

Let us now summarize our main results. Following the procedure of
\Ref{2004PhRvA..70b2102L}, we have introduced spin operators for
continuous-variable systems, from which a Bell operator can be
constructed. We have then calculated the expectation value of this
Bell operator in a two-mode squeezed state, allowing for a
non-vanishing squeezing angle $\varphi$.

This generalizes the previous results of \Ref{2004PhRvA..70b2102L} in a
direction that is necessary to follow if one wants to apply the
present construction to situations where the role of the continuous
variable is played by the (Fourier) amplitude of a quantum field, and
where $\varphi$ is a non-vanishing quantity one does not have
experimental control on. This is for instance the case of cosmic
inflationary perturbations, which we plan to study in future
publications~\cite{MV}. We have found that the observables involved in the
present calculation are highly sensitive on $\varphi$ and that,
compared to the situation $\varphi=0$, very different results can be
obtained even for tiny, non-vanishing values of $\varphi$. Actually,
if one needed to, this suggests that the spin operators discussed in
this paper might provide a way to measure $\varphi$ very accurately.
We have also optimized the polar angles defining the direction of the
Bell operator, and showed that in some cases, this procedure is
necessary to properly account for Bell's inequalities violation.

Depending on the values of the squeezing parameters $r$ and $\varphi$,
the numerical evaluation of the Bell operator expectation value can be
numerically very expensive, if not impossible. For this reason, we
have designed two dual schemes of approximation presented in
\Secs{sec:spinPheno} and~\ref{sec:varphi=pi/2} respectively, that
allow one to explore the entire squeezing parameters space, and to
gain some analytical insight on the results. In particular, a map of
the maximal Bell's operator expectation value was provided in
\Fig{fig:map}, that can serve as a useful guide to find the optimal
squeezing parameters values for a given experimental design. It was
found that Bell's inequalities violation occurs provided $r$ is
sufficiently large and $\varphi$ sufficiently small. More precisely,
it was shown that in the large squeezing limit, Bell's inequalities
violation is obtained if $\varphi \lesssim 0.34\, \ee^{-r}$.

At this stage, it is important to notice that although one does not
necessarily have experimental control on $\varphi$, one is a priori
free to choose the pseudo-spin operators with respect to another
direction in phase space than the position $Q$ considered so far. In
fact, in \Sec{sec:PhaseSpaceRotation}, it is shown that if one
performs a rotation in phase space with angle $\varphi$ and introduces
$\overline{Q}_i=\cos \varphi \bar{Q}_i+\sin \varphi P_i$ and
$\overline{P}_i=\cos\varphi P_i-\sin \varphi Q_i$, with $i=1,2$, then
the squeezing angle of the resulting wavefunction $\Psi_{\rm 2\,
  sq}(\overline{Q}_1,\overline{Q}_2)$ vanishes. As a consequence, if
one defines the pseudo-spin operators with respect to $\overline{Q}$
instead of $Q$ [that is to say, if one replaces $Q$ by $\overline{Q}$
in \Eqs{eq:projectoru:def}-(\ref{eq:defsplus})], one obtains the same
results as the ones derived above for $\varphi=0$. Since we have shown
that vanishing squeezing angles lead to maximal Bell's inequalities
violation, another important result of this work is therefore that the
choice of pseudo-spin operators orientation that maximises Bell's
inequalities violation is the one aligned with the wavefunction
squeezing angle. However, the squeezing angle is not necessarily known
to the observer and it can even be a complicated time varying quantity,
as is the case of cosmological perturbations during
inflation~\cite{Martin:2012pea,Martin:2012ua}. 

Finally, let us quickly sketch the procedure one would have to follow
to concretely measure the spin operators introduced in \Sec{sec:spin},
as it highlights another crucial difference coming from taking
$\varphi\neq 0$. Since $[\hat{Q},\hat{S}_z]=0$, the measurement of
$\hat{S}_z$ is rather straightforward and can be performed by
measuring the position operator $\hat{Q}$ itself. In practice indeed,
from a given realization $Q$ of $\hat{Q}$, one simply needs to
identify in which interval $[n\ell,(n+1)\ell]$ the number $Q$ lies,
and the result is given by $(-1)^n$. Another way of seeing that
$\hat{S}_z$ can be measured by measuring $\hat{Q}$ only is to look at
\Eq{eq:correlZ}, where $\langle
\hat{S}_z^{(1)}(\ell)\hat{S}_z^{(2)}(\ell) \rangle$ relies of the
modulus of the wavefunction only, $\vert\Psi(Q_1,Q_2)\vert$. By
repeating measurements of $(Q_1,Q_2)$, the squared modulus of the
wavefunction can be inferred, hence $\langle
\hat{S}_z^{(1)}(\ell)\hat{S}_z^{(2)}(\ell) \rangle$. Since
$[\hat{Q},\hat{S}_x]\neq 0$, measuring $S_x$ is more involved and
cannot, in general, be performed by measuring position only. This can
be seen at the level of \Eqs{eq:S+S+:Psi} and~(\ref{eq:S-S-:Psi})
where $\langle \hat{S}_x^{(1)}(\ell)\hat{S}_x^{(2)}(\ell) \rangle$
does not only depend on the wavefunction modulus, but also relies on
its relative phase between $(Q_1,Q_2)$ and
$(Q_1\pm\ell,Q_2\pm\ell)$. In practice, this means that position
measurements are not enough, and that phase information must be obtained
by measuring \eg the momentum, hence reconstructing the modulus of the
wavefunction's Fourier transform, or more generally using any state
tomography protocol~\cite{2011arXiv1112.3575L}. There is nonetheless one
exception, namely the case $\varphi=0$. From \Eq{eq:qstateposition},
one can check that the phase of the wavefunction is a constant in (and
only in) this situation. This shows that, if
$\varphi=0$, all spin correlators can be obtained from position
measurements only. In this case however, the wavefunction must be known to have a constant phase, and the practical verification of this assumption may not always be trivial.

This issue is important in situations where the information about the
momentum is hidden from us, as is the case for cosmic inflationary
perturbations for instance~\cite{Martin:2015qta}. The results presented in the present paper
show that in such situations, the value taken by $\varphi$ is crucial
for two reasons: first, it defines the possibility to carry out
Bell-type experiments from position measurements only, and second, it
determines whether Bell's inequalities can be violated and at which
level.
\section*{ACKNOWLEDGEMENTS}
This work is supported by STFC Grants No. ST/K00090X/1 and No. ST/L005573/1.\\
%
\appendix
\onecolumngrid
\section{Spin Operators Correlation Functions in a Two-Mode Squeezed State}
\label{sec:spinPheno}
In this first appendix, we explain how the correlation functions of
the spin operators introduced in \Sec{sec:spin} can be evaluated in a
two-mode squeezed state.  As explained in \Sec{sec:BellInequality}, we
consider a bipartite system the Hilbert space ${\cal E}$ of which is
of the form ${\cal E}={\cal E}_1\otimes {\cal E}_2$. Each subsystem is
a continuous variable system and the corresponding continuous
variables are noted $Q_1$ and $Q_2$. The quantum state in which this
bipartite system is placed is taken to be a two-mode squeezed state,
see \Eq{eq:qstate}. In position space, this can be expressed as
\begin{align}
\Psi_{\rm 2\, sq}\left(Q_1,Q_2\right)
& =\left \langle Q_1,Q_2\left \vert 
\sum _{n=0}^{\infty}
\ee^{-2in\varphi}\frac{\tanh ^n r}{\cosh r}  \right \vert n_1,n_2
\right \rangle 
= \frac{\ee^{-(Q_1^2+Q_2^2)/2}}{\sqrt{\pi} \cosh r}
\sum _{n=0}^{\infty}\frac{\ee^{-2in\varphi}}{2^nn!}
\tanh ^n (r)
{\rm H}_n(Q_1){\rm H}_n(Q_2),
\end{align}
where $r$ is the squeezing parameter, $\varphi$ the squeezing angle,
and ${\rm H}_n$ is a Hermite polynomial of order $n$. This expression
can be simplified,\footnote{ One can use the
  formula~\cite{Gradshteyn:1965aa}
\begin{align}
\sum _{n=0}^{\infty}\frac{w^n}{n!}{\rm H}_n(Q_1){\rm H}_n(Q_2)
=\frac{1}{\sqrt{1-4w^2}}
\exp\left\{\frac{2w\left[2w\left(Q_1^2+Q_2^2\right)
-2Q_1Q_2\right]}{4w^2-1}\right\},
\label{eq:2sq:simp}
\end{align}
with $w=\ee^{-2i\varphi}\tanh r /2$.
}
and one obtains
\begin{align}
\label{eq:qstateposition}
\Psi_{\rm 2\, sq}\left(Q_1,Q_2\right)
=&
\frac{\ee^{A\left(Q_1^2+Q_2^2\right)
-BQ_1Q_2}}{\cosh r\sqrt{\pi}\sqrt{1-\ee^{-4i\varphi}\tanh^2 r}}
,
\end{align}
where $A(r,\varphi)$ and $B(r,\varphi)$ are functions of $r$ and
$\varphi$ only, explicitly
\begin{align}
A(r,\varphi) &\equiv \frac{\ee^{-4i\varphi}\tanh^2r+1}{2(
\ee^{-4i\varphi}\tanh^2r-1)}, \quad
B(r,\varphi) \equiv \frac{2\ee^{-2i\varphi}\tanh r}
{\ee^{-4i\varphi}\tanh^2r-1}.
\end{align}
When there is no squeezing, $r=0$, one has $A=-1/2$ and $B=0$. In that
case, the state of the system becomes factorizable and the two
subsystems evolve independently, each one being placed in a Gaussian
state.
\subsection{Correlation Function $\langle \Psi_{2\, {\rm sq}}\vert \hat{S}^{(1)}_z(\ell) \hat{S}^{(2)}_z(\ell)\vert \Psi_{2\, {\rm sq}}\rangle$}
\label{sec:spinPheno:z}
\begin{figure}[t]
\begin{center}
\includegraphics[width=0.45\textwidth,clip=true]{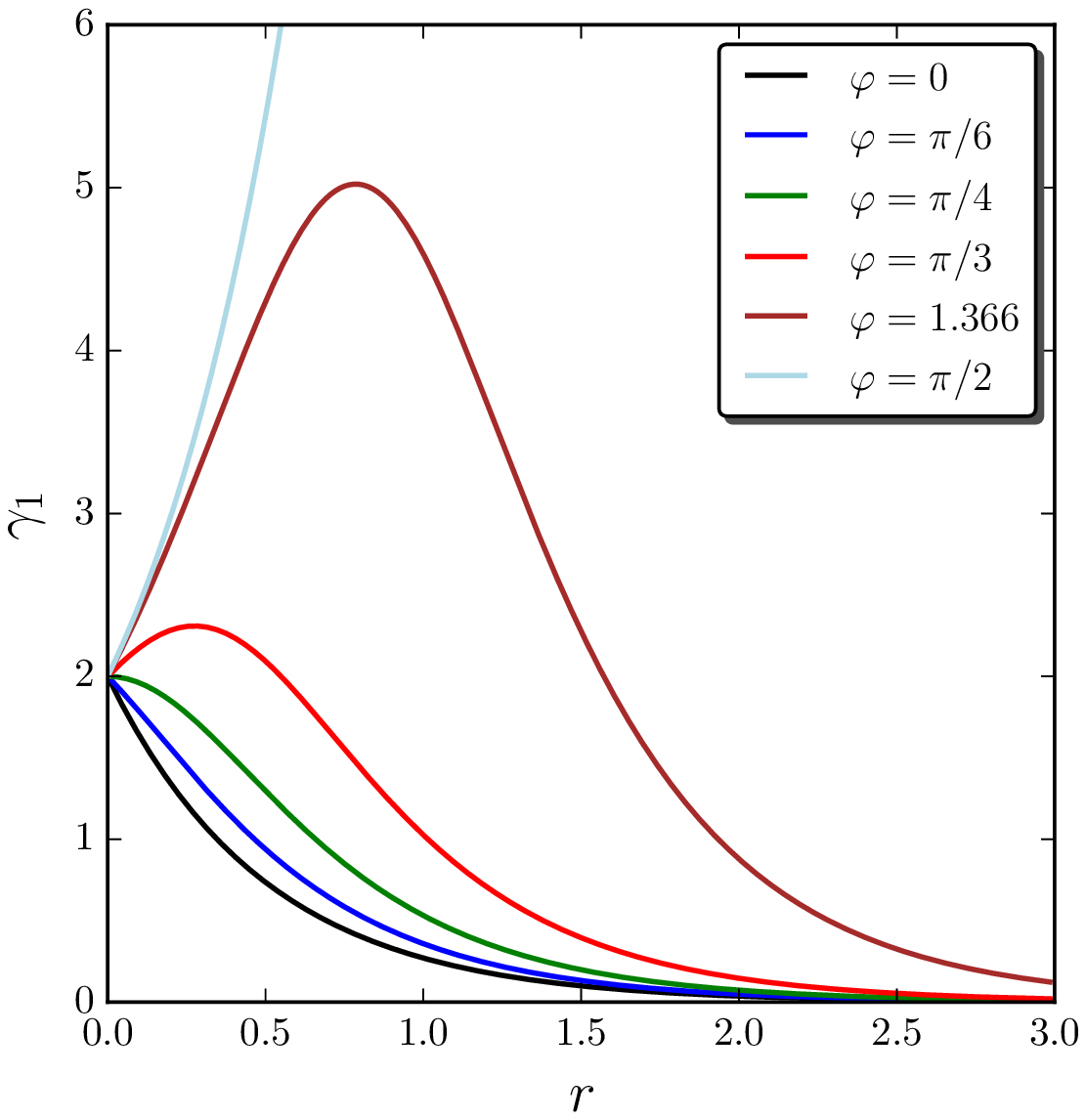}
\includegraphics[width=0.45\textwidth,clip=true]{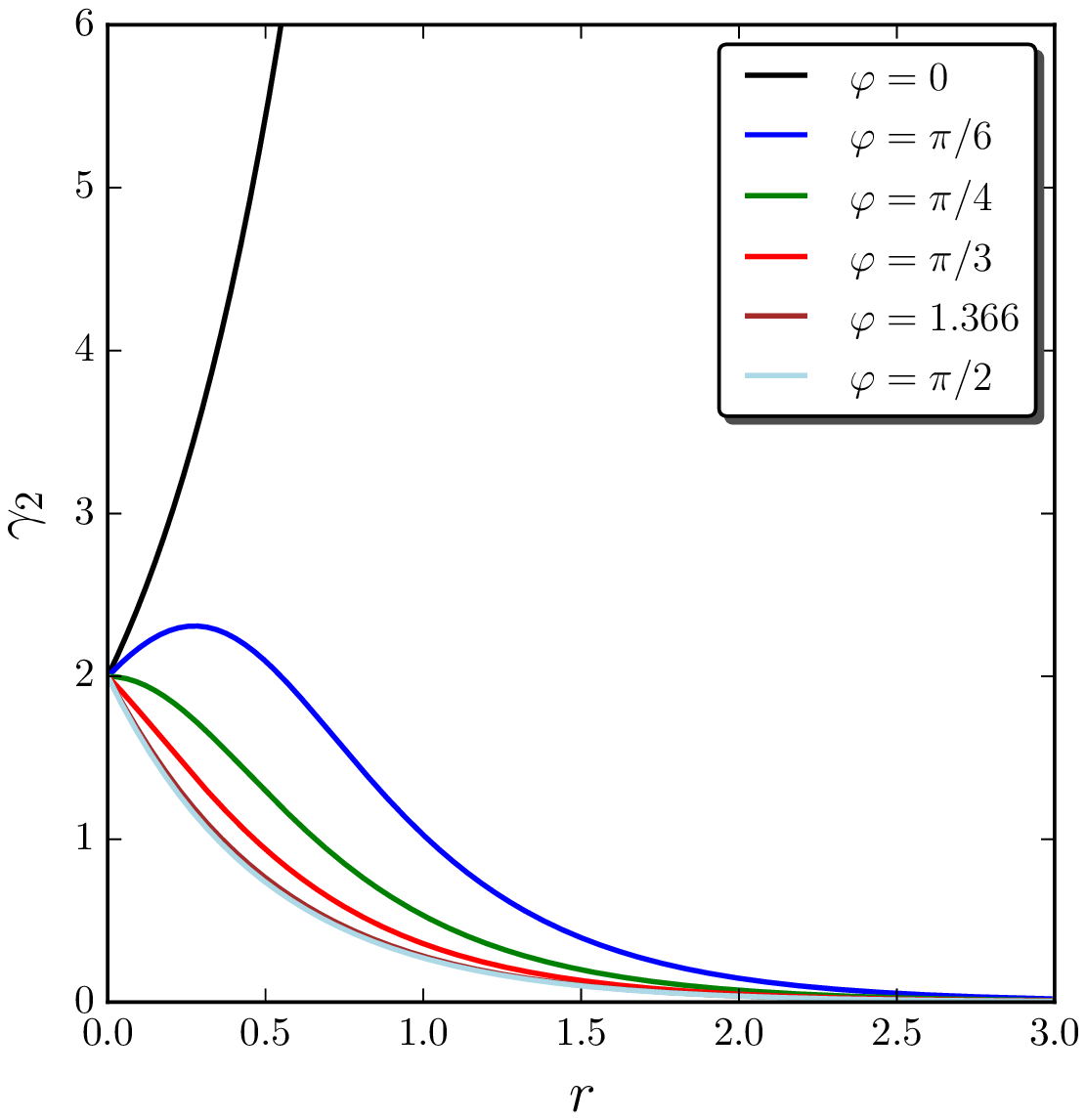}
\caption{$\gamma_1$ (left panel) and $\gamma_2$ (right panel) defined in \Eqs{eq:gamma1:def} and~(\ref{eq:gamma2:def}), as a function of $r$, for $\varphi=0$, $\pi/6$, $\pi/4$, $\pi/3$, $\pi/2.3$ and $\pi/2$.}
\label{fig:gamma12}
\end{center}
\end{figure}
We now turn to the calculation of the spin correlation functions.  Let
us first consider the operator~(\ref{eq:defsz2}) and calculate its
two-point correlation function in the two-mode squeezed
state~(\ref{eq:2sq:simp}). Straightforward manipulations lead to the
following expression
\begin{align}
  \langle \Psi_{2\, {\rm sq}}\vert \hat{S}^{(1)}_z(\ell)
  \hat{S}^{(2)}_z(\ell) \vert \Psi_{2\, {\rm sq}}\rangle &=\sum
  _{n=-\infty}^{\infty} \sum_{m=-\infty}^{\infty}(-1)^{n+m} \int
  _{n\ell}^{(n+1)\ell}{\rm d}Q_1 \int _{m\ell}^{(m+1)\ell}{\rm d}Q_2
  \vert \Psi_{\rm 2\, sq}\left(Q_1,Q_2\right)
  \vert ^2 \\
  & = \frac{1}{\pi \cosh^2 r} \frac{1}{\sqrt{ \tanh^4 r-2\tanh^2 r
      \cos(4\varphi)+1}}\sum _{n=-\infty}^{\infty}
  \sum_{m=-\infty}^{\infty}(-1)^{n+m}Z_{n,m},
\label{eq:correlZ}
\end{align}
where the quantity $Z_{n,m}$ is defined by
\begin{align}
\label{eq:defInm}
Z_{n,m}\equiv 
\int _{n\ell}^{(n+1)\ell}{\rm d}Q_1
\int _{m\ell}^{(m+1)\ell}{\rm d}Q_2
\ee^{(A+A^*)\left(Q_1^2+Q_2^2\right)
-(B+B^*)Q_1Q_2}\, .
\end{align}
Let us perform the change of integration variables: $Q_1=u+v$ and
$Q_2=u-v$. In this way, the double integration can be expressed as the
product of two one-dimensional integrals. It follows that the quantity
$Z_{n,m}$ is now given by
\begin{align}
\label{eq:Z1:def}
Z_{n,m} &=2
\int _{(n+m)\ell/2}^{(n+m+1)\ell/2}{\rm d}u
\ee^{-\gamma_1u^2}
\int _{n\ell -u}^{u-m\ell}{\rm d}v
\ee^{-\gamma_2v^2}
+
2
\int _{(n+m+1)\ell/2}^{(n+m+2)\ell/2}{\rm d}u
\ee^{-\gamma_1u^2}
\int _{u-(m+1)\ell}^{(n+1)\ell-u}{\rm d}v
\ee^{-\gamma_2 v^2}
\equiv Z_{n,m}^{(1)}+Z_{n,m}^{(2)},
\end{align}
with
\begin{align}
\label{eq:gamma1:def}
\gamma_1&\equiv-2\left(A+A^*\right)+\left(B+B^*\right)
=\frac{2}{\cosh(2 r)+\cos(2\varphi)\sinh(2r)}, \\
\gamma_2&\equiv -2\left(A+A^*\right)-\left(B+B^*\right)
=\frac{2}{\cosh(2 r)-\cos(2\varphi)\sinh(2r)}.
\label{eq:gamma2:def}
\end{align}
Let us notice that $\gamma_1(r,\varphi)$ and $\gamma_2(r,\varphi)$ are
always positive definite. They are displayed in \Fig{fig:gamma12} as a
function of $r$ and for different values of $\varphi$.  One can check
that $\gamma_{1,2}(r,\varphi)=\gamma_{1,2}(r,\varphi+\pi)$ but also
that $\gamma_{1,2}(r,\pi/2+\varphi)=\gamma_{1,2}(r,\pi/2-\varphi)$.
Since the two-point correlator of $\hat{S}_z$ only depends on
$\gamma_1$ and $\gamma_2$, it can be studied in the range
$\varphi\in[0,\pi/2]$ only as its value for any other $\varphi$ can be
inferred using these symmetries.

We have just seen that the quantities $Z_{n,m}^{(1)}$ and
$Z_{n,m}^{(2)}$ are given by the product of two quadratures. One of
them can be performed analytically and the result is expressed in
terms of the error function. Performing the change of integration
variable $z=2u/\ell-n-m$ in $Z_{n,m}^{(1)}$ and $z=2-2u/\ell+n+m$ in
$Z_{n,m}^{(2)}$, one obtains
\begin{align}
\label{eq:Z1:def:simp}
Z_{n,m}^{(1)}&=\frac{\ell}{2}\sqrt{\frac{\pi}{\gamma_2}}
\int _0^1 {\rm d}z \ee^{-\gamma_1\ell^2(z+n+m)^2/4}
\left\{{\erf}\left[\frac{\ell}{2}\sqrt{\gamma_2}\left(z+n-m\right)\right]
+{\erf}\left[\frac{\ell}{2}\sqrt{\gamma_2}\left(z-n+m\right)\right]\right\}\, ,
\\
Z_{n,m}^{(2)}&=\frac{\ell}{2}\sqrt{\frac{\pi}{\gamma_2}}
\int _0^1 {\rm d}z \ee^{-\gamma_1\ell^2(z-n-m-2)^2/4}
\left\{{\erf}\left[\frac{\ell}{2}\sqrt{\gamma_2}\left(z+n-m\right)\right]
+{\erf}\left[\frac{\ell}{2}\sqrt{\gamma_2}\left(z-n+m\right)\right]\right\}
=Z_{-n-1,-m-1}^{(1)}
\, .
\end{align}
Because the term $Z_{n,m}^{(2)}$ can be expressed in terms of
$Z_{n,m}^{(1)}$, it follows that one has $\sum_{n,m}
Z_{n,m}^{(2)}=\sum_{n,m} Z_{n,m}^{(1)}$, leading to
\begin{align}
\langle \Psi_{2\, {\rm sq}}\vert \hat{S}^{(1)}_z(\ell) \hat{S}^{(2)}_z(\ell)
\vert \Psi_{2\, {\rm sq}}\rangle
& = \frac{\sqrt{\gamma_1\gamma_2}}{\pi}\sum _{n=-\infty}^{\infty}
\sum_{m=-\infty}^{\infty}(-1)^{n+m}Z_{n,m}^{(1)}\, ,
\label{eq:SzSz:exact}
\end{align}
where $Z_{n,m}^{(1)}$ is given by \Eq{eq:Z1:def:simp}.
\begin{figure}[t]
\begin{center}
\includegraphics[width=0.445\textwidth,clip=true]{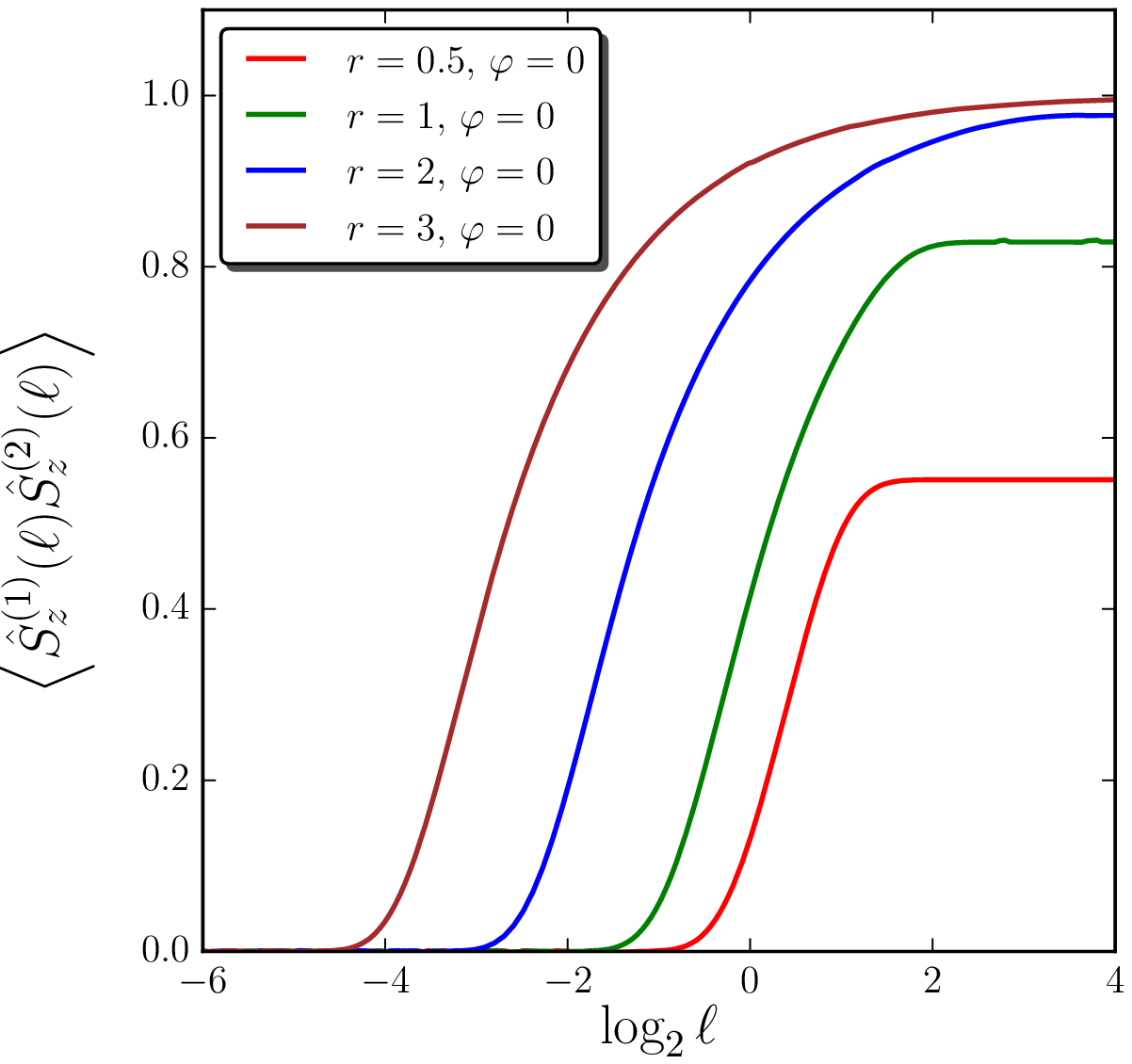}
\includegraphics[width=0.455\textwidth,clip=true]{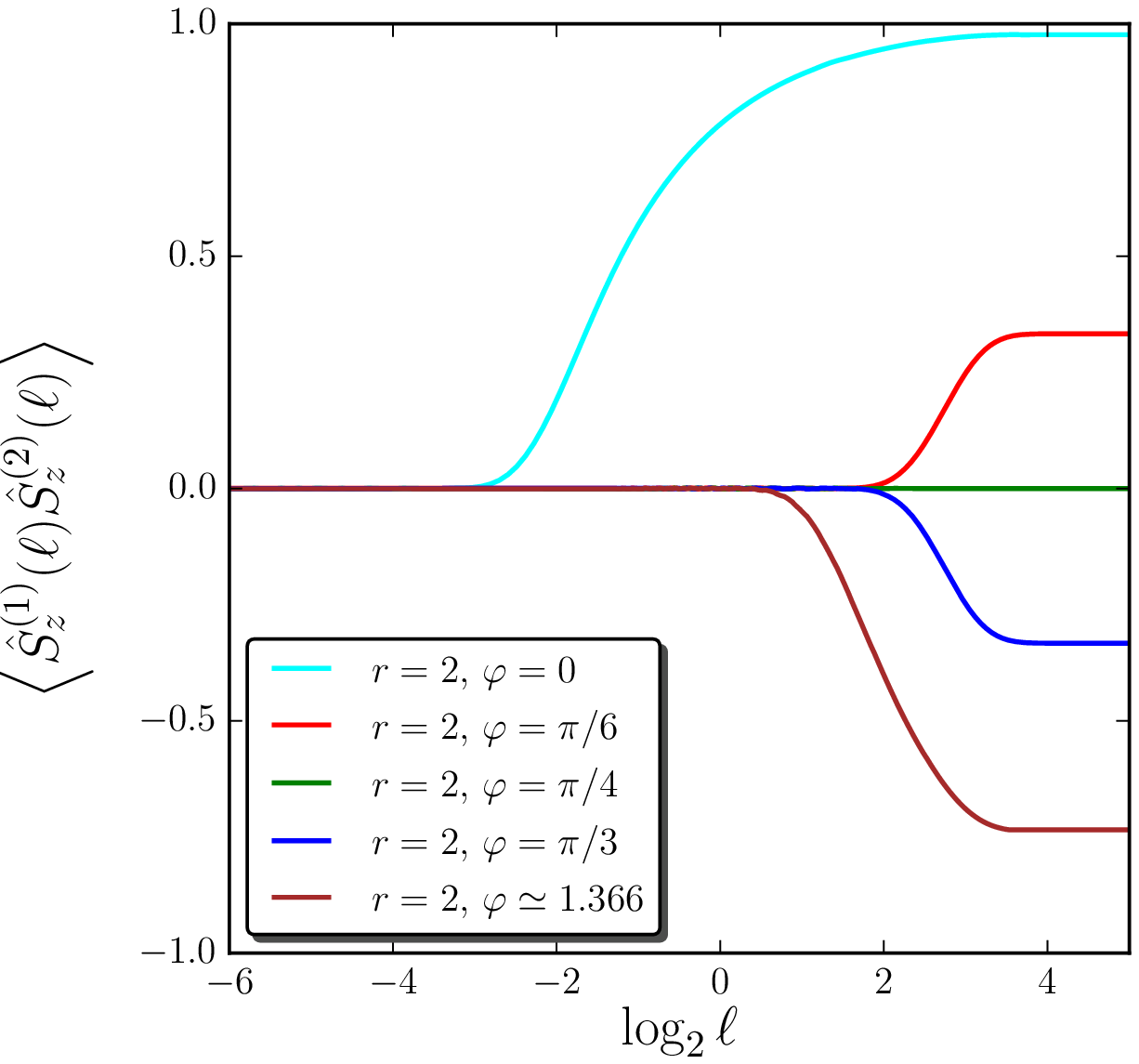}
\caption{Two-point correlator $\langle \Psi_{2\, {\rm sq}}\vert
  \hat{S}^{(1)}_z(\ell) \hat{S}^{(2)}_z(\ell)\vert \Psi_{2\, {\rm
      sq}}\rangle$ as a function of $\ell$, for $\varphi=0$ and a few
  values of $ r$ (left panel), and $r=2$ and a few values of $\varphi$
  (right panel).}
\label{fig:exactSzSz}
\end{center}
\end{figure}

Unfortunately, the second quadrature cannot be performed analytically
and has to be done numerically, and so has the sum appearing in
\Eq{eq:SzSz:exact}. The results are presented in
\Fig{fig:exactSzSz}. In the left panel, $\langle \Psi_{2\, {\rm sq}
}\vert \hat{S}^{(1)}_z(\ell) \hat{S}^{(2)}_z(\ell) \vert \Psi_{2\,
  {\rm sq}}\rangle$ is displayed versus $\log_2 \ell$ for different
values of the squeezing parameter and a vanishing squeezing angle. One
can check that these curves are consistent\footnote{ More precisely,
  even if the overall shape of the correlation functions is clearly
  similar as well as the numerical value of the plateau at large
  $\ell$, it seems that the curves of \Ref{2004PhRvA..70b2102L} are
  systematically shifted towards larger $\ell$ compared to ours. For
  instance, for $r=0.5$ and $\varphi=0$ (top right panel of Fig.~1 in
  \Ref{2004PhRvA..70b2102L}), the correlation function takes off from
  zero around $\log_2 \ell\simeq -2$ while in our case it is rather
  around $\log_2 \ell \simeq -1$. The same shift also appears for the
  other cases. The origin of this shift in the results of
  \Ref{2004PhRvA..70b2102L} is unclear to us.} with the results of
\Ref{2004PhRvA..70b2102L} (see Fig.~1 of that article).  In the right
panel, the same quantity is represented for $r=2$ and different values
of the squeezing angle. These results are completely new to our
knowledge. One can see that the overall structure of the correlation
function is preserved, namely it vanishes at small $\ell$ and exhibits
a plateau at large $\ell$, in agreement with the analytical limits of
\Secs{sec:largeelllimit} and~\ref{sec:smallelllimit}. However, when
$\varphi\neq 0$, the correlation function can become negative.

The case $\varphi=\pi/4$ is even more intriguing since the correlation
function vanishes regardless of the value of $\ell$. In fact, this can
be understood analytically and allows us to check the consistency of
our numerical calculations. Indeed, for $\varphi=\pi/4$, one has
$\gamma_1=\gamma_2=2/\cosh(2r)$. From \Eqs{eq:gamma1:def}
and~(\ref{eq:gamma2:def}), this implies that $A+A^*=-\gamma_1/2$ and
$B+B^*=0$. As a consequence, \Eq{eq:defInm} can be written as
\begin{equation}
Z_{n,m}\left(r,\varphi=\frac{\pi}{4}\right)
=\int_{n\ell}^{(n+1)\ell}{\rm d}Q_1 \ee^{-\gamma_1Q_1^2/2}
\int_{m\ell}^{(m+1)\ell}{\rm d}Q_2 \ee^{-\gamma_1Q_2^2/2}.
\label{eq:Znm:phi=pi/4}
\end{equation}
In other words, the two sub-systems are now ``decoupled'' and the
two-point correlation function of the bipartite system is in fact the
product of two one-point functions and must therefore vanish. Let us
see how it works in practice. From \Eq{eq:Znm:phi=pi/4}, $Z_{n,m}$ can
be calculated explicitly and reads
\begin{align}
Z_{n,m}\left(r,\varphi=\frac{\pi}{4}\right)
=&\frac{\pi}{2\gamma_1}\left\{\erf\left[\ell \sqrt{\frac{\gamma_1}{2}}
(n+1)\right]
-\erf\left(\ell \sqrt{\frac{\gamma_1}{2}}n\right)\right\}
\left\{\erf\left[\ell \sqrt{\frac{\gamma_1}{2}}
(m+1)\right]
-\erf\left(\ell \sqrt{\frac{\gamma_1}{2}}m\right)\right\}
\\
=& \frac{\pi}{2\gamma_1}z_nz_m,
\end{align}
with $z_n\equiv \mathfrak{z}_{n+1}-\mathfrak{z}_n$ and
$\mathfrak{z}_n\equiv \erf(n\ell\sqrt{\gamma_1/2})$. Then, from
\Eq{eq:correlZ}, it follows that
\begin{align}
\langle \Psi_{2\, {\rm sq}}\vert \hat{S}^{(1)}_z(\ell) \hat{S}^{(2)}_z(\ell)
\vert \Psi_{2\, {\rm sq}}\rangle &=\frac{1}{4}\sum _{n=-\infty}^{\infty}
(-1)^nz_n\sum_{m=-\infty}^{\infty}(-1)^{m}z_m.
\end{align}
But one has $\sum _{n=-\infty}^{\infty}(-1)^nz_n=\sum
_{n=-\infty}^{\infty}(-1)^n(\mathfrak{z}_{n+1}-\mathfrak{z}_n)=0$,
where in the last expression we have used the fact that
$\mathfrak{z}_n=-\mathfrak{z}_{-n}$. This explains why the correlation
function vanishes in the case $\varphi=\pi/4$.

In fact, this result can also be understood as the consequence of the
fact that the two-point correlator of $\hat{S}_z$ is odd with respect
to $\varphi=\pi/4$ (hence vanishes at $\varphi=\pi/4$). Indeed, in the
right panel of \Fig{fig:exactSzSz}, one can notice that the
correlation function for $\varphi=\pi/3$ is the opposite to that for
$\varphi=\pi/6$. This can be understood as follows. Let us go back to
\Eq{eq:defInm} which, using \Eqs{eq:gamma1:def}
and~(\ref{eq:gamma2:def}), can be rewritten as
\begin{align}
Z_{n,m}(r,\varphi)=
\int _{n\ell}^{(n+1)\ell}{\rm d}Q_1
\int _{m\ell}^{(m+1)\ell}{\rm d}Q_2
\ee^{-[\gamma_1(r,\varphi)+\gamma_2(r,\varphi)]\left(Q_1^2+Q_2^2\right)/4
-[\gamma_1(r,\varphi)-\gamma_2(r,\varphi)]Q_1Q_2/2}\, .
\end{align}
From \Eqs{eq:gamma1:def} and~(\ref{eq:gamma2:def}), one can see that
the functions $\gamma_1$ and $\gamma_2$ are related through
$\gamma_{1,2}(r,\varphi)=\gamma_{2,1}(r,\pi/2-\varphi)$.  This implies
that
\begin{align}
Z_{n,m}\left(r,\frac{\pi}{2}-\varphi\right)
&=
\int _{n\ell}^{(n+1)\ell}{\rm d}Q_1
\int _{m\ell}^{(m+1)\ell}{\rm d}Q_2
\ee^{-[\gamma_1(r,\varphi)+\gamma_2(r,\varphi)]
\left(Q_1^2+Q_2^2\right)/4
+[\gamma_1(r,\varphi)-\gamma_2(r,\varphi)]Q_1Q_2/2}\\
&=
-\int _{n\ell}^{(n+1)\ell}{\rm d}Q_1
\int _{-m\ell}^{-(m+1)\ell}{\rm d}Q_2
\ee^{-[\gamma_1(r,\varphi)+\gamma_2(r,\varphi)]
\left(Q_1^2+Q_2^2\right)/4
-[\gamma_1(r,\varphi)-\gamma_2(r,\varphi)]Q_1Q_2/2},
\end{align}
where, in the last expression, we have changed the integration variable $Q_2$ to
$-Q_2$. As a consequence, one can write that
$Z_{n,m}(r,\pi/2-\varphi)=Z_{n,-m-1}(r,\varphi)$ and it follows that
\begin{align}
\sum _{n=-\infty}^{\infty}
\sum_{m=-\infty}^{\infty}(-1)^{n+m}Z_{n,m}\left(r,\frac{\pi}{2}-\varphi\right)
=-\sum _{n=-\infty}^{\infty}
\sum_{m'=-\infty}^{\infty}(-1)^{n+m'}Z_{n,m'}\left(r,\varphi\right),
\end{align}
with $m'=-m-1$. We have thus established that the correlation function
evaluated with $r$ and $\varphi$ is minus the one calculated with $r$
and $\pi/2-\varphi$, and that these two configurations are therefore
``dual'' in a sense that will be further discussed in what follows,
notably in \Sec{sec:varphi=pi/2}. This also means that one can study
this correlation function in the interval $\varphi\in[0,\pi/4]$ only.
\subsection{Correlation Function $\langle \Psi_{2\, {\rm sq}}\vert \hat{S}^{(1)}_x(\ell) \hat{S}^{(2)}_x(\ell)\vert \Psi_{2\, {\rm sq}}\rangle$}
\label{sec:spinPheno:x}
\begin{figure}[t]
\begin{center}
\includegraphics[width=0.45\textwidth,clip=true]{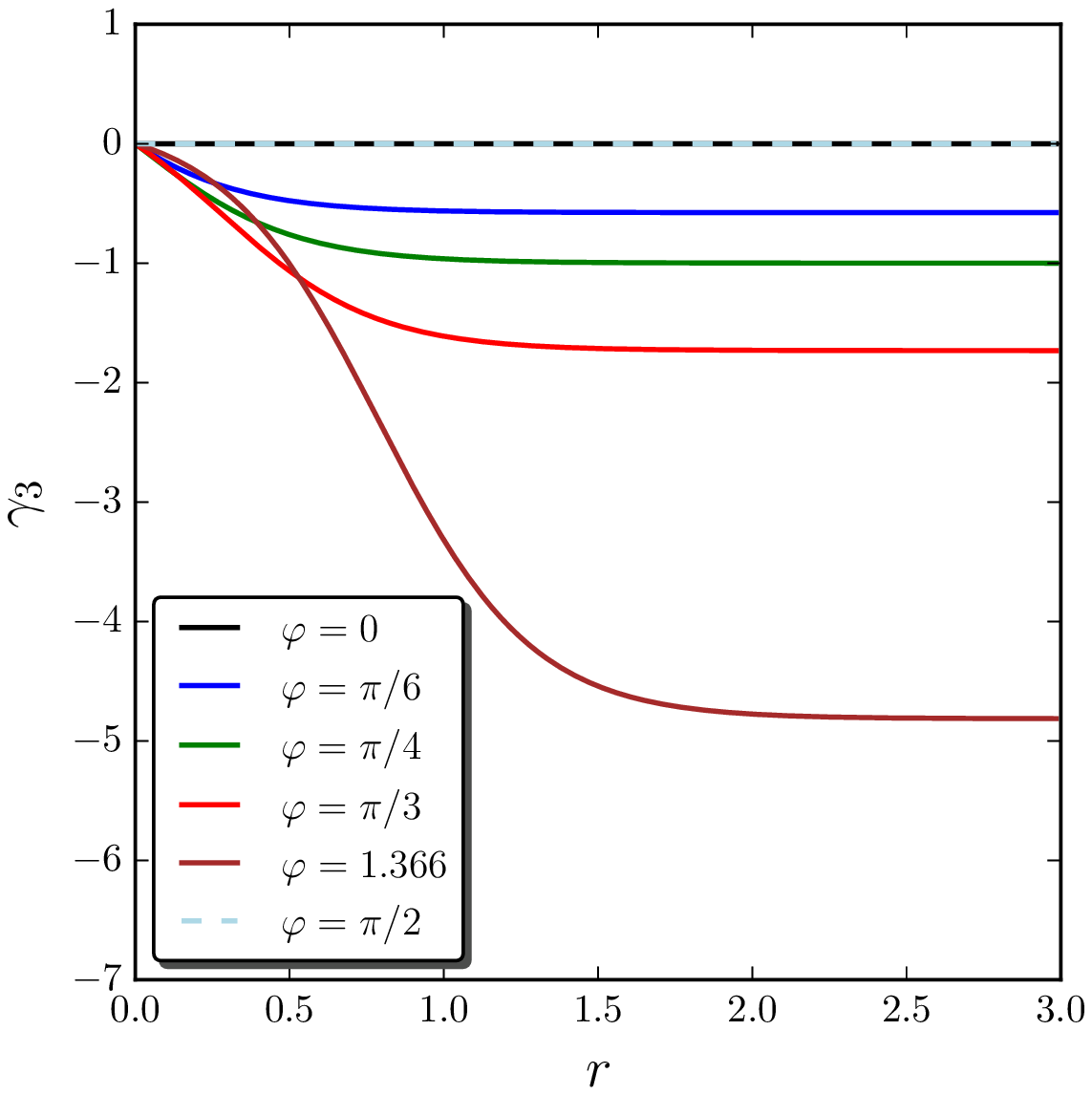}
\includegraphics[width=0.45\textwidth,clip=true]{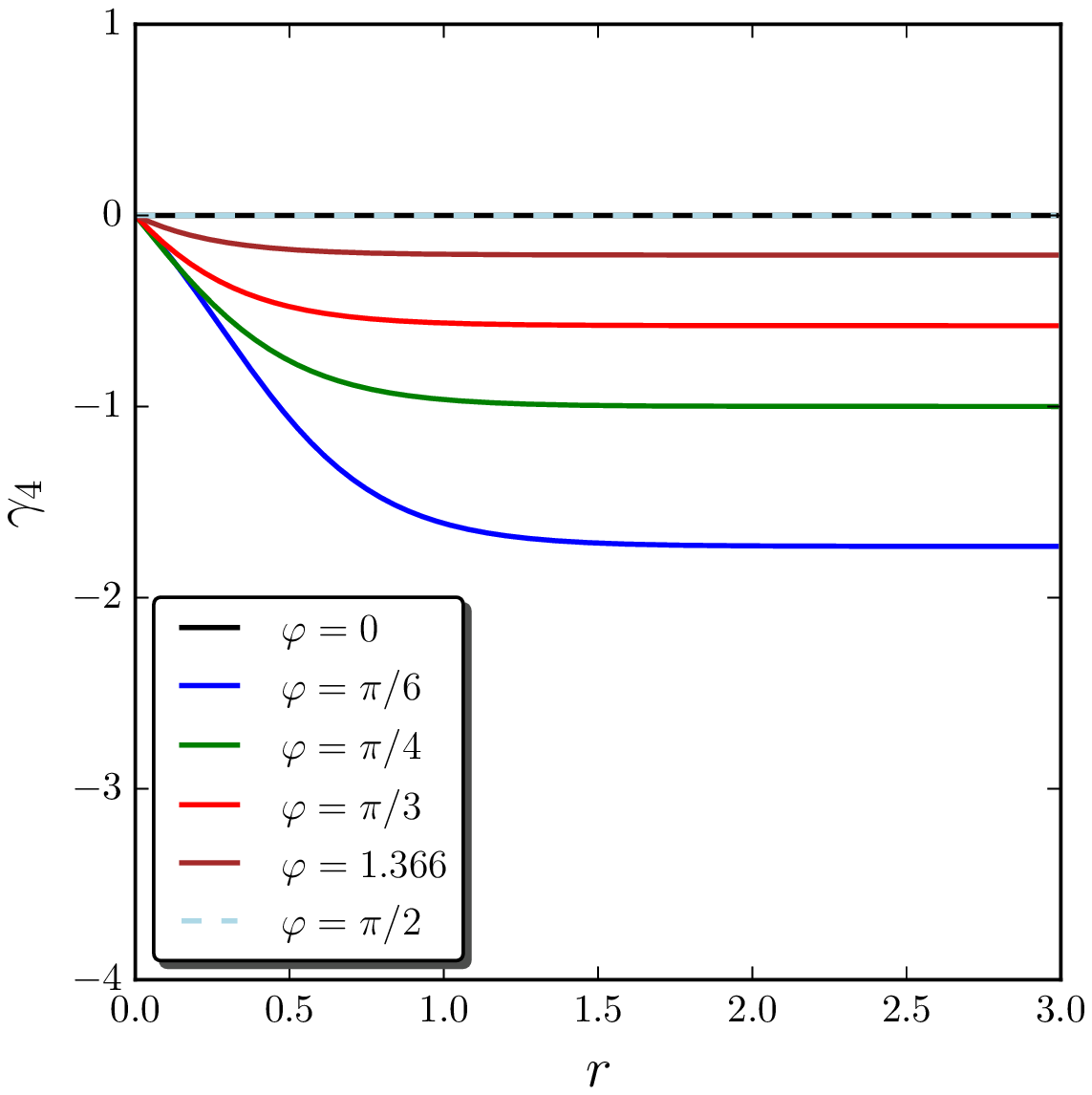}
\caption{$\gamma_3$ (left panel) and $\gamma_4$ (right panel) defined in \Eqs{eq:gamma3:def} and~(\ref{eq:gamma4:def}), as a function of $r$, for $\varphi=0$, $\pi/6$, $\pi/4$, $\pi/3$, $\pi/2.3$ and $\pi/2$.}
\label{fig:gamma34}
\end{center}
\end{figure}
Let us now calculate the two-point correlation function of the
operator $\hat{S}_x(\ell)$. Using its definition in terms of the spin
step operators, see \Eq{eq:defsx},
$\hat{S}_x(\ell)=\hat{S}_+(\ell)+\hat{S}_-(\ell)$, one has
\begin{align}
\langle \Psi_{2\, {\rm sq}}\vert \hat{S}^{(1)}_x(\ell) \hat{S}^{(2)}_x(\ell)
\vert \Psi_{2\, {\rm sq}}\rangle =& 
\langle \Psi_{2\, {\rm sq}}\vert \hat{S}^{(1)}_+(\ell) \hat{S}^{(2)}_+(\ell)
\vert \Psi_{2\, {\rm sq}}\rangle+
\langle \Psi_{2\, {\rm sq}}\vert \hat{S}^{(1)}_-(\ell) \hat{S}^{(2)}_-(\ell)
\vert \Psi_{2\, {\rm sq}}\rangle
\nonumber \\ & +
\langle \Psi_{2\, {\rm sq}}\vert \hat{S}^{(1)}_+(\ell) \hat{S}^{(2)}_-(\ell)
\vert \Psi_{2\, {\rm sq}}\rangle+
\langle \Psi_{2\, {\rm sq}}\vert \hat{S}^{(1)}_-(\ell) \hat{S}^{(2)}_+(\ell)
\vert \Psi_{2\, {\rm sq}}\rangle\\ 
= & 2 \Re\mathrm{e}\left[
\langle \Psi_{2\, {\rm sq}}\vert \hat{S}^{(1)}_+(\ell) \hat{S}^{(2)}_+(\ell)
\vert \Psi_{2\, {\rm sq}}\rangle+
\langle \Psi_{2\, {\rm sq}}\vert \hat{S}^{(1)}_+(\ell) \hat{S}^{(2)}_-(\ell)
\vert \Psi_{2\, {\rm sq}}\rangle\right]\, ,
\label{eq:SxSx:S+S-}
\end{align}
where we have used the relation
$\hat{S}_+(\ell)=\hat{S}_-^\dagger(\ell)$ and the fact that the
two-mode squeezed state is symmetric if one exchanges the sub-spaces
$(1)$ and $(2)$, see \Eq{eq:qstateposition}. Therefore, one has to
calculate two quantities. The first one is given by
\begin{align}
\label{eq:S+S+:Psi}
\langle \Psi_{2\, {\rm sq}}\vert \hat{S}^{(1)}_+(\ell) \hat{S}^{(2)}_+(\ell)
\vert \Psi_{2\, {\rm sq}}\rangle &=\sum _{n=-\infty}^{\infty}
\sum_{m=-\infty}^{\infty}
\int _{2n\ell}^{(2n+1)\ell}{\rm d}Q_1
\int _{2m\ell}^{(2m+1)\ell}{\rm d}Q_2
\Psi_{\rm 2\, sq}^*\left(Q_1,Q_2\right)
\Psi_{\rm 2\, sq}\left(Q_1+\ell,Q_2+\ell\right)\\
& = \frac{1}{\pi \cosh^2 r}
\frac{1}{\sqrt{
\tanh^4 r-2\tanh^2 r \cos(4\varphi)+1}}
\sum _{n=-\infty}^{\infty}
\sum_{m=-\infty}^{\infty}
J_{n,m}
\end{align}
with 
\begin{equation}
\label{eq:J:def}
J_{n,m}\equiv \int _{2n\ell}^{(2n+1)\ell}{\rm d}Q_1
\int _{2m\ell}^{(2m+1)\ell}{\rm d}Q_2
\ee^{(2A-B)\ell^2}
\ee^{(A+A^*)\left(Q_1^2+Q_2^2\right)
-(B+B^*)Q_1Q_2+(2A-B)\ell
\left(Q_1+Q_2\right)}\, .
\end{equation}
This integral has a structure similar to that of $Z_{n,m}$ except
that, in the argument of the exponential, there is now a term
proportional to $Q_1+Q_2$. The second quantity that needs to be
calculated reads
\begin{align}
\label{eq:S-S-:Psi}
\langle \Psi_{2\, {\rm sq}}\vert \hat{S}^{(1)}_+(\ell) \hat{S}^{(2)}_-(\ell)
\vert \Psi_{2\, {\rm sq}}\rangle &=\sum _{n=-\infty}^{\infty}
\sum_{m=-\infty}^{\infty}
\int _{(2n+1)\ell}^{(2n+2)\ell}{\rm d}Q_1
\int _{2m\ell}^{(2m+1)\ell}{\rm d}Q_2
\Psi_{\rm 2\, sq}^*\left(Q_1,Q_2\right)
\Psi_{\rm 2\, sq}\left(Q_1-\ell,Q_2+\ell\right)\\
& = \frac{1}{\pi \cosh^2 r}
\frac{1}{\sqrt{
\tanh^4 r-2\tanh^2 r \cos(4\varphi)+1}}
\sum _{n=-\infty}^{\infty}
\sum_{m=-\infty}^{\infty}
K_{n,m}
\end{align}
with 
\begin{equation}
\label{eq:K:def}
K_{n,m}=\int _{(2n+1)\ell}^{(2n+2)\ell}{\rm d}Q_1
\int _{2m\ell}^{(2m+1)\ell}{\rm d}Q_2
\ee^{(2A+B)\ell^2}
\ee^{(A+A^*)\left(Q_1^2+Q_2^2\right)
-(B+B^*)Q_1Q_2-(2A+B)\ell
\left(Q_1-Q_2\right)}\, .
\end{equation}
We notice that the argument of the exponential also contains a new
type of terms, this time proportional to $Q_1-Q_2$. In order to have the
same integral limits in \Eqs{eq:J:def} and~(\ref{eq:K:def}), it is
convenient to perform the change of integration variables $y_1=Q_1-\ell$ and
$y_2=Q_2$ in \Eq{eq:K:def}, which gives rise to the following
expression
\begin{align}
K_{n,m}=\int _{2n\ell}^{(2n+1)\ell}\dd y_1
\int _{2m\ell}^{(2m+1)\ell} \dd y_2
\ee^{(A^*+A)\ell^2}
\ee^{(A+A^*)\left(y_1^2+y_2^2\right)
-(B+B^*)y_1y_2+(2A^*-B)\ell y_1
+(2A-B^*)\ell y_2}\, .
\end{align}
Plugging the above results in \Eq{eq:SxSx:S+S-}, one can write the
correlation function as
\begin{align}
\label{eq:correlX}
\langle \Psi_{2\, {\rm sq}}\vert \hat{S}^{(1)}_x(\ell) \hat{S}^{(2)}_x(\ell)
\vert \Psi_{2\, {\rm sq}}\rangle &= \frac{1}{\pi \cosh^2 r}
\frac{1}{\sqrt{
\tanh^4 r-2\tanh^2 r \cos(4\varphi)+1}}\sum _{n=-\infty}^{\infty}
\sum_{m=-\infty}^{\infty}X_{n,m}\, ,
\end{align}
where the quantity $X_{n,m}$ is defined by
\begin{align}
X_{n,m}\equiv & 2 \int _{2n\ell}^{(2n+1)\ell}{\rm d}y_1
\int _{2m\ell}^{(2m+1)\ell}{\rm d}y_2
\ee^{(A+A^*)\left(y_1^2+y_2^2\right)
-(B+B^*)y_1y_2}
\nonumber \\ & 
\Re\mathrm{e}
\left[
\ee^{(2A-B)\ell^2+(2A-B)\ell
\left(y_1+y_2\right)}
+
\ee^{(A^*+A)\ell^2+(2A^*-B)\ell y_1
+(2A-B^*)\ell y_2}
\right].
\end{align}
At this stage, it is convenient to introduce the new parameters
$\gamma_3$ and $\gamma_4$, defined by
\begin{align}
\label{eq:gamma3:def}
2A-B=&-\frac{1}{\cosh(2 r)+\cos(2\varphi)\sinh(2r)}
-i\frac{2\tanh r\sin (2\varphi)}{
\tanh^2r+2\tanh r\cos(2\varphi)
+1}\equiv -\frac{\gamma_1}{2}+i\gamma_3 \\
2A^*-B =& -\frac{1}{\cosh(2 r)+\cos(2\varphi)\sinh(2r)}
-i\frac{2\tanh r\sin (2\varphi)}{
\tanh^2r-2\tanh r\cos(2\varphi)
+1}
\equiv -\frac{\gamma_1}{2}+i\gamma_4\, .
\label{eq:gamma4:def}
\end{align}
We see that the problem can be described in terms of four real
functions, $\gamma_1$, $\gamma_2$, $\gamma_3$ and $\gamma_4$, which is
consistent with the fact that the quantum state is given in terms of
two complex functions $A$ and $B$. In fact, one can show that
$A-A^*=i(\gamma_3-\gamma_4)/2$ and
$B-B^*=-i(\gamma_3+\gamma_4)$. Recalling that
$A+A^*=-(\gamma_1+\gamma_2)/4$ and $B+B^*=(\gamma_1-\gamma_2)/2$, one
can then express $X_{n,m}$ in terms of the $\gamma_i$ functions
only. The corresponding formula reads
\begin{align}
X_{n,m}=& 2 \int _{2n\ell}^{(2n+1)\ell}{\rm d}y_1
\int _{2m\ell}^{(2m+1)\ell}{\rm d}y_2
\ee^{-(\gamma_1+\gamma_2)\left(y_1^2+y_2^2\right)/4
-(\gamma_1-\gamma_2)y_1y_2/2}
\biggl\{
\ee^{-\gamma_1\left[\ell^2+\ell
\left(y_1+y_2\right)\right]/2}
\cos \left[\gamma_3\ell^2+\gamma_3\ell
\left(y_1+y_2\right)\right]
\nonumber \\ &
+\ee^{-(\gamma_1+\gamma_2)\ell^2/4 -\gamma_1/2\ell
\left(y_1+y_2\right)}
\cos \left[\gamma_4\ell
\left(y_1-y_2\right)\right]\biggr\}\, .
\label{eq:Xwithgamma}
\end{align}
At this stage, it is also interesting to notice that $\gamma_3$ and
$\gamma_4$ satisfy
$\gamma_{3,4}(r,\varphi)=\gamma_{3,4}(r,\varphi+\pi)$ and
$\gamma_{3,4}(r,\pi/2+\varphi)=-\gamma_{3,4}(r,\pi/2-\varphi)$, the
last minus sign being the only difference with the otherwise similar
symmetry properties given in \Sec{sec:spinPheno:z} for
$\gamma_1$ and $\gamma_2$. But since \Eq{eq:Xwithgamma} is unchanged
when one flips the sign of $\gamma_3$ and $\gamma_4$, this means that,
as in \Sec{sec:spinPheno:z}, one can study the correlation
function in the interval $\varphi\in[0,\pi/2]$ and use these
symmetries to extend the result to other values of $\varphi$. The next
step consists in performing the same change of integration variables
as in \Sec{sec:spinPheno:z}, namely $y_1=u+v$ and $y_2=u-v$,
since this allows us to perform one of the two quadratures. This leads
to
\begin{align}
X_{n,m} &=
4\ee^{-\gamma_1\ell^2/2}
\int _{(n+m)\ell}^{(n+m+1/2)\ell}{\rm d}u
\ee^{-\gamma_1u^2-\gamma_1\ell u}
\cos\left[\gamma_3\ell\left(\ell+2u\right)\right]
\int _{2n\ell -u}^{u-2m\ell}{\rm d}v
\ee^{-\gamma_2v^2}
\nonumber \\ &
+4\ee^{-(\gamma_1+\gamma_2)\ell^2/4}
\int _{(n+m)\ell}^{(n+m+1/2)\ell}{\rm d}u
\ee^{-\gamma_1u^2-\gamma_1\ell u}
\int _{2n\ell -u}^{u-2m\ell}{\rm d}v
\ee^{-\gamma_2v^2}
\cos\left(2\gamma_4\ell v\right)
\nonumber \\ &
+4 \ee^{-\gamma_1\ell^2/2}
\int _{(n+m+1/2)\ell}^{(n+m+1)\ell}{\rm d}u
\ee^{-\gamma_1u^2-\gamma_1\ell u/2}
\cos\left[\gamma_3\ell\left(\ell+2u\right)\right]
\int _{u-(2m+1)\ell}^{(2n+1)\ell-u}{\rm d}v
\ee^{-\gamma_2v^2}
\nonumber \\ &
+4 \ee^{-(\gamma_1+\gamma_2)\ell^2/4}
\int _{(n+m+1/2)\ell}^{(n+m+1)\ell}{\rm d}u
\ee^{-\gamma_1u^2-\gamma_1\ell u}
\int _{u-(2m+1)\ell}^{(2n+1)\ell-u}{\rm d}v
\ee^{-\gamma_2v^2}
\cos\left(2\gamma_4\ell v\right)
\label{eq:X1X2X3X4:def} \\ & 
\equiv X_{n,m}^{(1)}+X_{n,m}^{(2)}+X_{n,m}^{(3)}+X_{n,m}^{(4)}\, .
\end{align}
Again the structure of the integrals $X_{n,m}^{(i)}$ is very similar
to that of $Z_{n,m}$. The only differences originate from the fact
that the arguments of the exponentials now contain a term linear in
$u$, and a cosine function is present in the first and
third integrals. As before, the integrals over $v$ can be performed
by means of error functions and one obtains
\begin{align}
\label{eq:X1:def}
X_{n,m}^{(1)} &=
\ell\sqrt{\frac{\pi}{\gamma_2}}
\ee^{-\gamma_1\ell^2/2}
\int _0^1 {\rm d}z
\ee^{-\gamma_1 \ell^2(z+2n+2m)^2/4-\gamma_1\ell^2(z+2n+2m)/2}
\cos\left[\gamma_3\ell^2(z+2n+2m+1)\right]
\nonumber \\ & \times
\left\{{\rm erf}\left[\frac{\ell}{2}\sqrt{\gamma_2}
(z-2n+2m)\right]
+{\rm erf}\left[\frac{\ell}{2}\sqrt{\gamma_2}
(z+2n-2m)\right]\right\}\, ,\\
X_{n,m}^{(2)} &=
\ell\sqrt{\frac{\pi}{\gamma_2}}
\ee^{-\left[(\gamma_1+\gamma_2)/4+\gamma_4^2/\gamma_2\right]\ell^2}
\int _0^1 {\rm d}z
\ee^{-\gamma_1 \ell^2(z+2n+2m)^2/4-\gamma_1\ell^2(z+2n+2m)/2}
\nonumber \\ & \times
\Re\mathrm{e}\biggl\{{\rm erf}\left[\frac{\ell}{2}\sqrt{\gamma_2}
(z-2n+2m)+i\frac{\gamma_4}{\sqrt{\gamma_2}}\ell\right]
+{\rm erf}\left[\frac{\ell}{2}\sqrt{\gamma_2}
(z+2n-2m)+i\frac{\gamma_4}{\sqrt{\gamma_2}}\ell\right]
\biggr\}
\, ,
\label{eq:X2:def}
\\
X_{n,m}^{(3)} &=
\label{eq:X3:def}
\ell\sqrt{\frac{\pi}{\gamma_2}}
\ee^{-\gamma_1\ell^2/2}
\int _0^1 {\rm d}z
\ee^{-\gamma_1 \ell^2(z-2n-2m-2)^2/4+\gamma_1\ell^2(z-2n-2m-2)/2}
\cos\left[\gamma_3\ell^2(z-2n-2m-3)\right]
\nonumber \\ & \times
\left\{{\rm erf}\left[\frac{\ell}{2}\sqrt{\gamma_2}
(z-2n+2m)\right]
+{\rm erf}\left[\frac{\ell}{2}\sqrt{\gamma_2}
(z+2n-2m)\right]\right\}\, ,
\\
X_{n,m}^{(4)} &=
\ell \sqrt{\frac{\pi}{\gamma_2}}
\ee^{-\left[(\gamma_1+\gamma_2)/4+\gamma_4^2/\gamma_2\right]\ell^2}
\int _0^1 {\rm d}z
\ee^{-\gamma_1 \ell^2(z-2n-2m-2)^2/4+\gamma_1\ell^2(z-2n-2m-2)/2}
\nonumber \\ & \times
\Re\mathrm{e}
\biggl\{{\rm erf}\left[\frac{\ell}{2}\sqrt{\gamma_2}
(z-2n+2m)+i\frac{\gamma_4}{\sqrt{\gamma_2}}\ell\right]
+{\rm erf}\left[\frac{\ell}{2}\sqrt{\gamma_2}
(z+2n-2m)+i\frac{\gamma_4}{\sqrt{\gamma_2}}\ell\right]
\biggr\}\, .
\label{eq:X4:def}
\end{align}
Finally, the expression of the spin correlation function can be
written as
\begin{align}
\langle \Psi_{2\, {\rm sq}}\vert \hat{S}^{(1)}_x(\ell) \hat{S}^{(2)}_x(\ell)
\vert \Psi_{2\, {\rm sq}}\rangle &=  \frac{\sqrt{\gamma_1\gamma_2}}{2\pi}
\sum _{n=-\infty}^{\infty}
\sum_{m=-\infty}^{\infty}
\left[X^{(1)}_{n,m}+X^{(2)}_{n,m}+X^{(3)}_{n,m}+X^{(4)}_{n,m}\right]
\, .
\label{eq:SxSx:exact}
\end{align}
\begin{figure}[t]
\begin{center}
\includegraphics[width=0.445\textwidth,clip=true]{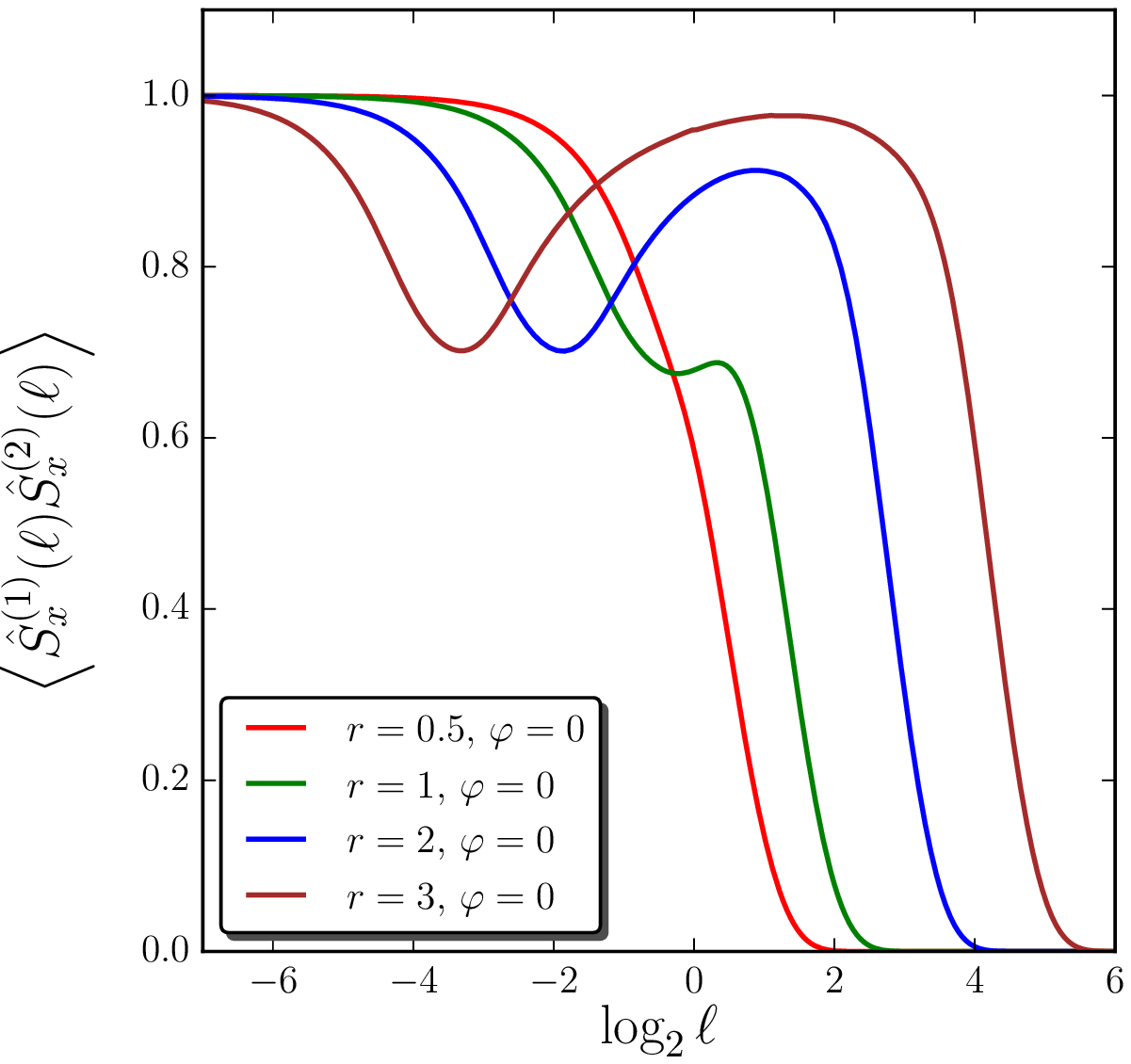}
\includegraphics[width=0.455\textwidth,clip=true]{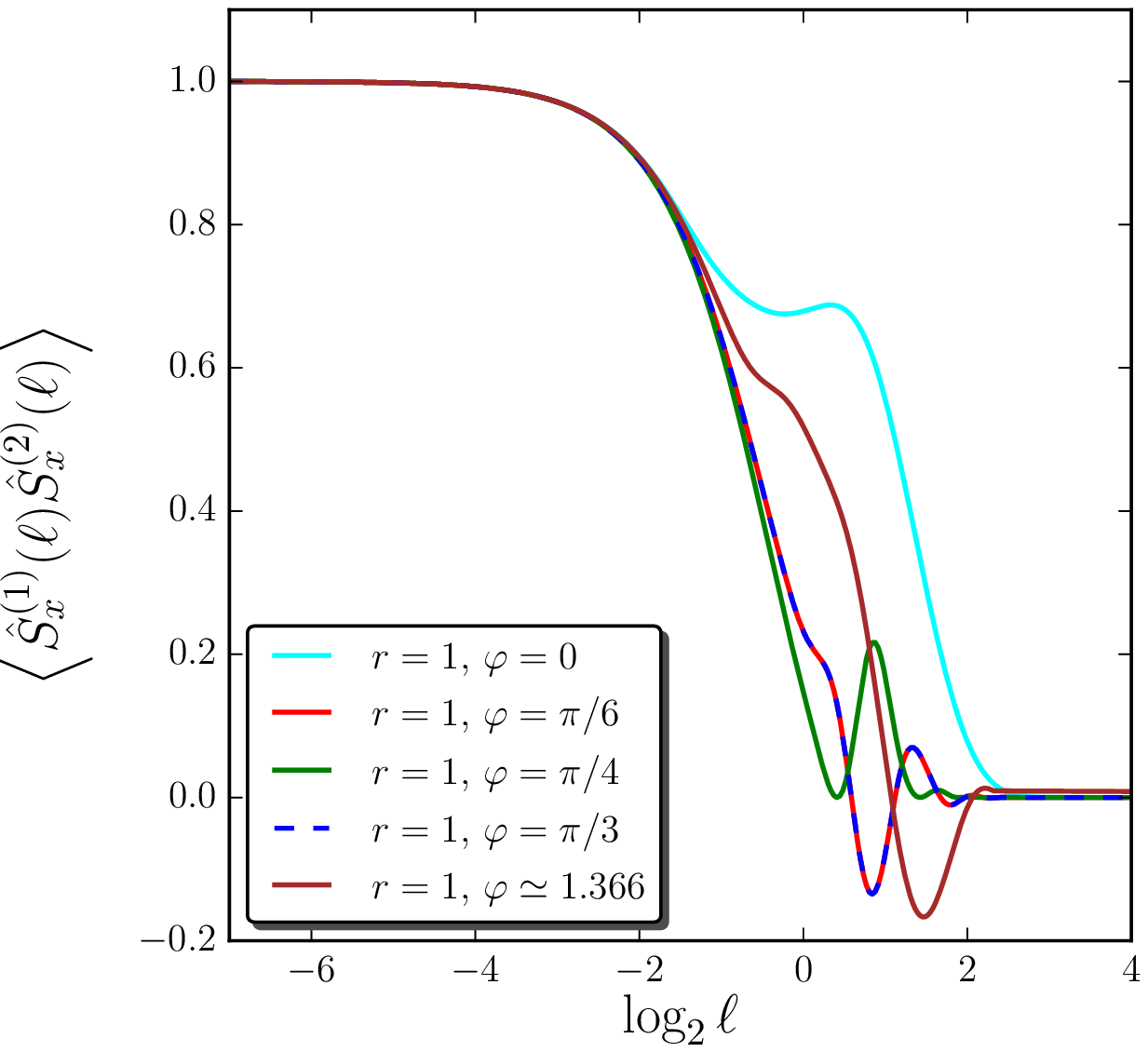}
\caption{Two-point correlator $\langle \Psi_{2\, {\rm sq}}\vert
  \hat{S}^{(1)}_x(\ell) \hat{S}^{(2)}_x(\ell)\vert \Psi_{2\, {\rm
      sq}}\rangle$ as a function of $\ell$, for $\varphi=0$ and a few
  values of $ r$ (left panel), and $r=2$ and a few values of $\varphi$
  (right panel).}
\label{fig:exactSxSx}
\end{center}
\end{figure}
As it was already the case for the integrals $Z_{n,m}$, the remaining
integrals need to be performed numerically, and so do the sums over
$n$ and $m$. The result is displayed in \Fig{fig:exactSxSx}. In the
left panel, the correlation function is given for different values of
$r$ and a vanishing squeezing angle. Our curves are consistent with
those of \Ref{2004PhRvA..70b2102L} even if the systematic shift
already observed for the correlation function $\langle \Psi_{2\, {\rm
    sq} }\vert \hat{S}^{(1)}_z(\ell) \hat{S}^{(2)}_z(\ell) \vert
\Psi_{2\, {\rm sq}}\rangle$ is still present. In the right panel,
results for $r=1$ and different squeezing angles are displayed. We
notice that the small and large $\ell$ limits (namely one and zero,
respectively) are not affected by the fact that $\varphi\neq 0$, see
the analytical results of \Secs{sec:largeelllimit}
and~\ref{sec:smallelllimit}. Only the structure between these two
regimes is changed. In particular, we see some oscillatory patterns
originating from the fact that the integrals $X_{n,m}^{(i)}$ contain
cosine functions and complex error functions.

We also notice that $\langle \Psi_{2\, {\rm sq} }\vert
\hat{S}^{(1)}_x(\ell) \hat{S}^{(2)}_x(\ell) \vert \Psi_{2\, {\rm
    sq}}\rangle$ for $\varphi=\pi/6$ (solid blue line) and
$\varphi=\pi/3$ (solid red line) are exactly equal. This is a
consequence of the fact that this correlation function is even with
respect to $\varphi=\pi/4$. Indeed, from \Eqs{eq:gamma3:def}
and~(\ref{eq:gamma4:def}), one can check that the functions $\gamma_3$
and $\gamma_4$ satisfy the same additional symmetry as $\gamma_1$ and
$\gamma_2$, namely
$\gamma_{3,4}(r,\varphi)=\gamma_{4,3}(r,\pi/2-\varphi)$. Using this
property in \Eq{eq:Xwithgamma}, one obtains
\begin{align}
X_{n,m}\left(r,\frac{\pi}{2}-\varphi\right)=& 2 
\int _{2n\ell}^{(2n+1)\ell}{\rm d}y_1
\int _{2m\ell}^{(2m+1)\ell}{\rm d}y_2
\ee^{-(\gamma_1+\gamma_2)\left(y_1^2+y_2^2\right)/4
+(\gamma_1-\gamma_2)y_1y_2/2}
\biggl\{
\ee^{-\gamma_2\left[\ell^2+\ell
\left(y_1+y_2\right)\right]/2}
\nonumber \\ & \times
\cos \left[\gamma_4\ell^2+\gamma_4\ell
\left(y_1+y_2\right)\right]
+\ee^{-(\gamma_1+\gamma_2)\ell^2/4 -\gamma_2/2\ell
\left(y_1+y_2\right)}
\cos \left[\gamma_3\ell
\left(y_1-y_2\right)\right]\biggr\}\, ,
\end{align}
with all $\gamma_i$ functions evaluated at $r$ and $\varphi$. Let us
then perform the change of integration variable $y_2\rightarrow
-y_2-\ell$. After straightforward manipulations, it is easy to show
that
$X_{n,m}\left(r,\pi/2-\varphi\right)=X_{n,-m-1}\left(r,\varphi\right)$.
As a consequence, one has $\sum _{n=-\infty}^{\infty}
\sum_{m=-\infty}^{\infty}X_{n,m}(r,\pi/2-\varphi)=\sum
_{n=-\infty}^{\infty}
\sum_{m=-\infty}^{\infty}X_{n,m}(r,\varphi)$. This confirms that
$\langle \Psi_{2\, {\rm sq} }\vert \hat{S}^{(1)}_x(\ell)
\hat{S}^{(2)}_x(\ell) \vert \Psi_{2\, {\rm sq}}\rangle$ is even with
respect to $\varphi=\pi/4$, hence one can restrict the present
analysis to $\varphi\in[0,\pi/4]$. This also checks the validity of
our numerical computation in the cases $\varphi=\pi/6$ and
$\varphi=\pi/3$.
\subsection{Correlation Function $\langle \Psi_{2\, {\rm sq} }\vert \hat{S}^{(1)}_y(\ell) \hat{S}^{(2)}_y(\ell) \vert \Psi_{2\, {\rm sq}}\rangle$.}
\label{sec:spinPheno:y}

\begin{figure}[t]
\begin{center}
\includegraphics[width=0.45\textwidth,clip=true]{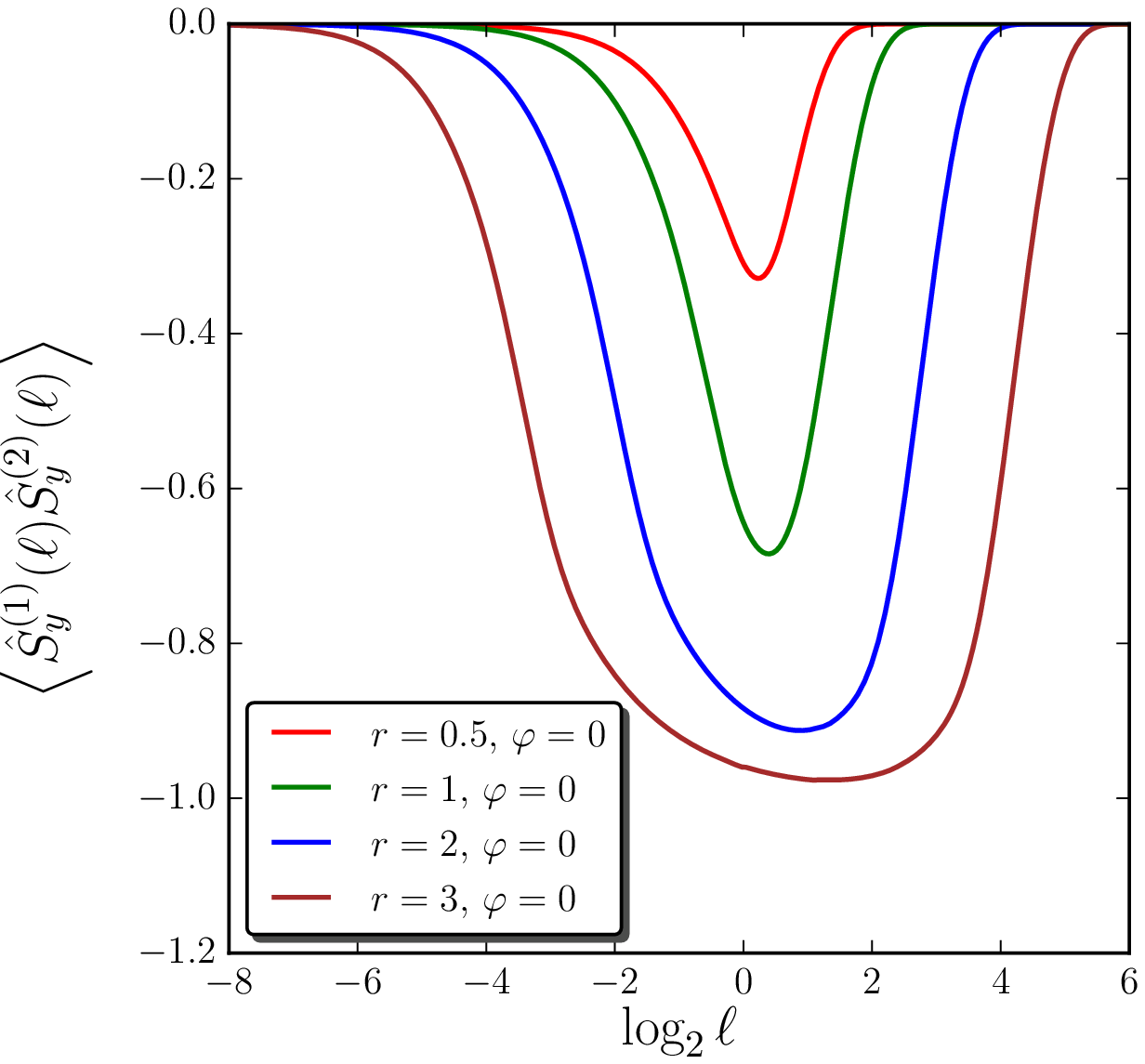}
\includegraphics[width=0.45\textwidth,clip=true]{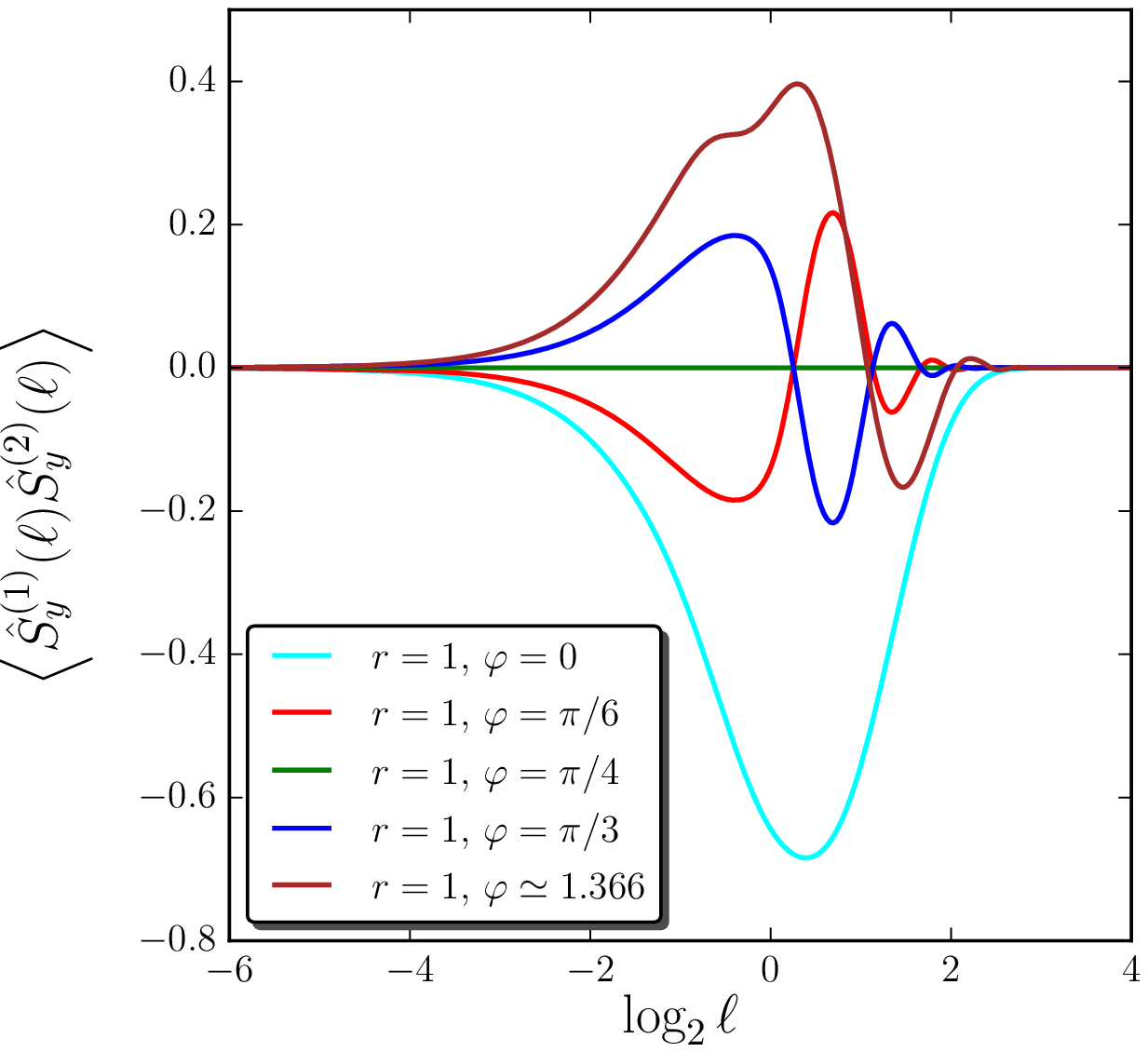}
\caption{Two-point correlator $\langle \Psi_{2\, {\rm sq}}\vert
  \hat{S}^{(1)}_y(\ell) \hat{S}^{(2)}_y(\ell)\vert \Psi_{2\, {\rm
      sq}}\rangle$ as a function of $\ell$, for $\varphi=0$ and a few
  values of $ r$ (left panel), and $r=2$ and a few values of $\varphi$
  (right panel).}
\label{fig:exactSySy}
\end{center}
\end{figure}

Let us then calculate the two-point correlation function of the
operator $\hat{S}_y(\ell)$. Since, see \Eq{eq:defsy}, one has
$\hat{S}_y(\ell)=-i[\hat{S}_+(\ell)-\hat{S}_-(\ell)]$, one can write
\begin{align}
\langle \Psi_{2\, {\rm sq}}\vert \hat{S}^{(1)}_y(\ell) \hat{S}^{(2)}_y(\ell)
\vert \Psi_{2\, {\rm sq}}\rangle =& 
-\langle \Psi_{2\, {\rm sq}}\vert \hat{S}^{(1)}_+(\ell) \hat{S}^{(2)}_+(\ell)
\vert \Psi_{2\, {\rm sq}}\rangle-
\langle \Psi_{2\, {\rm sq}}\vert \hat{S}^{(1)}_-(\ell) \hat{S}^{(2)}_-(\ell)
\vert \Psi_{2\, {\rm sq}}\rangle
\nonumber \\ & +
\langle \Psi_{2\, {\rm sq}}\vert \hat{S}^{(1)}_+(\ell) \hat{S}^{(2)}_-(\ell)
\vert \Psi_{2\, {\rm sq}}\rangle+
\langle \Psi_{2\, {\rm sq}}\vert \hat{S}^{(1)}_-(\ell) \hat{S}^{(2)}_+(\ell)
\vert \Psi_{2\, {\rm sq}}\rangle\\ 
= & 2 \Re\mathrm{e}\left[\langle \Psi_{2\, {\rm sq}}\vert 
\hat{S}^{(1)}_+(\ell) \hat{S}^{(2)}_-(\ell)
\vert \Psi_{2\, {\rm sq}}\rangle-
\langle \Psi_{2\, {\rm sq}}\vert \hat{S}^{(1)}_+(\ell) \hat{S}^{(2)}_+(\ell)
\vert \Psi_{2\, {\rm sq}}\rangle
\right]\, ,
\end{align}
where, as in \Sec{sec:spinPheno:x}, we have used the relation
$\hat{S}_+(\ell)=\hat{S}_-^\dagger(\ell)$ and the fact that the
two-mode squeezed state is symmetric in $(1)\leftrightarrow
(2)$. Comparing this formula with \Eq{eq:SxSx:S+S-}, one can see that
the calculation one has to perform is exactly the same, up to two sign
differences and one obtains
\begin{align}
\langle \Psi_{2\, {\rm sq}}\vert \hat{S}^{(1)}_y(\ell) \hat{S}^{(2)}_y(\ell)
\vert \Psi_{2\, {\rm sq}}\rangle &=
\frac{\sqrt{\gamma_1\gamma_2}}{2\pi}
\sum _{n,m}\left[-X^{(1)}_{n,m}-X^{(3)}_{n,m}+X^{(2)}_{n,m}+X^{(4)}_{n,m}\right]
\, .
\label{eq:SySy:exact}
\end{align}
This correlation function is displayed in \Fig{fig:exactSySy}. The
left panel represents $\langle \Psi_{2\, {\rm sq}}\vert
\hat{S}^{(1)}_y(\ell) \hat{S}^{(2)}_y(\ell) \vert \Psi_{2\, {\rm
    sq}}\rangle$ for different values of $r$ and a vanishing squeezing
angle while the right panel is for $r=1$ and different squeezing
angles. The correlation function vanishes at small and large $\ell$,
in agreement with the analytical results of \Secs{sec:largeelllimit}
and~\ref{sec:smallelllimit}. Otherwise, the same remarks as the ones
made for $\langle \Psi_{2\, {\rm sq}}\vert \hat{S}^{(1)}_x(\ell)
\hat{S}^{(2)}_x(\ell) \vert \Psi_{2\, {\rm sq}}\rangle$ are still
valid in the present case and, therefore, need not be repeated
here. In particular, the same correspondence between $\varphi$ and
$\pi/2-\varphi$ takes place, so that these two configurations are
``dual'' and connected through the formulas
\begin{align}
\label{eq:dual:corr:SzSz}
\langle \Psi_{2\, {\rm sq}}\vert \hat{S}^{(1)}_z(\ell) 
\hat{S}^{(2)}_z(\ell)\vert \Psi_{2\, {\rm sq}}\rangle 
\left(r,\pi/2-\varphi\right)&=-\langle 
\Psi_{2\, {\rm sq}}\vert \hat{S}^{(1)}_z(\ell) 
\hat{S}^{(2)}_z(\ell)\vert \Psi_{2\, {\rm sq}}
\rangle \left(r,\varphi\right)\, ,\\
\label{eq:dual:corr:SxSx}
\langle \Psi_{2\, {\rm sq}}\vert \hat{S}^{(1)}_x(\ell) 
\hat{S}^{(2)}_x(\ell)\vert \Psi_{2\, {\rm sq}}\rangle 
\left(r,\pi/2-\varphi\right)&=\langle \Psi_{2\, {\rm sq}}
\vert \hat{S}^{(1)}_x(\ell) \hat{S}^{(2)}_x(\ell)\vert 
\Psi_{2\, {\rm sq}}\rangle \left(r,\varphi\right)\, ,\\
\langle \Psi_{2\, {\rm sq}}\vert \hat{S}^{(1)}_y(\ell) 
\hat{S}^{(2)}_y(\ell)\vert \Psi_{2\, {\rm sq}}\rangle 
\left(r,\pi/2-\varphi\right)&=-\langle \Psi_{2\, {\rm sq}}
\vert \hat{S}^{(1)}_y(\ell) \hat{S}^{(2)}_y(\ell)\vert 
\Psi_{2\, {\rm sq}}\rangle \left(r,\varphi\right)\, .
\label{eq:dual:corr:SySy}
\end{align}

\subsection{Correlation Function $\langle \Psi_{2\, {\rm sq} }\vert
  \hat{S}^{(1)}_x(\ell) \hat{S}^{(2)}_z(\ell) \vert \Psi_{2\, {\rm
      sq}}\rangle$.}
\label{sec:spinPheno:xz}
Let us finally calculate the cross correlation function of the
operators $\hat{S}_x(\ell)$ and $\hat{S}_z(\ell)$. Since
$\hat{S}_x(\ell)=\hat{S}_+(\ell)+\hat{S}_-(\ell)$, see \Eq{eq:defsx},
one has
\begin{align}
\langle \Psi_{2\, {\rm sq}}\vert \hat{S}^{(1)}_x(\ell) \hat{S}^{(2)}_z(\ell)
\vert \Psi_{2\, {\rm sq}}\rangle =& 
\langle \Psi_{2\, {\rm sq}}\vert \hat{S}^{(1)}_+(\ell) \hat{S}^{(2)}_z(\ell)
\vert \Psi_{2\, {\rm sq}}\rangle
+\langle \Psi_{2\, {\rm sq}}\vert \hat{S}^{(1)}_-(\ell) \hat{S}^{(2)}_z(\ell)
\vert \Psi_{2\, {\rm sq}}\rangle\\
=& 
\langle \Psi_{2\, {\rm sq}}\vert \hat{S}^{(1)}_+(\ell) \hat{S}^{(2)}_z(\ell)
\vert \Psi_{2\, {\rm sq}}\rangle
+\langle \Psi_{2\, {\rm sq}}\vert \hat{S}^{(1)}_z(\ell) \hat{S}^{(2)}_-(\ell)
\vert \Psi_{2\, {\rm sq}}\rangle
\label{eq:SxSz:S+Sz:1}
\\
=& 2\Re\mathrm{e}\left[\langle \Psi_{2\, {\rm sq}}\vert \hat{S}^{(1)}_+(\ell) \hat{S}^{(2)}_z(\ell)
\vert \Psi_{2\, {\rm sq}}\rangle\right]\, ,
\label{eq:SxSz:S+Sz}
\end{align}
where, in \Eq{eq:SxSz:S+Sz:1}, one has used the fact that the two-mode
squeezed state is symmetric if one exchanges the sub-spaces $(1)$ and
$(2)$, see \Eq{eq:qstateposition}, and where in \Eq{eq:SxSz:S+Sz}, one
has used that $S_+(\ell)=S_-^\dagger(\ell)$. One therefore has to
calculate the quantity
\begin{align}
\label{eq:S+S+:Psi}
\langle \Psi_{2\, {\rm sq}}\vert \hat{S}^{(1)}_+(\ell) \hat{S}^{(2)}_z(\ell)
\vert \Psi_{2\, {\rm sq}}\rangle &=\sum _{n=-\infty}^{\infty}
\sum_{m=-\infty}^{\infty}
(-1)^n
\int _{n\ell}^{(n+1)\ell}{\rm d}Q_1
\int _{2m\ell}^{(2m+1)\ell}{\rm d}Q_2
\Psi_{\rm 2\, sq}^*\left(Q_1,Q_2\right)
\Psi_{\rm 2\, sq}\left(Q_1+\ell,Q_2\right)\\
& = \frac{1}{\pi \cosh^2 r}
\frac{1}{\sqrt{
\tanh^4 r-2\tanh^2 r \cos(4\varphi)+1}}
\sum _{n=-\infty}^{\infty}
\sum_{m=-\infty}^{\infty}
(-1)^n
L_{n,m}
\end{align}
with
\begin{align}
L_{n,m}=\int _{n\ell}^{(n+1)\ell}{\rm d}Q_1
\int _{2m\ell}^{(2m+1)\ell}{\rm d}Q_2
\ee^{A\ell^2}\ee^{(A+A_*)(Q_1^2+Q_2^2)-(B+B_*)Q_1Q_2+(2A Q_1-BQ_2)\ell}\, .
\end{align}
Let us now perform the same change of integration variable as in
\Sec{sec:spinPheno:z}, namely $Q_1=u+v$ and $Q_2=u-v$. As a
consequence, one obtains
\begin{align}
L_{n,m}& = 2\ee^{A\ell^2} \int_{(n+2m)\ell/2}^{(n+2m+1)\ell/2}\dd u
\ee^{-\gamma_1 u^2 -\left(\frac{\gamma_1}{2}-i\gamma_3\right)\ell u}
\int_{n\ell-u}^{u-2m\ell}\dd v
\ee^{-\gamma_2 v^2-\left(\frac{\gamma_2}{2}+i\gamma_4\right)\ell v}
\nonumber\\ 
& +2\ee^{A\ell^2} \int_{(n+2m+1)\ell/2}^{(n+2m+2)\ell/2} \dd u
\ee^{-\gamma_1 u^2 -\left(\frac{\gamma_1}{2}-i\gamma_3\right)\ell u}
\int_{u-(2m+1)\ell}^{(n+1)\ell-u}\dd v
\ee^{-\gamma_2 v^2-\left(\frac{\gamma_2}{2}+i\gamma_4\right)\ell v}
\equiv L_{n,m}^{(1)}+L_{n,m}^{(2)}\, .
\end{align}
Similarly as before, the change of integration variable
$z=2u/\ell-n-2m$ in $L_{n,m}^{(1)}$ and $z=-2u/\ell+n+2m+2$ in
$L_{n,m}^{(2)}$ leads to
\begin{align}
L_{n,m}^{(1)} &= \ell \ee^{A\ell^2} \int_{0}^{1}\dd z
\ee^{-\gamma_1 u^2 -\left(\frac{\gamma_1}{2}-i\gamma_3\right)\ell u}
\int_{(n-2m-z)\ell/2}^{(n-2m+z)\ell/2}\dd v
\ee^{-\gamma_2 v^2-\left(\frac{\gamma_2}{2}+i\gamma_4\right)\ell v}
\\
L_{n,m}^{(2)} &= \ell \ee^{A\ell^2} \int_{0}^{1} \dd z
\ee^{-\gamma_1 u^2 -\left(\frac{\gamma_1}{2}-i\gamma_3\right)\ell u}
\int^{(n-2m-z)\ell/2}_{(n-2m+z)\ell/2}\dd v
\ee^{-\gamma_2 v^2-\left(\frac{\gamma_2}{2}+i\gamma_4\right)\ell v}\, .
\end{align}
These two expressions are the same, except that the lower and upper
bounds of the integral over $v$ are inverted, hence
$L_{n,m}^{(1)}=-L_{n,m}^{(2)}$ and $L_{n,m}=0$. We have thus shown
that $\langle \Psi_{2\, {\rm sq}}\vert \hat{S}^{(1)}_+(\ell)
\hat{S}^{(2)}_z(\ell) \vert \Psi_{2\, {\rm sq}}\rangle=0$, and
consequently
\begin{align}
\langle \Psi_{2\, {\rm sq}}\vert \hat{S}^{(1)}_x(\ell) \hat{S}^{(2)}_z(\ell)
\vert \Psi_{2\, {\rm sq}}\rangle =0\, ,
\end{align}
that is to say measurements along orthogonal directions are
uncorrelated for the two-mode squeezed state.

Having established the exact (numerical) form of the spin correlation
functions, we now turn to the question of finding analytical
approximations.

\section{Large and Small $\ell$ Limits}
\label{sec:Appr}
In \Figs{fig:exactSzSz}, \ref{fig:exactSxSx} and~\ref{fig:exactSySy},
one can see that the two-point correlation functions of the spin
operators reach constant values at small and large $\ell$. In this
section, we derive the analytical expressions of the corresponding
asymptotic values.
\subsection{The large $\ell$ limit}
\label{sec:largeelllimit}
Let us first consider the asymptotic behavior of the correlation
functions at large $\ell$. We start by treating the correlation
function of the $z$-component of the spin. One can notice that the
integration domains appearing in $Z_{n,m}^{(1)}$ defined in
\Eq{eq:Z1:def} are of two kinds. Either they contain the point
$(u=0,v=0)$, close to which the integrand is maximal, either they do
not. In the second case, when $\ell\gg 1$, the integrand is
exponentially suppressed and the corresponding value for
$Z_{n,m}^{(1)}$ negligibly contributes to the overall
sum~(\ref{eq:SzSz:exact}). Therefore, in the $\ell\gg 1$ limit, it is
enough to keep the contributions from the first kind of integrals
only. It is easy to see that three terms are of this first kind, namely
$Z_{0,0}^{(1)}$, $Z_{-1,0}^{(1)}$ and $Z_{0,-1}^{(1)}$. In the limit
$\ell\rightarrow\infty$, they are given by
\begin{align}
Z_{0,0}^{(1)}& \simeq 2 \int_0^\infty\dd u\ee^{-\gamma_1 u^2} 
\int_{-u}^u\dd v\ee^{-\gamma_2 v^2}\, ,\quad 
Z_{-1,0}^{(1)} \simeq 2 \int_{-\infty}^0\dd u\ee^{-\gamma_1 u^2}
\int_{-\infty}^u\dd v\ee^{-\gamma_2 v^2}\, ,
\\
Z_{0,-1}^{(1)}& \simeq 2 \int_{-\infty}^0\dd u\ee^{-\gamma_1 u^2}
\int_{-u}^{\infty}\dd v\ee^{-\gamma_2 v^2}\, .
\end{align}
As a consequence, the sum appearing in \Eq{eq:SzSz:exact} can be
written as
\begin{align}
\sum _{n,m}(-1)^{n+m}Z_{n,m}^{(1)} &\simeq Z_{0,0}^{(1)} - Z_{-1,0}^{(1)} - Z_{0,-1}^{(1)}
= 2 \int_0^\infty\dd u\ee^{-\gamma_1 u^2} \left(\int_{-u}^u\dd v\ee^{-\gamma_2 v^2} - 2 \int_{u}^\infty\dd v\ee^{-\gamma_2 v^2}\right)
\\
&= 2\sqrt{\frac{\pi}{\gamma_2}} \int_0^\infty\dd u\ee^{-\gamma_1 u^2} \left[ 2\erf\left(u\sqrt{\gamma_2}\right)-1\right]
= \frac{2}{\sqrt{\gamma_1\gamma_2}}\left[2\arctan\left(\sqrt{\frac{\gamma_2}{\gamma_1}}\right)-\frac{\pi}{2}\right]\, .
\end{align}
This formula can be simplified and it is convenient to rewrite it
as\footnote{By inverting the relation $\tan(x+\pi/4)=(1+\tan
  x)/(1-\tan x)$, one obtains a relation between $\arctan(x)$ and
  $\arctan[(x-1)/(x+1)]$. One can therefore express the result with a
  single $\arctan$ function only:
\begin{align}
  \sum _{n,m}(-1)^{n+m}Z_{n,m}^{(1)} =
  \frac{4}{\sqrt{\gamma_1\gamma_2}}\arctan\left(\frac{\sqrt{\gamma_2}-\sqrt{\gamma_1}}{\sqrt{\gamma_2}+\sqrt{\gamma_1}}\right)\,
  .
\end{align}
Then one can use the generic relation $\arctan(\tanh x) = 1/2 \arctan(\sinh 2x)$ to write $\arctan x = 1/2 \arctan[2x/(1-x^2)]$. The later is valid only when $-1<x<1$, but this condition is verified by $(\sqrt{\gamma_2}-\sqrt{\gamma_1})/(\sqrt{\gamma_2}+\sqrt{\gamma_1})$.
}
\begin{align}
\sum _{n,m}(-1)^{n+m}Z_{n,m}^{(1)} \simeq \frac{2}{\sqrt{\gamma_1\gamma_2}}\arctan\left(\frac{\gamma_2-\gamma_1}{2\sqrt{\gamma_1\gamma_2}}\right)\, .
\end{align}
Making use of \Eqs{eq:SzSz:exact}, (\ref{eq:gamma1:def})
and~(\ref{eq:gamma2:def}), the last equation can expressed explicitly
in terms of the squeezing parameter $r$ and squeezing parameter angle
$\varphi$. One eventually obtains
\begin{align}
& \langle \Psi_{2\, {\rm sq}}\vert \hat{S}^{(1)}_z(\ell) \hat{S}^{(2)}_z(\ell)
\vert \Psi_{2\, {\rm sq}}\rangle \simeq
\frac{2}{\pi}\arctan\left[\frac{\cos(2\varphi)\sinh(2r)}{\sqrt{\cosh^2(2r)-\cos^2(2\varphi)\sinh^2(2r)}}\right]\, .
\label{eq:largeell:SzSz}
\end{align}
In the case where $\varphi=0$, one obtains $\langle \Psi_{2\, {\rm
    sq}}\vert \hat{S}^{(1)}_z(\ell) \hat{S}^{(2)}_z(\ell) \vert
\Psi_{2\, {\rm sq}}\rangle \simeq 2 \arctan[\sinh(2r)]/\pi$, in
agreement with Eq.~(17) of \Ref{2004PhRvA..70b2102L}, but the
expression derived here is more general.  The asymptotic plateau given
by \Eq{eq:largeell:SzSz} is compared to the numerical curve obtained
from \Eq{eq:SzSz:exact} in \Fig{fig:approxSzSz} (black line), and
one can check that the agreement is indeed excellent.

\par

The same strategy can be employed to approximate the correlation
functions of $\hat{S}_x(\ell)$ and $\hat{S}_y(\ell)$ in the large
$\ell$ limit. From \Eq{eq:X1X2X3X4:def}, it is clear that no
integration domain in $X_{n,m}$ is of the first kind [\ie contains the
point $(u,v)=(0,0)$] and therefore, $\langle \Psi_{2\, {\rm sq}}\vert
\hat{S}^{(1)}_x(\ell) \hat{S}^{(2)}_x(\ell) \vert \Psi_{2\, {\rm
    sq}}\rangle \simeq \langle \Psi_{2\, {\rm sq}}\vert
\hat{S}^{(1)}_y(\ell) \hat{S}^{(2)}_y(\ell) \vert \Psi_{2\, {\rm
    sq}}\rangle \simeq 0 $ in this limit.
\subsection{The small $\ell$ limit}
\label{sec:smallelllimit}
Let us now consider the asymptotic behavior of the correlation
functions in the opposite limit, $\ell\ll 1$. As before, we first
consider the correlation function for the $z$-component of the
spin. Since $\ell$ is small, one can simplify the integrand in
\Eq{eq:Z1:def:simp} by expanding the error functions around
$\ell\sqrt{\gamma_2}(n-m)/2$:
\begin{align}
{\erf}\left[\frac{\ell}{2}\sqrt{\gamma_2}\left(z+n-m\right)\right]
+{\erf}\left[\frac{\ell}{2}\sqrt{\gamma_2}\left(z-n+m\right)\right]&=
{\erf}\left[\frac{\ell}{2}\sqrt{\gamma_2}\left(n-m+z\right)\right]-
{\erf}\left[\frac{\ell}{2}\sqrt{\gamma_2}\left(n-m-z\right)\right]\\&\simeq
2\ell\sqrt{\frac{\gamma_2}{\pi}}\ee^{-\ell^2\gamma_2(n-m)^2/4} z\, .
\end{align}
One then has
\begin{align}
Z_{n,m}^{(1)}&\simeq \ell^2 \ee^{-\ell^2\gamma_2(n-m)^2/4} \int_0^1\dd z\ee^{-\gamma_1 \ell^2 (z+n+m)^2/4} z\, .
\end{align}
This integral can be expressed in terms of the error
function. Alternatively, one can notice that when $\ell\ll 1$, the
argument of the exponential in the integrand vanishes, except if
$\vert n+m\vert \gg 1$. In the later case, $z+n+m\simeq n+m$ and one
can therefore replace $z$ by $0$ in the argument of the exponential in
this limit. As a consequence, one obtains
\begin{align}
Z_{n,m}^{(1)}&\simeq \frac{\ell^2}{2} \ee^{-\frac{\ell^2}{4}\left[\gamma_2(n-m)^2+\gamma_1(n+m)^2\right]}\, .
\end{align}
In this expression, only the combinations $n-m$ and $n+m$ are
involved. Defining $p\equiv n+m$ and $q\equiv n-m$, it follows that
the sum appearing in \Eq{eq:SzSz:exact} can be written as
\begin{align}
\sum_{n,m=-\infty}^\infty (-1)^{n+m}Z_{n,m}^{(1)}=
\sum_{q=-\infty}^\infty\mathcal{Z}_q^{(1)}\sum_{p\in\mathcal{P}(q)}\mathcal{Z}_p^{(1)}\, ,
\label{eq:smallell:SzSz:pq}
\end{align}
where the quantities $\mathcal{Z}_q^{(1)}$ and $\mathcal{Z}_p^{(1)}$
are defined by
\begin{align}
\mathcal{Z}_q^{(1)} = \frac{\ell^2}{2} \ee^{-\gamma_2 q^2\frac{\ell^2}{4}}
\, ,\quad\quad
\mathcal{Z}_p^{(1)} = (-1)^p \ee^{-\gamma_1 p^2\frac{\ell^2}{4}}\, .
\label{eq:smallell:Zqp}
\end{align}
In \Eq{eq:smallell:SzSz:pq}, the symbol $\mathcal{P}(q)$ stands for
all integer numbers having the same parity as $q$. Therefore, if $q$
is even, the sum over $p$ can be expressed as
\begin{align}
\sum_{p\in\mathcal{P}(q)}\mathcal{Z}_p^{(1)} = 
\sum_{p^\prime=-\infty}^\infty\mathcal{Z}_{2p^\prime}^{(1)}
=\sum_{p^\prime=-\infty}^\infty\ee^{-\gamma_1 {p^\prime}^2\ell^2}
=\vartheta_3\left(0,\ee^{-\gamma_1\ell^2}\right)
\label{eq:smalll:sumZp:even}
\end{align}
where $\vartheta_3$ is the third Jacobi
function~\cite{Gradshteyn:1965aa}. On the other hand, if $q$ is odd,
this sum can be written as
\begin{align}
\sum_{p\in\mathcal{P}(q)}\mathcal{Z}_p^{(1)} = \sum_{p^\prime=-\infty}^\infty\mathcal{Z}_{2p^\prime+1}^{(1)}=-\sum_{p^\prime=-\infty}^\infty\ee^{-\gamma_1 \left(p^\prime+\frac{1}{2}\right)^2\ell^2}=-\vartheta_2\left(0,\ee^{-\gamma_1\ell^2}\right)
\label{eq:smalll:sumZp:odd}
\end{align}
where $\vartheta_2$ is the second Jacobi
function~\cite{Gradshteyn:1965aa}. When $\ell\ll 1$, the behavior of
the two Jacobi functions is similar
\footnote{\label{footnote:theta:asympt}Here, we make use of the
  asymptotic formula~\cite{Gradshteyn:1965aa} $
  \theta_2(b,\ee^{-a})\sim \theta_3(b,\ee^{-a})\sim
  \sqrt{\frac{\pi}{a}}\ee^{-\frac{b^2}{a}}\, , $ valid when when $a,\,
  b\ll 1$.}  and one obtains
\begin{align}
\sum_{n,m=-\infty}^\infty (-1)^{n+m}Z_{n,m}^{(1)}&=
\sqrt{\frac{\pi}{\gamma_1}}\frac{\ell}{2}\left[
\sum_{q=-\infty}^\infty \ee^{-\gamma_2 q^2\ell^2}
-\sum_{q=-\infty}^\infty \ee^{-\gamma_2 \left(q+\frac{1}{2}\right)^2\ell^2}
\right]
\\ &
= \sqrt{\frac{\pi}{\gamma_1}}\frac{\ell}{2}\left[
\vartheta_3\left(0,\ee^{-\gamma_2\ell^2}\right)
-\vartheta_2\left(0,\ee^{-\gamma_2\ell^2}\right)
\right]
\simeq
\frac{2\pi}{\sqrt{\gamma_1\gamma_2}}\ee^{-\frac{\pi^2}{\gamma_2\ell^2}},
\end{align}
where in the last expression one has used the limit $\ell\ll 1$ again.
Since $\gamma_2>0$, this means that $\langle \Psi_{2\, {\rm sq}}\vert
\hat{S}^{(1)}_z(\ell) \hat{S}^{(2)}_z(\ell) \vert \Psi_{2\, {\rm
    sq}}\rangle $ vanishes when $\ell\rightarrow 0$ in accordance with
what is observed in \Figs{fig:exactSzSz}.

\par

The same calculation can be performed for the correlation functions of
the $x$- and $y$- components of the spin. Let us first express
$X_{n,m}^{(1)}$, $X_{n,m}^{(2)}$, $X_{n,m}^{(3)}$ and $X_{n,m}^{(4)}$
in the $\ell\ll 1$ limit. In \Eqs{eq:X1:def} and~(\ref{eq:X3:def}),
the error functions can be expanded around $\ell\sqrt{\gamma_2}(n-m)$
and one obtains
\begin{align}
\erf\left[\frac{\ell}{2}\sqrt{\gamma_2}
(z-2n+2m)\right]
+\erf\left[\frac{\ell}{2}\sqrt{\gamma_2}
(z+2n-2m)\right]\simeq
2\ell\sqrt{\frac{\gamma_2}{\pi}}\ee^{-\ell^2\gamma_2(n-m)^2} z\, .
\end{align}
One then has
\begin{align}
X_{n,m}^{(1)} &\simeq 
2\ell^2
\ee^{-\gamma_1\ell^2/2-\gamma_2(n-m)^2 \ell^2}
\int _0^1 {\rm d}z
\ee^{-\gamma_1 \ell^2(z+2n+2m)^2/4-\gamma_1\ell^2(z+2n+2m)/2}
\cos\left[\gamma_3\ell^2(z+2n+2m+1)\right] z\, .
\end{align}
For the same reasons as the ones explained before, in the $\ell\ll 1$
limit, one can put $z=0$ in the exponentials and in the cosine of the
integrand and $X_{n,m}^{(1)}$ takes the following form
\begin{align}
X_{n,m}^{(1)} &\simeq 
\ell^2
\ee^{-\gamma_1\ell^2/2-\gamma_2(n-m)^2 \ell^2-\gamma_1 (n+m)^2\ell^2-\gamma_1(n+m)\ell^2}
\cos\left[\gamma_3\ell^2(2n+2m+1)\right]\, .
\end{align}
The same trick can be used for $X_{n,m}^{(3)}$ resulting in
$X_{n,m}^{(3)}\simeq X^{(1)}_{n+1/2,m+1/2}$.

Let us now consider the quantity $X_{n,m}^{(2)}$. In the same manner,
in \Eqs{eq:X2:def}-(\ref{eq:X4:def}), the error functions can be
expanded at leading order in $\ell$ around $\ell\sqrt{\gamma_2}(n-m)$
and one obtains
\begin{align}
& \Re\mathrm{e}\left\lbrace {\rm erf}\left[\frac{\ell}{2}\sqrt{\gamma_2}
(z-2n+2m)+i\frac{\gamma_4}{\sqrt{\gamma_2}}\ell\right]
+{\rm erf}\left[\frac{\ell}{2}\sqrt{\gamma_2}
(z+2n-2m)+i\frac{\gamma_4}{\sqrt{\gamma_2}}\ell\right]
\right\rbrace
\simeq 2 \ell\sqrt{\frac{\gamma_2}{\pi}}\ee^{-\ell^2\gamma_2(n-m)^2}z\, .
\end{align}
It follows that $X_{n,m}^{(2)}$ can be expressed as
\begin{align}
X_{n,m}^{(2)} &\simeq
2\ell^2
\ee^{-\left[(\gamma_1+\gamma_2)/4+\gamma_4^2/\gamma_2+\gamma_2(n-m)^2\right]\ell^2}
\int _0^1 {\rm d}z
\ee^{-\gamma_1 \ell^2(z+2n+2m)^2/4-\gamma_1\ell^2(z+2n+2m)/2}z\, .
\end{align}
As before, in the $\ell\ll 1$ limit, $z$ can be replaced with $0$ in
the argument of the exponential function of the integrand and one
obtains
\begin{align}
\label{eq:x2inter}
X_{n,m}^{(2)} &\simeq
\ell^2
\ee^{-\left[(\gamma_1+\gamma_2)/4+\gamma_4^2/\gamma_2+\gamma_2(n-m)^2+\gamma_1(n+m)^2+\gamma_1(n+m)\right]\ell^2}\, ,
\end{align}
and $X_{n,m}^{(4)}\simeq X_{n+1/2,m+1/2}^{(2)}$.

The next step is to calculate the following sum:
$\sum_{n,m}X^{(1)}_{n,m}+X^{(1)}_{n+1/2,m+1/2}+X^{(2)}_{n,m}+X^{(2)}_{n+1/2,m+1/2}$.
We notice that, again, the terms of this sum depends on the previously
defined $p=n+m$ and $q=n-m$ only. Therefore, one can use the same
techniques to perform the calculation. The first term is given by
\begin{align}
\sum_{n,m=-\infty}^\infty X_{n,m}^{(1)} = 
\sum_{q=-\infty}^\infty\mathcal{X}_q^{(1)}\sum_{p\in\mathcal{P}(q)}\mathcal{X}_p^{(1)}\, ,
\end{align}
where the quantities $\mathcal{X}_q^{(1)}$ and $\mathcal{X}_p^{(1)}$
are defined by
\begin{align}
\label{eq:smallell:X1pq:def}
\mathcal{X}_q^{(1)} =  \ell^2
\ee^{-\gamma_1\ell^2/2-\gamma_2 q^2 \ell^2}\, ,
\quad\quad
\mathcal{X}_p^{(1)} = 
\ee^{-\gamma_1 p^2\ell^2-\gamma_1 p\ell^2}
\cos\left[\gamma_3\ell^2(2p+1)\right]\, ,
\end{align}
and where the meaning of the symbol ${\cal P}(q)$ is the same as
before. As a consequence, if $q$ is even, one finds that the sum over
$p$ can be expressed as
\begin{align}
\sum_{p\in\mathcal{P}(q)}\mathcal{X}_p^{(1)} &=
\sum_{p^\prime=-\infty}^\infty\mathcal{X}_{2p^\prime}^{(1)}=
\sum_{p^\prime=-\infty}^\infty
\ee^{-4 \gamma_1 {p^\prime}^2\ell^2-2\gamma_1 p^\prime\ell^2}
\cos\left[\gamma_3\ell^2(4p^\prime+1)\right]\\
&=\Re\mathrm{e}\left[
\ee^{i\gamma_3\ell^2}\sum_{p^\prime=-\infty}^\infty
\ee^{-4\gamma_1\ell^2{p^\prime}^2+4\left(-\gamma_1/2+i\gamma_3\right)\ell^2 p^\prime}
 \right]
 =\Re\mathrm{e}\left\lbrace
\ee^{i\gamma_3\ell^2}\vartheta_3\left[2\ell^2
\left(\gamma_3+i\frac{\gamma_1}{2}\right),\ee^{-4\gamma_1\ell^2}\right]
 \right\rbrace
 \\&\simeq 
 \Re\mathrm{e}\left\lbrace
\ee^{i\gamma_3\ell^2}\sqrt{\frac{\pi}{\gamma_1}}
\frac{1}{2\ell}\ee^{-\frac{\left(\gamma_3+i\gamma_1/2\right)^2}{\gamma_1}\ell^2}
 \right\rbrace
\simeq \sqrt{\frac{\pi}{\gamma_1}}\frac{1}{2\ell}\, .
\end{align}
where the asymptotic formula given in
footnote~\ref{footnote:theta:asympt} has been used again. On the other
hand, if $q$ is odd, the same kind of manipulations lead to
\begin{align}
\sum_{p\in\mathcal{P}(q)}\mathcal{X}_p^{(1)} &=
\sum_{p^\prime=-\infty}^\infty\mathcal{X}_{2p^\prime+1}^{(1)}=
\sum_{p^\prime=-\infty}^\infty
\ee^{-4 \gamma_1 \left({p^\prime}+\frac{1}{2}\right)^2\ell^2-2\gamma_1
\left(p^\prime+\frac{1}{2}\right)\ell^2}
\cos\left[\gamma_3\ell^2(4p^\prime+3)\right]\\
&=\Re\mathrm{e}\left[\ee^{\left(3i\gamma_3-2\gamma_1+\right)\ell^2}
\sum_{p^\prime=-\infty}^\infty\ee^{-4\gamma_1\ell^2 {p^\prime}^2 
+4\left(i\gamma_3-3\gamma_1/2\right)\ell^2 p^\prime}
\right]
=\Re\mathrm{e}\left\lbrace\ee^{\left(3i\gamma_3-2\gamma_1\right)\ell^2}
\vartheta_3\left[2\ell^2\left(\gamma_3+3i\frac{\gamma_1}{2}\right),
\ee^{-4\gamma_1\ell^2}\right]
\right\rbrace
\\ &
\simeq\Re\mathrm{e}\left[\ee^{\left(3i\gamma_3-2\gamma_1\right)\ell^2}
\sqrt{\frac{\pi}{\gamma_1}}\frac{1}{2\ell}
\ee^{-\frac{\ell^2}{\gamma_1}\left(\gamma_3+3i\gamma_1/2\right)^2}
\right]
\simeq 
\sqrt{\frac{\pi}{\gamma_1}}\frac{1}{2\ell}\, .
\end{align}
We see that the result is in fact independent of the parity of $p$. As
a consequence, the first sum is given by the following expression
\begin{align}
\sum_{n,m=-\infty}^\infty X_{n,m}^{(1)} &= \sqrt{\frac{\pi}{\gamma_1}}
\frac{1}{2\ell} \sum_{q=-\infty}^\infty \mathcal{X}_q^{(1)}
=\sqrt{\frac{\pi}{\gamma_1}}\frac{\ell}{2}\ee^{-\gamma_1\ell^2/2}
\sum_{q=-\infty}^\infty \ee^{-\gamma_2\ell^2 q^2}
\\& = 
\sqrt{\frac{\pi}{\gamma_1}}\frac{\ell}{2}\ee^{-\gamma_1/2\ell^2}
\vartheta_3\left(0,\ee^{-\gamma_2\ell^2}\right)\simeq
\frac{\pi}{2\sqrt{\gamma_1\gamma_2}}\, .
\end{align}
Then, let us quickly treat the third sum since this one leads to a
result identical to the first one. Indeed, one has
\begin{align}
\sum_{n,m=-\infty}^\infty X_{n,m}^{(3)}=
\sum_{n,m=-\infty}^\infty X^{(1)}_{n+1/2,m+1/2}=\sum_{q=-\infty}^\infty
\mathcal{X}_q^{(1)}\sum_{p\in\mathcal{P}(q)}\mathcal{X}_{p+1}^{(1)}
=\sum_{q=-\infty}^\infty\mathcal{X}_q^{(1)}\sum_{p\in\bar{\mathcal{P}}(q)}
\mathcal{X}_{p}^{(1)}\, .
\end{align}
As just mentioned, the calculation one has to perform is therefore
very similar to the first sum, the only difference being that $p$ is
now summed over $\bar{\mathcal{P}}(q)$, \ie over integer numbers
having the opposite parity as $q$. But we have just seen that for the
first sum, the summation over $p$ gives a result that is, at leading
order in $\ell$, independent of the parity of $q$. As announced, the
first and third sums are therefore the same.

\par

The next step is to calculate the second sum. In \Eq{eq:x2inter}, we
see that the term $X_{n,m}^{(2)}$ also depends on $p$ and $q$ only. Therefore, 
one can write that
\begin{align}
\sum_{n,m=-\infty}^\infty X_{n,m}^{(2)} = 
\sum_{q=-\infty}^\infty\mathcal{X}_q^{(2)}\sum_{p\in\mathcal{P}(q)}\mathcal{X}_{p}^{(2)}
\end{align}
where the quantities $\mathcal{X}_q^{(2)}$ and $\mathcal{X}_p^{(2)}$
are defined by the following expressions
\begin{align}
\label{eq:smallell:X2pq:def}
\mathcal{X}_q^{(2)}=
\ell^2
\ee^{-\left[(\gamma_1+\gamma_2)/4+\gamma_4^2/\gamma_2+\gamma_2q^2\right]\ell^2}
\, ,\quad\quad
\mathcal{X}_p^{(2)}=
\ee^{-\gamma_1p^2\ell^2-\gamma_1 p\ell^2}\, .
\end{align}
Then, one can apply the same techniques as before and distinguish the
cases where $q$ is even and odd. If $q$ is even, the sum over $p$
takes the form
\begin{align}
\sum_{p\in\mathcal{P}(q)}\mathcal{X}_p^{(2)} &=
\sum_{p^\prime=-\infty}^\infty\mathcal{X}_{2p^\prime}^{(2)}=
\sum_{p^\prime=-\infty}^\infty\ee^{-4\gamma_1\ell^2{p^\prime}^2-2\gamma_1\ell^2 p^\prime}
\\
&=\vartheta_3\left(i\gamma_1\ell^2,\ee^{-4\gamma_1\ell^2}\right)
\simeq \sqrt{\frac{\pi}{\gamma_1}}\frac{1}{2\ell}\ee^{\frac{\gamma_1}{4}\ell^2}
\simeq \sqrt{\frac{\pi}{\gamma_1}}\frac{1}{2\ell}\, ,
\end{align}
while, if $q$ is odd, one obtains the following result
\begin{align}
\sum_{p\in\mathcal{P}(q)}\mathcal{X}_p^{(2)} &=
\sum_{p^\prime=-\infty}^\infty\mathcal{X}_{2p^\prime+1}^{(2)}=
\sum_{p^\prime=-\infty}^\infty\ee^{-4\gamma_1\ell^2({p^\prime}
+\frac{1}{2})^2-2\gamma_1\ell^2 (p^\prime+\frac{1}{2})}
\\
&=\ee^{-2\gamma_1\ell^2}
\vartheta_3\left[3\ell^2i\gamma_1,\ee^{-4\gamma_1\ell^2}\right]
\simeq \ee^{-2\gamma_1\ell^2}
\sqrt{\frac{\pi}{\gamma_1}}\frac{1}{2\ell}
\ee^{\frac{9}{4}\gamma_1\ell^2}
\simeq \sqrt{\frac{\pi}{\gamma_1}}\frac{1}{2\ell}\, .
\end{align}
Again, we notice that the result (at least in the limit considered
here) does not depend on the parity of $q$. As a consequence, one
finds that the second sum is given by
\begin{align}
\sum_{n,m=-\infty}^\infty X_{n,m}^{(2)} &= 
\sqrt{\frac{\pi}{\gamma_1}}\frac{1}{2\ell} \sum_{q=-\infty}^\infty 
\mathcal{X}_q^{(2)}
=\sqrt{\frac{\pi}{\gamma_1}}\frac{\ell}{2}
\ee^{-(\gamma_1+\gamma_2)\ell^2-\gamma_4^2\ell^2/\gamma_2}
\sum_{q=-\infty}^\infty \ee^{-\gamma_2\ell^2 q^2}
\\&= \sqrt{\frac{\pi}{\gamma_1}}\frac{\ell}{2}
\ee^{-(\gamma_1+\gamma_2)\ell^2-\gamma_4^2\ell^2/\gamma_2}
\vartheta_3\left(0,\ee^{-\gamma_2\ell^2}\right)
\simeq\frac{\pi}{2\sqrt{\gamma_1\gamma_2}}\, .
\end{align}
Calculating the fourth sum remains to be done. As the third sum was
equal to the first one, it is clear that the fourth one will be
identical to the second one we have just evaluated. Straightforward
manipulations confirm this guess, namely
\begin{align}
\sum_{n,m=-\infty}^\infty X_{n,m}^{(4)}=
\sum_{n,m=-\infty}^\infty X^{(2)}_{n+1/2,m+1/2}
=\sum_{q=-\infty}^\infty\mathcal{X}_q^{(2)}
\sum_{p\in\mathcal{P}(q)}\mathcal{X}_{p+1}^{(2)}
=\sum_{q=-\infty}^\infty\mathcal{X}_q^{(2)}
\sum_{p\in\bar{\mathcal{P}}(q)}\mathcal{X}_{p}^{(2)}\, .
\end{align}
As announced above, in the same manner as before, the fourth and
second sums are very similar, the only difference being that $p$ is
now summed over integer numbers having the opposite parity as $q$. But
since for the second sum, the summation over $p$ gives a result that
is, at leading order in $\ell$, independent of the parity of $q$, the
second and fourth sums are the same. We conclude that the four sums
are in fact equal in the small $\ell$ limit.

\par

The above considerations allow us to derive the asymptotic behavior
of the correlation functions. Making use of \Eqs{eq:SxSx:exact}
and~(\ref{eq:SySy:exact}), one obtains that
\begin{align}
\langle \Psi_{2\, {\rm sq}}\vert \hat{S}^{(1)}_x(\ell) \hat{S}^{(2)}_x(\ell)
\vert \Psi_{2\, {\rm sq}}\rangle &\simeq 1\, ,\quad 
\langle \Psi_{2\, {\rm sq}}\vert \hat{S}^{(1)}_y(\ell) \hat{S}^{(2)}_y(\ell)
\vert \Psi_{2\, {\rm sq}}\rangle \simeq 0\, ,
\end{align}
in excellent agreement with what is observed in \Figs{fig:exactSxSx}
and~\ref{fig:exactSySy}.

\section{Approximation Scheme}
\label{sec:appr:scheme}
In \Sec{sec:spinPheno}, explicit formulas for calculating the
correlation functions of the spin operators~(\ref{eq:defsz}),
(\ref{eq:defsx}) and~(\ref{eq:defsy}) in the two-mode squeezed
state~(\ref{eq:qstate}) were derived. These formulas are rather
involved as they rely on two-dimensional infinite sums of integrals
that need to be computed numerically, and are therefore not easy to
interpret. Moreover, it can be difficult to numerically evaluate them
when the squeezing parameter $r$ is large and the $\gamma_i$
parameters introduced above take extreme values. This is why in this
section, we develop approximation schemes in order to gain some
analytical insight on the physics at play. This will also allow us to
numerically evaluate the correlation functions in regimes where direct
computations are intractable otherwise.

In \Sec{sec:smallelllimit}, the small $\ell$ limit was
calculated by noticing that, in this regime, the integration variable
in the argument of the exponential function present in the integrand
could be set to $0$, thus making the integral explicitly
calculable. In this section, we use this same idea to design a more
general approximation scheme.
\subsection{Validity regime}
The argument of the exponentials appearing in the integrals of
\Sec{sec:spinPheno} are of the form $\gamma_i(z\pm n\pm m)$, where $z$
is the integration variable to be varied between $0$ and $1$. When
$\gamma_i\ll 1$, either $\pm n\pm m \ll 1/\gamma_i$ in which case the
argument is very small and one can take $z=0$ without any harm, either
$\pm n\pm m \gg 1/\gamma_i \gg 1$ in which case $z\pm n\pm m \simeq
\pm n\pm m$ and taking $z=0$ is also a good approximation. This
defines the regime of validity of this approximation. From
\Eqs{eq:Z1:def:simp}, (\ref{eq:X1:def})-(\ref{eq:X4:def}), this means
that one must have $\gamma_1, \gamma_3 \ll 1$.  From the discussion
around \Fig{fig:gamma12}, one can see that $\gamma_1\ll 1$ corresponds
to $r\gg 1$ when $\varphi\not\simeq\pi/2$ (which is why the dual case
$\varphi\simeq\pi/2$ is treated separately in
\Sec{sec:varphi=pi/2}). More precisely, from \Eq{eq:gamma1:def}, one
can see that $\gamma_1\ll 1$ is equivalent to
\begin{align}
\label{eq:cond:1}
r\gg 1\quad\mathrm{and}\quad\cos \varphi \gg \ee^{-r}\, .
\end{align}
For $\gamma_3$, in the limit $r\gg 1$ one has $\gamma_3\simeq
-\tan\varphi$, hence the correlation functions of $\hat{S}_x$ and
$\hat{S}_y$ can be accurately reproduced if the condition
\begin{align}
\label{eq:cond:2}
\varphi\ll 1
\end{align}
is also fulfilled. These two relations strictly define the conditions
of validity of the approximation scheme derived in this section.

However, let us notice that $z\pm n\pm m \not\simeq \pm n\pm m$ only
when $\pm n\pm m$ vanishes or is of order $1$, that is to say only for
a small subset of terms. This is why, in the following, we will see
that the approximated formulas derived in this section can be used
even if the two above conditions are relaxed. In other words, they
usually have a broader range of applicability. The only limitation is
that, since we have shown in \Sec{sec:smallelllimit} that the terms
such that $\pm n\pm m$ vanishes or is of order $1$ are precisely those
that dominate in the $\ell\gg 1$ limit, we expect our approximation to
fail at large $\ell$ when used outside the regime strictly defined by
\Eqs{eq:cond:1}-(\ref{eq:cond:2}) (which is also the reason why this
regime was separately studied in \Sec{sec:smallelllimit}).
\subsection{Approximating the Correlation Function $\langle \Psi_{2\,
    {\rm sq}}\vert \hat{S}^{(1)}_z(\ell) \hat{S}^{(2)}_z(\ell)\vert
  \Psi_{2\, {\rm sq}}\rangle$}

We start with approximating the correlation function $\langle
\Psi_{2\, {\rm sq}}\vert \hat{S}^{(1)}_z(\ell)
\hat{S}^{(2)}_z(\ell)\vert \Psi_{2\, {\rm sq}}\rangle$. Let us
therefore consider \Eq{eq:Z1:def:simp} again and, according to the
above considerations, neglect the integration variable $z$ in the
exponential term. It follows that
\begin{align}
Z_{n,m}^{(1)}&\simeq \frac{\ell}{2}\sqrt{\frac{\pi}{\gamma_2}}
\ee^{-\gamma_1\ell^2(n+m)^2/4}
\int _0^1 {\rm d}z 
\left\{{\erf}\left[\frac{\ell}{2}\sqrt{\gamma_2}\left(z+n-m\right)\right]
+{\erf}\left[\frac{\ell}{2}\sqrt{\gamma_2}\left(z-n+m\right)\right]\right\}\, .
\end{align}
The integral over $z$ can now be performed explicitly\footnote{Here,
  we make use of the relation $\int_0^x\erf(az)\dd z = x
  \erf(ax)+(\ee^{-a^2 x^2}-1)/(a\sqrt{\pi})$.} and one obtains
\begin{align}
Z_{n,m}^{(1)} & \simeq \frac{ \ee^{-\gamma_1 \ell^2 (n+m)^2/4}}{\gamma_2}\left[\ee^{-\gamma_2 \ell^2(n-m+1)^2/4}+\ee^{-\gamma_2 \ell^2(n-m-1)^2/4}-2\ee^{-\gamma_2 \ell^2(n-m)^2/4}\right]
\nonumber\\ &
+\ell\sqrt{\frac{\pi}{\gamma_2}}\ee^{-\gamma_1 \ell^2 (n+m)^2/4}\left[ u_{(n-m+1)/2}+u_{(n-m-1)/2} - 2 u_{(n-m)/2} \right] \, ,
\end{align}
where we have introduced the notation $u_q\equiv
q\erf(\ell\sqrt{\gamma_2}q)$.  In this expression, as we have already
seen in \Sec{sec:smallelllimit}, only the combinations $q=n-m$ and
$p=n+m$ are involved. As a consequence, following the same strategy as
before, one can write the sum over $n$ and $m$ as a sum over $q$ and
$p$, $p$ having the same parity as $q$. This leads to
\begin{align}
\sum_{n,m=-\infty}^\infty(-1)^{n+m}Z_{n,m}^{(1)} = \sum_{q=-\infty}^\infty  \mathcal{Z}_q^{(1)} \sum_{p\in \mathcal{P}(q)} \mathcal{Z}_p^{(1)} \, ,
\end{align}
where the quantities $\mathcal{Z}_q^{(1)}$ and $\mathcal{Z}_p^{(1)}$
are defined by
\begin{align}
\label{eq:Zq:def}
\mathcal{Z}_q^{(1)} = &
\frac{1}{\gamma_2}\left[\ee^{-\gamma_2 \ell^2(q+1)^2/4}+\ee^{-\gamma_2 \ell^2(q-1)^2/4}-2\ee^{-\gamma_2 \ell^2 q^2/4}\right]
+\ell\sqrt{\frac{\pi}{\gamma_2}}\left[ u_{(q+1)/2}+u_{(q-1)/2}-2u_{q/2} \right]\, ,
\end{align}
and 
\begin{align}
\mathcal{Z}_p^{(1)} = (-1)^p \ee^{-\gamma_1 \ell^2 p^2/4}\, .
\label{eq:appr:Zp}
\end{align}
In the above expression, let us stress again that $\mathcal{P}(q)$
stands for all integer numbers having the same parity as $q$. Since
\Eq{eq:appr:Zp} coincides with \Eq{eq:smallell:Zqp} for
$\mathcal{Z}_p^{(1)}$, we have already shown that
$\sum_{p\in\mathcal{P}(q)}\mathcal{Z}_p^{(1)}=\vartheta_3\left(0,\ee^{-\gamma_1\ell^2}\right)$
if $q$ is even and
$\sum_{p\in\mathcal{P}(q)}\mathcal{Z}_p^{(1)}=-\vartheta_2\left(0,\ee^{-\gamma_1\ell^2}\right)$
if $q$ is odd, 
see \Eqs{eq:smalll:sumZp:even} and~(\ref{eq:smalll:sumZp:odd}). One then has
\begin{align}
\label{eq:interZapprox}
\sum_{n,m=-\infty}^\infty(-1)^{n+m}Z_{n,m}^{(1)}
=\vartheta_3\left(0,\ee^{-\gamma_1\ell^2}\right)
\sum_{q=-\infty}^\infty \mathcal{Z}_{2q}^{(1)}
-\vartheta_2\left(0,\ee^{-\gamma_1\ell^2}\right)
\sum_{q=-\infty}^\infty \mathcal{Z}_{2q+1}^{(1)}\, .
\end{align}
Let us now calculate the two sums over $q$. In \Eq{eq:Zq:def}, two
types of terms are present, the exponential ones and the error
function ones. The exponential terms can be resumed explicitly and one
obtains, for the odd sum,
\begin{align}
\label{eq:interZapproxodd}
\sum _{q=-\infty}^{\infty}\mathcal{Z}_{2q+1}^{(1)} & =
-\frac{2}{\gamma_2}\left[\vartheta_2\left(0,\ee^{-\gamma_2\ell^2}\right)
-\vartheta_3\left(0,\ee^{-\gamma_2\ell^2}\right)\right]
+\ell\sqrt{\frac{\pi}{\gamma_2}}
\sum _{q=-\infty}^{\infty}
\left(u_{q+1}+u_q-2u_{q+1/2}\right)
\, .
\end{align}
In the same way, one can estimate the even sum and one obtains
\begin{align}
\label{eq:interZapproxeven}
\sum _{q=-\infty}^{\infty}\mathcal{Z}_{2q}^{(1)} & =
\frac{2}{\gamma_2}\left[\vartheta_2\left(0,\ee^{-\gamma_2\ell^2}\right)
-\vartheta_3\left(0,\ee^{-\gamma_2\ell^2}\right)\right]
+\ell\sqrt{\frac{\pi}{\gamma_2}}
\sum _{q=-\infty}^{\infty}
\left(u_{q+1/2}+u_{q-1/2}-2u_q\right)\, .
\end{align}
One notices that these two last expressions are symmetric under the
permutation $q \leftrightarrow q-1/2$. Using \Eqs{eq:interZapproxodd}
and~(\ref{eq:interZapproxeven}) in \Eq{eq:interZapprox}, one then
obtains the following expression
\begin{align}
\sum_{n,m=-\infty}^\infty(-1)^{n+m}Z_{n,m}^{(1)}
=& \frac{2}{\gamma_2}\left[\vartheta_2\left(0,\ee^{-\gamma_1\ell^2}\right)
+\vartheta_3\left(0,\ee^{-\gamma_1\ell^2}\right)\right]
\left[\vartheta_2\left(0,\ee^{-\gamma_2\ell^2}\right)
-\vartheta_3\left(0,\ee^{-\gamma_2\ell^2}\right)\right]
\nonumber \\ &
-\vartheta_2\left(0,\ee^{-\gamma_1\ell^2}\right)
\ell \sqrt{\frac{\pi}{\gamma_2}}
\sum_{q=-\infty}^\infty \left(u_{q+1}+u_q-2u_{q+1/2}\right)
\nonumber \\ &
+\vartheta_3\left(0,\ee^{-\gamma_1\ell^2}\right)
\ell \sqrt{\frac{\pi}{\gamma_2}}
\sum_{q=-\infty}^\infty \left(u_{q+1/2}+u_{q-1/2}-2u_{q}\right).
\end{align}
Finally, this expression can be simplified by making use of the
formulas relating the various Jacobi
functions\footnote{\label{footnote:theta:234}Concretely, we use the
  relations~\cite{Gradshteyn:1965aa} $\vartheta_2\left(2a,b^4\right)
  -\vartheta_3\left(2a,b^4\right)=-\vartheta_4(a,b)$ and
  $\vartheta_2\left(2a,b^4\right)
  +\vartheta_3\left(2a,b^4\right)=\vartheta_3(a,b)$.} and by using the
symmetry $u_q=u_{-q}$, which has the advantage of decreasing the
number of terms in the series. Inserting the above equation in
\Eq{eq:SzSz:exact}, one obtains the following expression for the
correlation function
\begin{align}
\langle \Psi_{2\, {\rm sq}}\vert \hat{S}^{(1)}_z(\ell) \hat{S}^{(2)}_z(\ell)
\vert \Psi_{2\, {\rm sq}}\rangle
& \simeq 
-\frac{2}{\pi}\sqrt{\frac{\gamma_1}{\gamma_2}}\vartheta_3
\left(0,\ee^{-\gamma_1\ell^2/4}\right)
\vartheta_4\left(0,\ee^{-\gamma_2\ell^2/4}\right)
-\ell\sqrt{\frac{\gamma_1}{\pi}}\vartheta_3\left(0,\ee^{-\gamma_1\ell^2}\right)
{\rm erf}
\left(\sqrt{\gamma_2}\frac{\ell}{2}\right)
\nonumber \\ &
+2\ell\sqrt{\frac{\gamma_1}{\pi}}\vartheta_3\left(0,\ee^{-\gamma_1\ell^2}\right)
\sum_{q=0}^\infty \left(u_{q+1/2}+u_{q-1/2}-2u_{q}\right)
\nonumber \\ &
-2\ell\sqrt{\frac{\gamma_1}{\pi}}\vartheta_2\left(0,\ee^{-\gamma_1\ell^2}\right)
\sum_{q=0}^\infty \left(u_{q+1}+u_q-2u_{q+1/2}\right)\, .
\label{eq:SzSz:appr}
\end{align}
This formula is not yet ``analytical'' in the sense that the series
still contain an infinite number of terms. But, as we now discuss,
they can be truncated. For practical purpose, let us evaluate how many
terms $q_\mathrm{lim}$ must be computed in the above infinite sums to
reach an accuracy sufficient to match well the numerical results. When
$q\gg 1$, one has $u_{q+1}+u_q-2u_{q+1/2}\simeq
u_{q+1/2}+u_{q-1/2}-2u_{q}\simeq -\ee^{-\ell^2\gamma_2q^2}/(\ell
\sqrt{\pi \gamma_2})$.
This implies that, when $q\gg 1/(\ell \sqrt{\gamma_2})$, the two terms
of the series rapidly go to zero. Therefore, $1/(\ell
\sqrt{\gamma_2})$ gives the order of magnitude of the number of terms
one should compute.
\begin{figure}[t]
\begin{center}
\includegraphics[width=0.45\textwidth,clip=true]{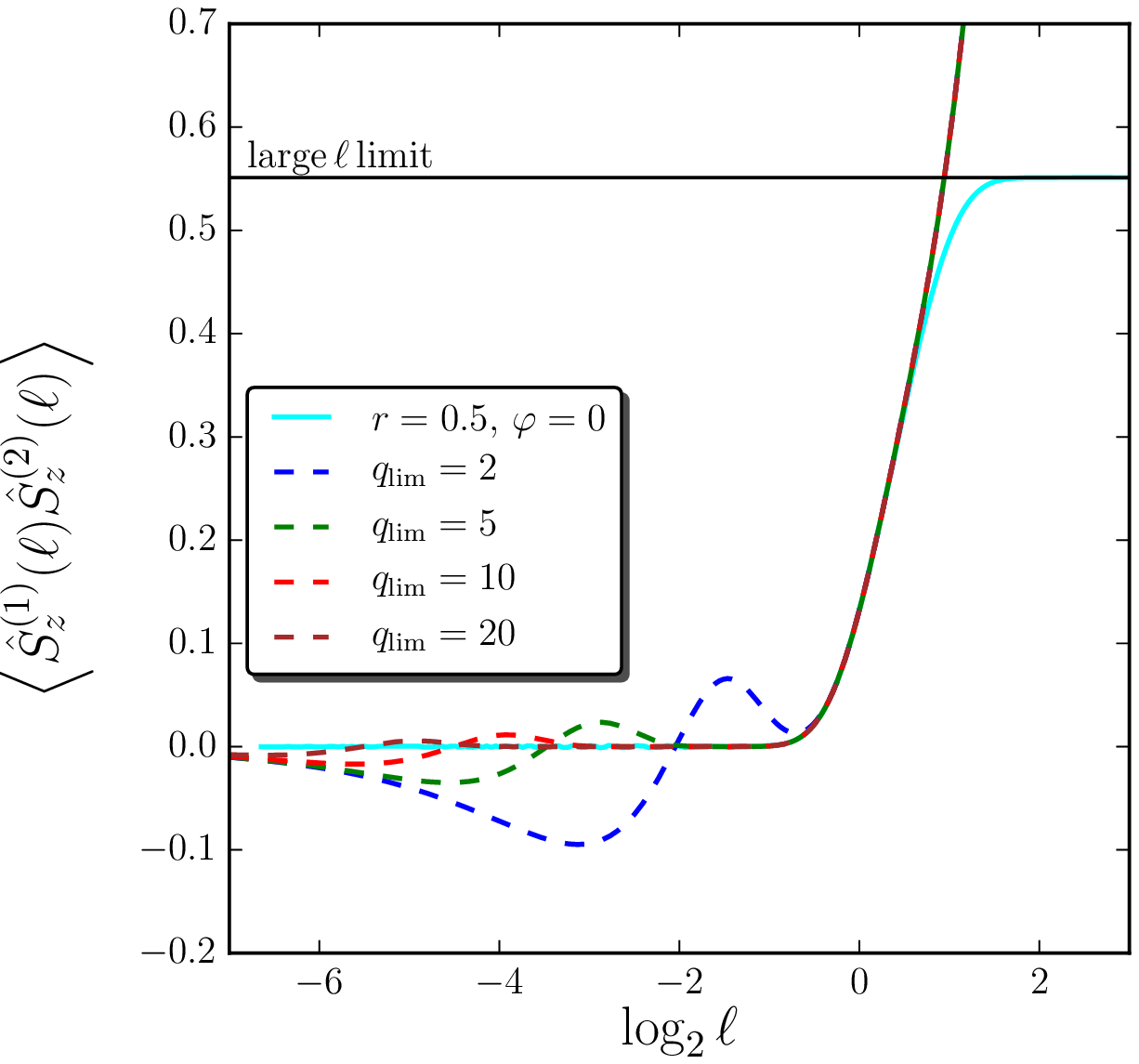}
\includegraphics[width=0.45\textwidth,clip=true]{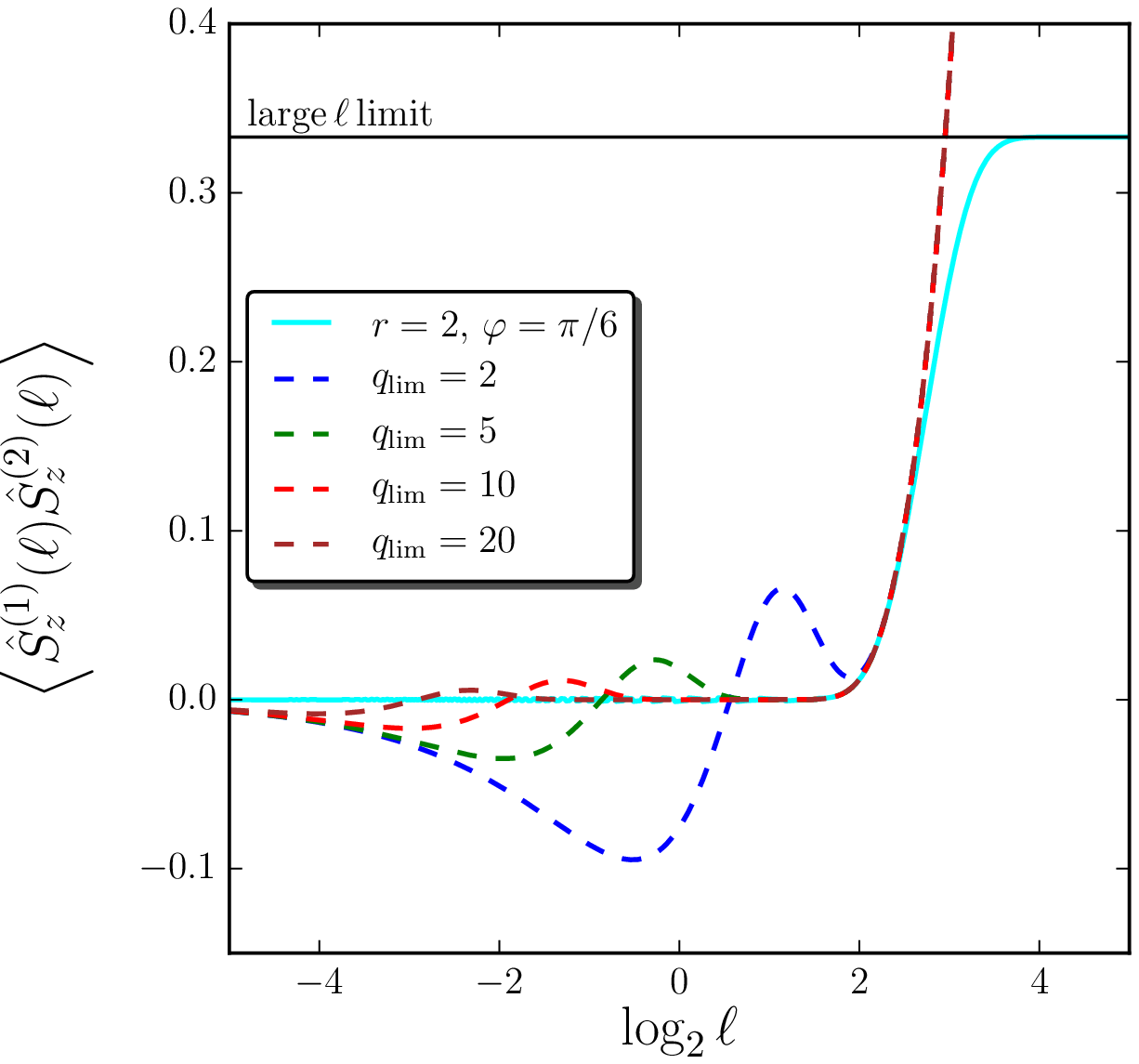}
\caption{Two-point correlation function of the $\hat{S}_z$ operator
  for $r=0.5$ and $\varphi=0$ (left panel), and $r=2$ and
  $\varphi=\pi/6$ (right panel). The blue solid line is the exact
  result~(\ref{eq:SzSz:exact}), the dashed colored lines correspond to
  the approximation~(\ref{eq:SzSz:appr}) with different numbers of
  terms $q_\mathrm{lim}$ in the two involved sums, and the black
  horizontal line corresponds to the large $\ell$
  limit~(\ref{eq:largeell:SzSz}).}
\label{fig:approxSzSz}
\end{center}
\end{figure}
In \Fig{fig:approxSzSz}, the approximation~(\ref{eq:SzSz:appr}) is
displayed and compared to the exact formula~(\ref{eq:SzSz:exact}). One
can check that the agreement is good if a sufficient number of terms
$q_\mathrm{lim}$ is kept. When $\ell$ is large, the approximation
fails to reproduce the exact result as expected and as discussed at
the beginning of this section. In this regime, however,
\Eq{eq:largeell:SzSz} gives the correct value for the asymptotic
plateau. When $\ell\ll 1$, more terms need to be summed over, as
expected from the fact that the generic term of the sum becomes
negligible only when $q\gg 1/(\ell \sqrt{\gamma_2})$. In between, one
can see that the approximated formula provides an excellent fit to the
numerical curve, even though $r$ is of order one and the strict
conditions~(\ref{eq:cond:1}) and~(\ref{eq:cond:2}) are not met.
\subsection{Approximating the Correlation Functions $\langle \Psi_{2\, {\rm sq}}\vert \hat{S}^{(1)}_x(\ell) \hat{S}^{(2)}_x(\ell)\vert \Psi_{2\, {\rm sq}}\rangle$ and $\langle \Psi_{2\, {\rm sq}}\vert \hat{S}^{(1)}_y(\ell) \hat{S}^{(2)}_y(\ell)\vert \Psi_{2\, {\rm sq}}\rangle$}
\label{sec:appr:xy}
Let us now approximate the two-point correlators of $\hat{S}_x(\ell)$
and $\hat{S}_y(\ell)$. Using the same techniques as before and,
therefore, neglecting the integration variable $z$ in the exponential
and cosine terms of \Eqs{eq:X1:def}-(\ref{eq:X4:def}), one has
\begin{align}
X_{n,m}^{(1)} &\simeq
\ell\sqrt{\frac{\pi}{\gamma_2}}
\ee^{-\gamma_1\ell^2(p^2+p+1/2)}
\cos\left[\gamma_3\ell^2(2p+1)\right]
\int_0^1 {\rm d}z
\left\{{\rm erf}\left[\frac{\ell}{2}\sqrt{\gamma_2}
(z-2q)\right]
+{\rm erf}\left[\frac{\ell}{2}\sqrt{\gamma_2}
(z+2q)\right]\right\}\, ,\\
X_{n,m}^{(2)} &\simeq
\ell\sqrt{\frac{\pi}{\gamma_2}}
\ee^{-\left[\frac{\gamma_2}{4}+\frac{\gamma_4^2}{\gamma_2}+\gamma_1(p^2+p+1/4)\right]\ell^2}
\int _0^1 \dd z\,
\Re\mathrm{e}\biggl\{{\rm erf}\left[\frac{\ell}{2}\sqrt{\gamma_2}
(z-2q)+i\frac{\gamma_4}{\sqrt{\gamma_2}}\ell\right]
+{\rm erf}\left[\frac{\ell}{2}\sqrt{\gamma_2}
(z+2q)+i\frac{\gamma_4}{\sqrt{\gamma_2}}\ell\right]
\biggr\}
\, ,
\end{align}
and $X_{n,m}^{(3)} \simeq X_{n+1/2,m+1/2}^{(1)}$, $X_{n,m}^{(4)}
\simeq X_{n+1/2,m+1/2}^{(3)}$. Notice that the above formulas are
expressed in terms of $p=n+m$ and $q=n-m$ again. As before, the
remaining integrals can be performed explicitly, namely the error
functions can be integrated, and one obtains
\begin{align}
X_{nm}^{(1)} &\simeq 
2\ell\sqrt{\frac{\pi}{\gamma_2}}
\ee^{-\gamma_1\ell^2(p^2+p+1/2)}
\cos\left[\gamma_3 \ell^2(1+2p)\right]
\biggl\{
u_{q-1/2}
+u_{q+1/2}
-2u_q
\nonumber \\ &
+\frac{1}{\ell \sqrt{\pi \gamma_2}}\left[\ee^{-\gamma_2\ell^2(q-1/2)^2}
+\ee^{-\gamma_2\ell^2(q+1/2)^2}
-2\ee^{-\gamma_2\ell^2q^2}
\right]\biggr\}\, ,\\
 X_{nm}^{(2)} & \simeq 
2\ell\sqrt{\frac{\pi}{\gamma_2}}
\ee^{-\left[\frac{\gamma_2}{4}+\frac{\gamma_4^2}{\gamma_2}+\gamma_1(p^2+p+1/4)\right]\ell^2}
\Re\mathrm{e}
\biggl\{u_{q-1/2-i\gamma_4/\gamma_2}
+u_{q+1/2+i\gamma_4/\gamma_2}
-2u_{q+i\gamma_4/\gamma_2}
\nonumber \\ &
+\frac{1}{\ell \sqrt{\pi \gamma_2}}
\biggl[\ee^{-\gamma_2\ell^2(q-1/2+i\gamma_4/\gamma_2)^2}
+\ee^{-\gamma_2\ell^2(q+1/2+i\gamma_4/\gamma_2)^2}
-2\ee^{-\gamma_2\ell^2(q+i\gamma_4/\gamma_2)^2}
\biggr]\biggr\}\, .
\end{align}
\begin{figure}[t]
\begin{center}
\includegraphics[width=0.445\textwidth,clip=true]{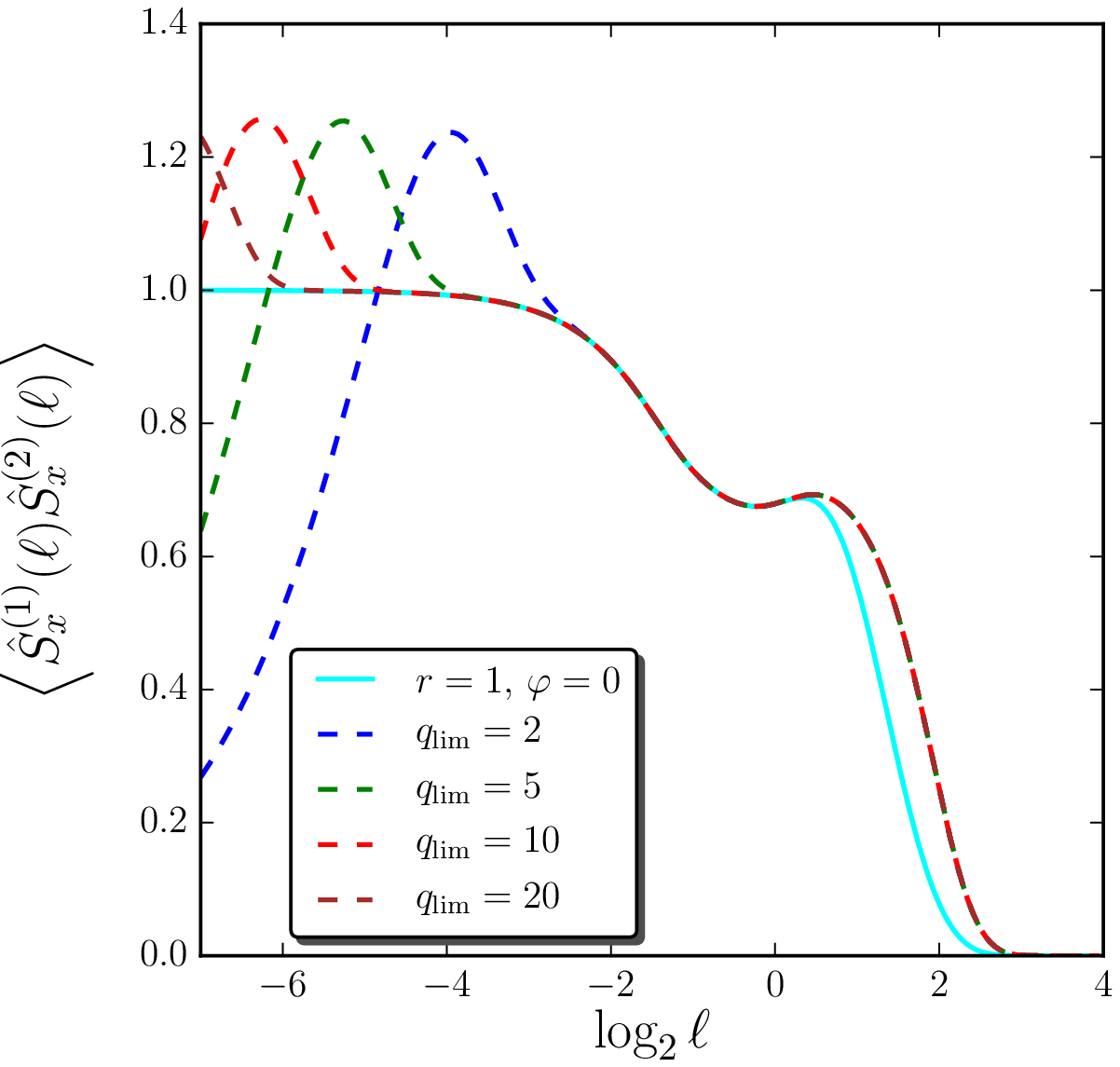}
\includegraphics[width=0.455\textwidth,clip=true]{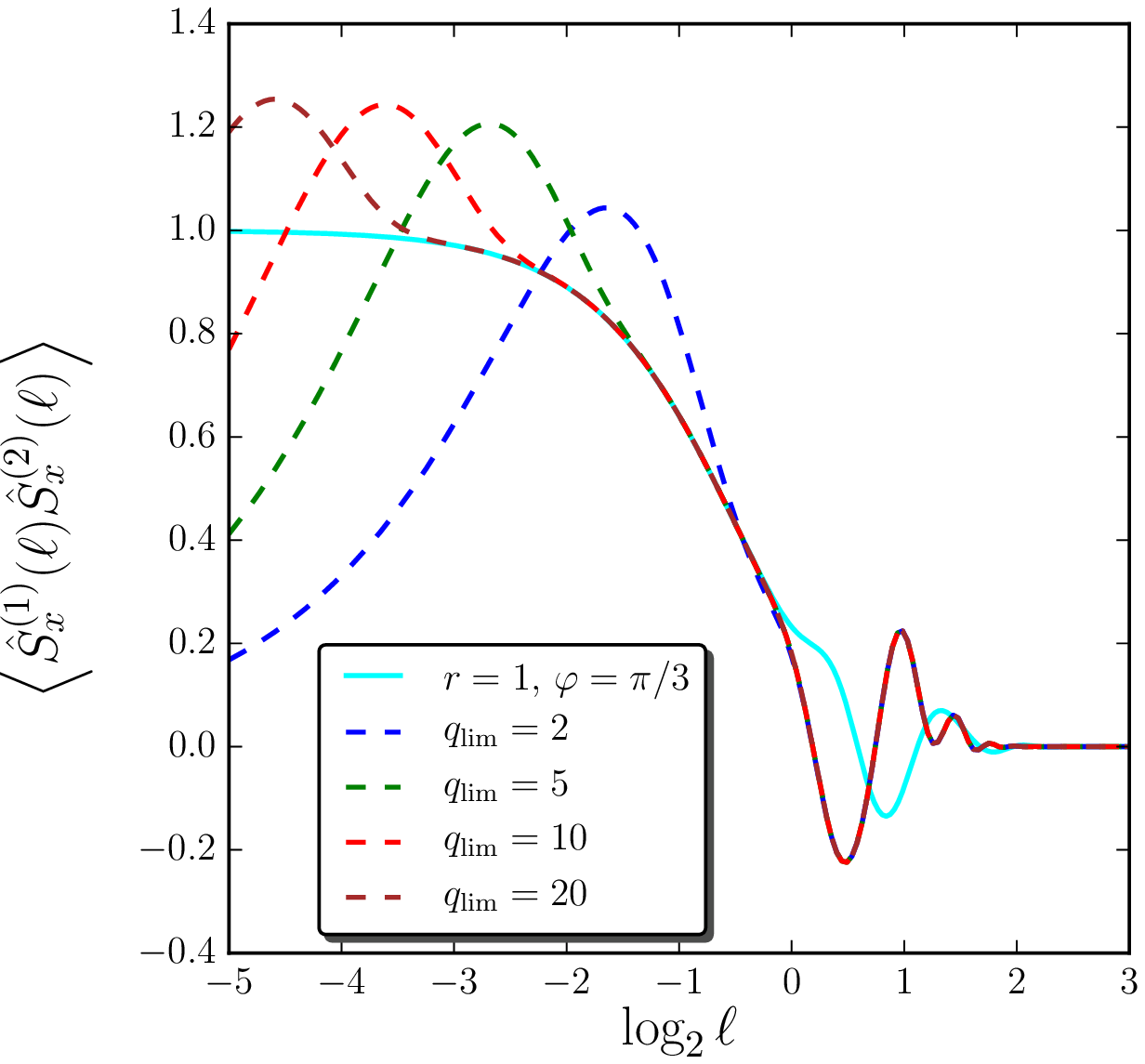}
\caption{Two-point correlator of the $\hat{S}_x$ operator for $r=1$
  and $\varphi=0$ (left panel), and $r=1$ and $\varphi=\pi/3$ (right
  panel). The blue solid line is the exact
  result~(\ref{eq:SxSx:exact}) and the dashed colored lines correspond
  to the approximation derived in \Sec{sec:appr:xy} with different
  numbers of terms $q_{\mathrm{lim}}$ kept in the different sums.}
\label{fig:approxSxSx}
\end{center}
\end{figure}

The next step consists in calculating the sums explicitly. Since, once
more, the above equations show that $X_{n,m}^{(1,2)}$ only depend on
$p$ and $q$, the sum over $n$ and $m$ can be performed as a sum over
$p$ and $q$, as we have now done several times. Concretely, the first
sum reads
\begin{align}
\sum_{n,m=-\infty}^\infty X_{n,m}^{(1)} = \sum_{q=-\infty}^\infty  
\mathcal{X}_q^{(1)} \sum_{p\in \mathcal{P}(q)} \mathcal{X}_p^{(1)} \, ,
\end{align}
with the following definitions for the quantities
$\mathcal{X}_q^{(1)}$ and $\mathcal{X}_q^{(2)}$
\begin{align}
\label{eq:appr:X1p:def}
\mathcal{X}_{p}^{(1)} &= 
\ell\sqrt{\frac{\pi}{\gamma_2}}
\ee^{-\gamma_1\ell^2\left(p^2+p+1/2\right)}
\cos\left[\gamma_3 \ell^2(2p+1)\right],\\
\mathcal{X}_{q}^{(1)} &=
2\left(u_{q-1/2}+u_{q+1/2}-2u_q\right)
+\frac{2}{\ell \sqrt{\pi \gamma_2}}\left[\ee^{-\gamma_2\ell^2(q-1/2)^2}
+\ee^{-\gamma_2\ell^2(q+1/2)^2}
-2\ee^{-\gamma_2\ell^2 q^2}
\right]\, .
\end{align}
One notices that \Eq{eq:appr:X1p:def} is the same as
\Eq{eq:smallell:X1pq:def} for $\mathcal{X}_{p}^{(1)}$, up to a
$p$-independent prefactor. The corresponding sum has
already be performed in \Sec{sec:smallelllimit}. Borrowing the 
corresponding result leads to
\begin{align}
\sum_{n,m=-\infty}^\infty X_{n,m}^{(1)}&=
\ell\sqrt{\frac{\pi}{\gamma_2}}
\Re\mathrm{e}\left\lbrace\ee^{(-\gamma_1/2+i\gamma_3)\ell^2}\vartheta_3
\left[2\ell^2\left(\gamma_3+i\frac{\gamma_1}{2}\right),\ee^{-4\gamma_1\ell^2}\right] 
\right\rbrace
\sum_{q=-\infty}^\infty\mathcal{X}_{2q}^{(1)}
\nonumber\\ &
+\ell\sqrt{\frac{\pi}{\gamma_2}}
\Re\mathrm{e}\left\lbrace\ee^{(-\gamma_1/2+i\gamma_3)\ell^2}\vartheta_2
\left[2\ell^2\left(\gamma_3+i\frac{\gamma_1}{2}\right),\ee^{-4\gamma_1\ell^2}\right] 
\right\rbrace
\sum_{q=-\infty}^\infty\mathcal{X}_{2q+1}^{(1)}\, .
\end{align}
The sums over $q$ remain to be done. Making use of the formulas given
in footnote~\ref{footnote:theta:234} and of additional properties of
the Jacobi functions\footnote{\label{footnote:theta23}Here, we make
  use of the relations~\cite{Gradshteyn:1965aa},
  $\vartheta_2(z,\ee^{i\tau})=\ee^{i\tau/4+i
    z}\vartheta_3\left(z+\tau/2,\ee^{i\tau}\right)$ and
  $\vartheta_{2,3}(z,q)=\vartheta_{2,3}(-z,q)$.}, one arrives at
\begin{align}
\sum_{q=-\infty}^\infty \mathcal{X}_{2q+1}^{(1)}&=2\sum_{q=-\infty}^\infty
(u_{2q+1/2}+u_{2q+3/2}-2u_{2q+1})
+\frac{2}{\ell \sqrt{\pi \gamma_2}}
\left[
\vartheta_2\left(0,\ee^{-\gamma_2\ell^2}\right)
-2\vartheta_2\left(0,\ee^{-4\gamma_2\ell^2}\right)
\right]
\\
\sum_{q=-\infty}^\infty \mathcal{X}_{2q}^{(1)}&=2\sum_{q=-\infty}^\infty
(u_{2q-1/2}+u_{2q+1/2}-2u_{2q})
+\frac{2}{\ell \sqrt{\pi \gamma_2}}
\left[\vartheta_2\left(0,\ee^{-\gamma_2\ell^2}\right)
-2\vartheta_3\left(0,\ee^{-4\gamma_2\ell^2}\right)\right]\, ,
\end{align}
which completes the calculation of the first sum. 

\par

For the second sum, the same logic can be applied again. Noticing
that $X_{n,m}^{(2)}$ only depends on $p$ and $q$, one can write that
\begin{align}
\sum_{n,m=-\infty}^\infty X_{n,m}^{(2)} = \sum_{q=-\infty}^\infty  
\mathcal{X}_q^{(2)} \sum_{p\in \mathcal{P}(q)} \mathcal{X}_p^{(2)} \, ,
\end{align}
with the following definitions
\begin{figure}[t]
\begin{center}
\includegraphics[width=0.45\textwidth,clip=true]{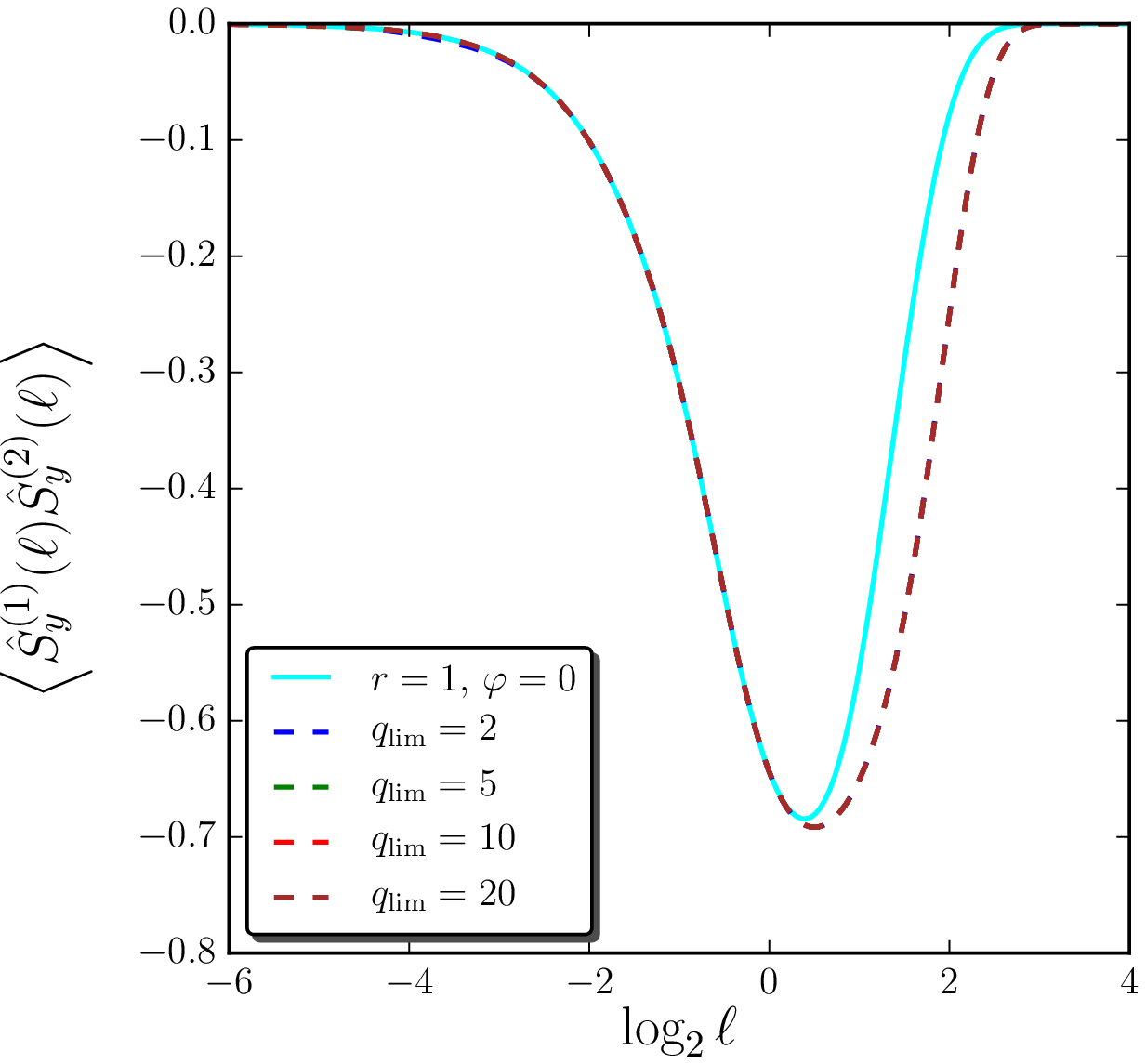}
\includegraphics[width=0.45\textwidth,clip=true]{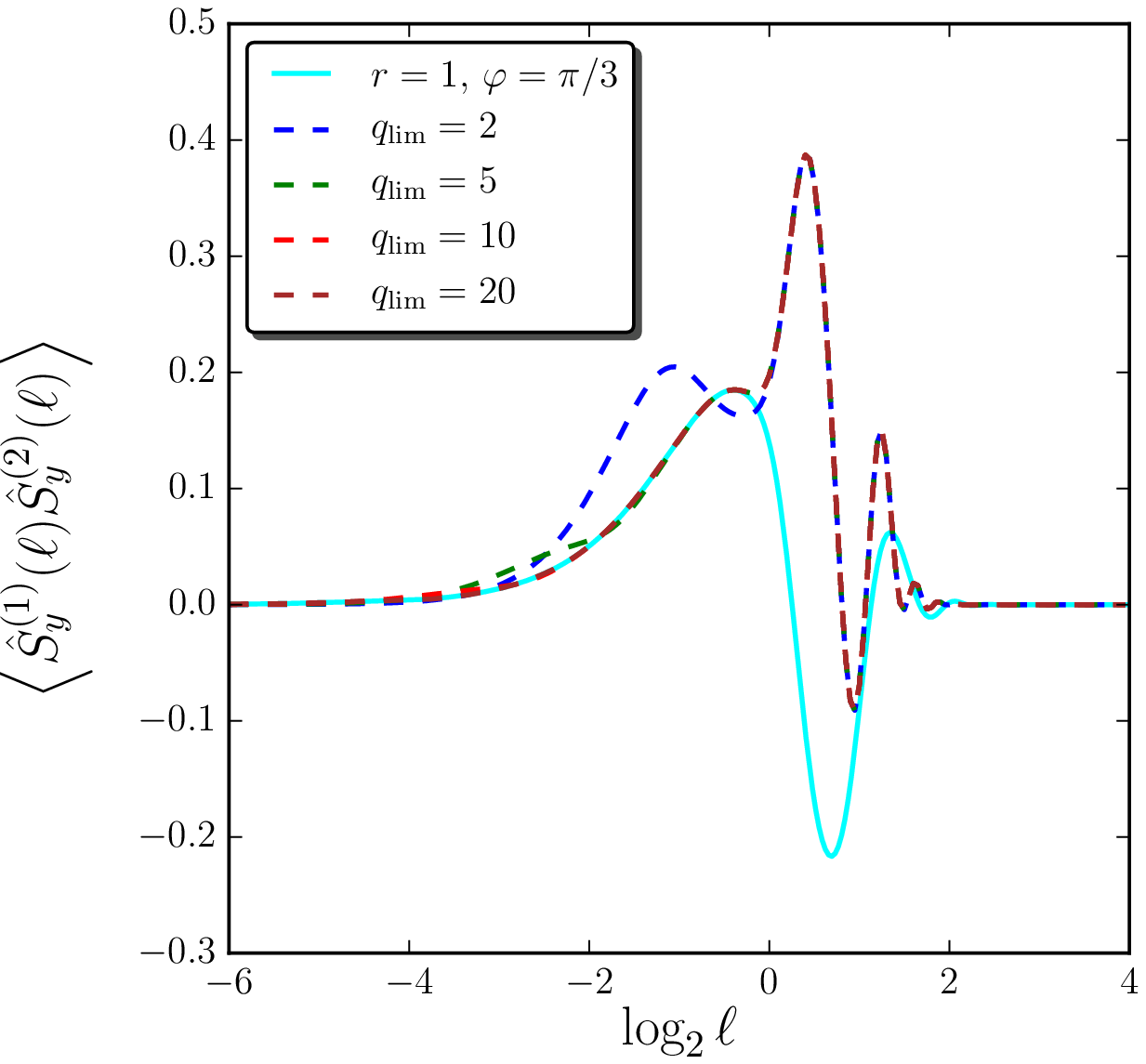}
\caption{Two-point correlator of the $\hat{S}_y$ operator for $r=1$
  and $\varphi=0$ (left panel), and $r=1$ and $\varphi=\pi/3$ (right
  panel). The blue solid line is the exact
  result~(\ref{eq:SySy:exact}) and the dashed colored lines correspond
  to the approximation derived in \Sec{sec:appr:xy} with different
  numbers of terms $q_{\mathrm{lim}}$ kept in the different sums.}
\label{fig:approxSySy}
\end{center}
\end{figure}
\begin{align}
\label{eq:appr:X2p:def}
\mathcal{X}_{p}^{(2)} &= 
\frac{\ell}{2}\sqrt{\frac{\pi}{\gamma_2}}
\ee^{-\left[\gamma_1(p^2+p+1/4)+\frac{\gamma_2}{4}+\frac{\gamma_4^2}{\gamma_2}\right]\ell^2}
,\\
\mathcal{X}_{q}^{(2)} &=
2\Re\mathrm{e}
\biggl\lbrace 
\left(u_{q-1/2+i\gamma_4/\gamma_2}
+u_{q+1/2+i\gamma_4/\gamma_2}
-2u_{q+i\gamma_4/\gamma_2}\right)
\nonumber  \\ & 
+\frac{1}{\ell \sqrt{\pi \gamma_2}}
\left[
+\ee^{-\gamma_2\ell^2(q-1/2+i\gamma_4/\gamma_2)^2}
+\ee^{-\gamma_2\ell^2(q+1/2+i\gamma_4/\gamma_2)^2}
-2\ee^{-\gamma_2\ell^2(q+i\gamma_4/\gamma_2)^2}
\right]
\biggr\rbrace 
\, .
\end{align}
As before, one notices that \Eq{eq:appr:X2p:def} is the same as
\Eq{eq:smallell:X2pq:def} for $\mathcal{X}_{p}^{(2)}$, up to a
$p$-independent prefactor and the corresponding sum has already been
performed in \Sec{sec:smallelllimit}. Using this result, one
can write that
\begin{align}
\sum_{n,m=-\infty}^\infty X_{nm}^{(2)}
=&
\frac{\ell}{2}\sqrt{\frac{\pi}{\gamma_2}}
\ee^{-(\gamma_1/4+\gamma_2/4+\gamma_4^2/\gamma_2)\ell^2}
\biggl[
\vartheta_2\left(i\gamma_1\ell^2,\ee^{-4\gamma_1\ell^2}\right)
\sum_{q=-\infty}^\infty \mathcal{X}_{2q+1}^{(2)}
+\vartheta_3\left(i\gamma_1\ell^2,\ee^{-4\gamma_1\ell^2}\right)
\sum_{q=-\infty}^\infty \mathcal{X}_{2q}^{(2)}\biggr]\, .
\end{align}
The two remaining sums in the above expression can be evaluated in a
similar way as before, in particular by making use of the formulas
given in footnote~\ref{footnote:theta23}. The result reads
\begin{align}
\sum_{q=-\infty}^\infty \mathcal{X}_{2q+1}^{(2)}
&= 4\Re\mathrm{e}\Bigg\{
\sum_{q=-\infty}^\infty
\left(u_{2q+1/2+i\gamma_4/\gamma_2}
+u_{2q+3/2+i\gamma_4/\gamma_2}
-2u_{2q+1+i\gamma_4/\gamma_2}
\right)
\nonumber \\ &
+\frac{2\ee^{\gamma_4^2\ell^2/\gamma_2}}{\ell\sqrt{\pi \gamma_2}}
\biggl[\ee^{-(\gamma_2/4+i\gamma_4)\ell^2}
\vartheta_2\left(2\gamma_4\ell^2-i\gamma_2\ell^2,\ee^{-4\gamma_2\ell^2}\right)
-\vartheta_2\left(2\gamma_4\ell^2,
\ee^{-4\gamma_2\ell^2}\right)\biggr]\Bigg\}\, ,
\\
\sum_{q=-\infty}^\infty \mathcal{X}_{2q}^{(2)}
&=4\Re\mathrm{e}\Bigg\{
\sum_{q=-\infty}^\infty
\left(u_{2q-1/2+i\gamma_4/\gamma_2}
+u_{2q+1/2+i\gamma_4/\gamma_2}
-2u_{2q+i\gamma_4/\gamma_2}
\right)
\nonumber \\ &
+\frac{2\ee^{\gamma_4^2\ell^2/\gamma_2}}{\ell\sqrt{\pi \gamma_2}}
\biggl[
\ee^{-(\gamma_2/4+i\gamma_4)\ell^2}
\vartheta_3\left(2\gamma_4\ell^2-i\gamma_2\ell^2,\ee^{-4\gamma_2\ell^2}\right)
-\vartheta_3\left(2\gamma_4\ell^2,
\ee^{-4\gamma_2\ell^2}\right)\biggr]\Bigg\}\, .
\end{align}
This completes the calculation of the second sum.

\par

For the third sum, one simply has
\begin{align}
\sum_{n,m=-\infty}^\infty X^{(3)}_{n,m}\simeq
\sum_{n,m=-\infty}^\infty X^{(1)}_{n+1/2,m+1/2}=\sum_{q=-\infty}^\infty
\mathcal{X}_q^{(1)}\sum_{p\in\mathcal{P}(q)}\mathcal{X}_{p+1}^{(1)}
=\sum_{q=-\infty}^\infty\mathcal{X}_q^{(1)}\sum_{p\in\bar{\mathcal{P}}(q)}
\mathcal{X}_{p}^{(1)}\, .
\end{align}
The calculation one has to perform is therefore very similar to the
first sum, the only difference being that $p$ is now summed over
integer numbers having the opposite parity as $q$, rather than the
same parity as usual. Therefore, we have already calculated all the
necessary quantities. The final result is simply a different
combination of them, namely
\begin{align}
\sum_{n,m=-\infty}^\infty X_{n,m}^{(3)}&=
\ell\sqrt{\frac{\pi}{\gamma_2}}
\Re\mathrm{e}\left\lbrace\ee^{(i\gamma_3-\gamma_1/2)\ell^2}\vartheta_3
\left[\left(2\gamma_3+i\gamma_1\right)\ell^2,\ee^{-4\gamma_1\ell^2}\right] 
\right\rbrace
\sum_{q=-\infty}^\infty\mathcal{X}_{2q+1}^{(1)}
\nonumber\\ &
+\ell\sqrt{\frac{\pi}{\gamma_2}}
\Re\mathrm{e}\left\lbrace\ee^{(i\gamma_3-\gamma_1/2)\ell^2}\vartheta_2
\left[\left(2\gamma_3+i\gamma_1\right)\ell^2,\ee^{-4\gamma_1\ell^2}\right] 
\right\rbrace
\sum_{q=-\infty}^\infty\mathcal{X}_{2q}^{(1)}\, .
\end{align}

Finally, the calculation of the fourth sum proceeds along the same
lines. We have 
\begin{align}
\sum_{n,m=-\infty}^\infty X^{(4)}_{n,m}\simeq
\sum_{n,m=-\infty}^\infty X^{(2)}_{n+1/2,m+1/2}=\sum_{q=-\infty}^\infty
\mathcal{X}_q^{(2)}\sum_{p\in\mathcal{P}(q)}\mathcal{X}_{p+1}^{(2)}
=\sum_{q=-\infty}^\infty\mathcal{X}_q^{(2)}\sum_{p\in\bar{\mathcal{P}}(q)}
\mathcal{X}_{p}^{(2)}\, ,
\end{align}
which leads to
\begin{figure*}[t]
\begin{center}
\includegraphics[width=0.45\textwidth,clip=true]{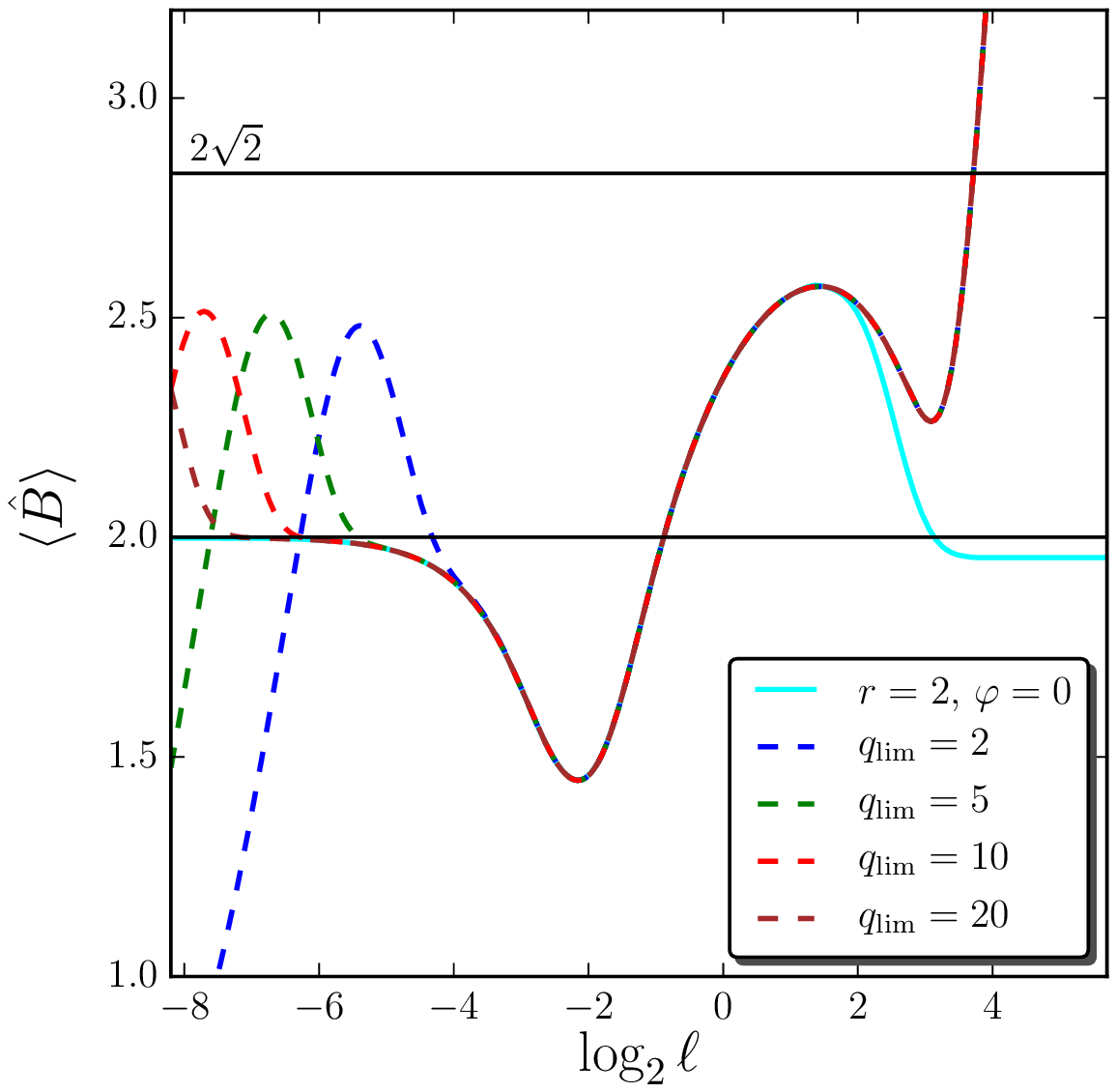}
\includegraphics[width=0.45\textwidth,clip=true]{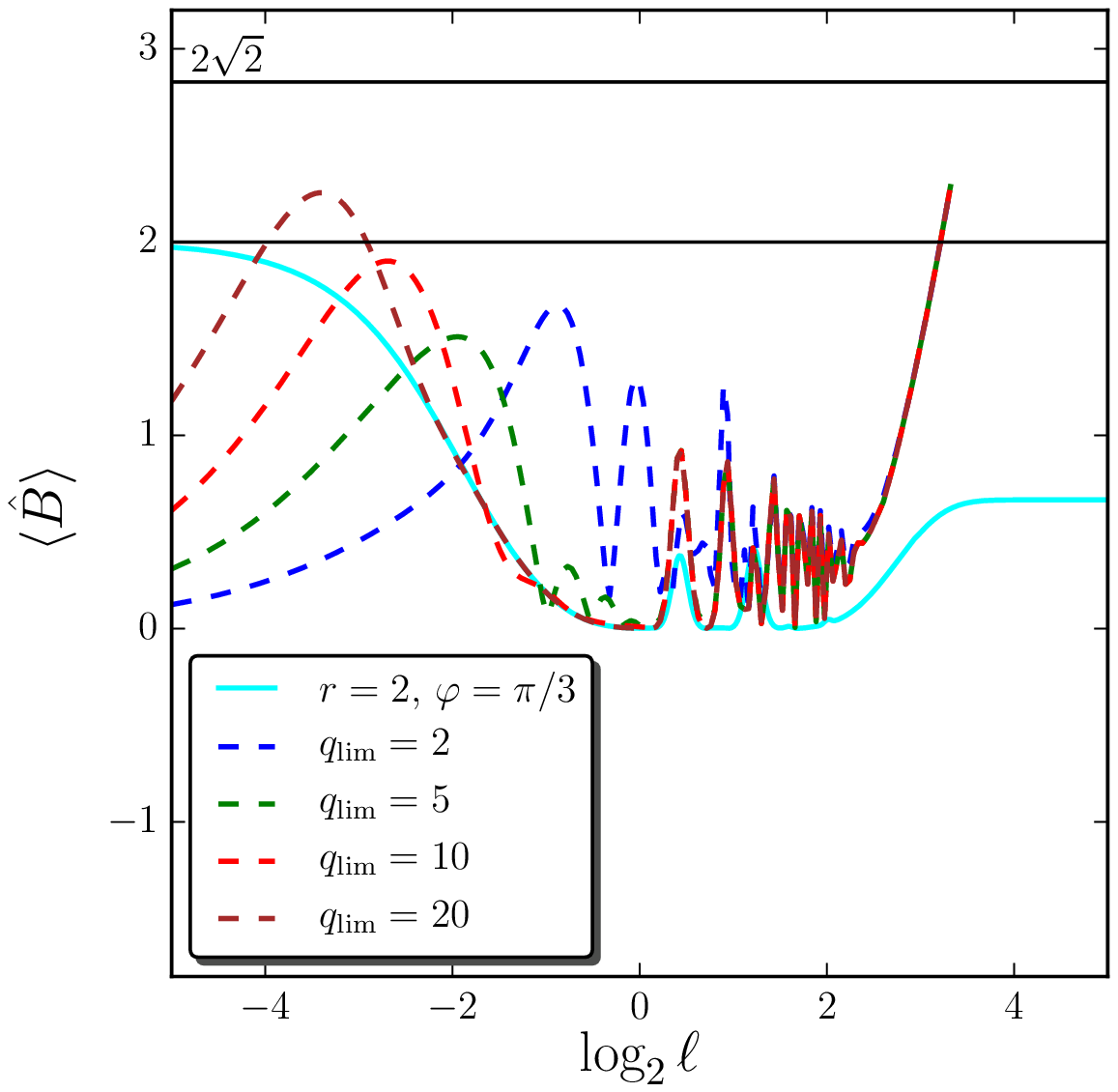}
\caption{Expectation value of the Bell
  operator~(\ref{eq:B:optimizedAngles}) as a function of $\ell$ for
  $r=2$ and $\varphi=0$ (left panel), and for $r=2$ and
  $\varphi=\pi/3$ (right panel). The blue solid lines stand for the
  result of the exact, numerical result while the dashed lines
  correspond to our approximation with different numbers of terms
  $q_\mathrm{lim}$ kept.}
\label{fig:approxBell}
\end{center}
\end{figure*}
\begin{align}
\sum_{n,m=-\infty}^\infty X_{nm}^{(4)}
=&
\frac{\ell}{2}\sqrt{\frac{\pi}{\gamma_2}}
\ee^{-(\gamma_1/4+\gamma_2/4+\gamma_4^2/\gamma_2)\ell^2}
\biggl[
\vartheta_2\left(i\gamma_1\ell^2,\ee^{-4\gamma_1\ell^2}\right)
\sum_{q=-\infty}^\infty \mathcal{X}_{2q}^{(2)}
+\vartheta_3\left(i\gamma_1\ell^2,\ee^{-4\gamma_1\ell^2}\right)
\sum_{q=-\infty}^\infty \mathcal{X}_{2q+1}^{(2)}\biggr]\, .
\end{align}
At this stage, we have now successfully calculated the four sums. It
should be obvious that the structure of the result is very similar to
that obtained for the correlation function $\langle \Psi_{2\, {\rm
    sq}}\vert \hat{S}^{(1)}_z(\ell) \hat{S}^{(2)}_z(\ell)\vert
\Psi_{2\, {\rm sq}}\rangle$. In particular, in order to obtain an
explicit expression, the remaining sums must be truncated and, of
course, the accuracy of the approximation will depend on the number of
terms kept in the series.

\par

In \Figs{fig:approxSxSx} and~\ref{fig:approxSySy}, the approximations
derived in the present section are displayed and compared with the
exact formulas~(\ref{eq:SxSx:exact}) and~(\ref{eq:SySy:exact}). One
one can check that the agreement is good if one sums over a sufficient
number of terms $q_\mathrm{lim}$. When $\ell$ is large, the
approximation fails to reproduce the exact result as discussed at the
beginning of this section and similarly to what happens for $\langle
\Psi_{2\, {\rm sq}}\vert \hat{S}^{(1)}_z(\ell)
\hat{S}^{(2)}_z(\ell)\vert \Psi_{2\, {\rm sq}}\rangle$. When $\ell\ll
1$, more terms need to be summed over, as expected from the fact that
the generic term of the sums becomes negligible when $q\gg 1/(\ell
\sqrt{\gamma_2})$, again as for the two-point correlation function of
$\hat{S}_z$. In between, there is a range of $\ell$ values where the
approximation provides a good fit to the exact result, even outside
the (strict) domain of validity defined by \Eqs{eq:cond:1}
and~(\ref{eq:cond:2}). Finally, in \Fig{fig:approxBell}, we have
displayed the corresponding expectation value of the Bell operator
given by \Eq{eq:B:optimizedAngles}, for $r=2$ and $\varphi=0$ (left
panel), and for $r=2$ and $\varphi=\pi/3$ (right panel). One can see
that, in case Bell's inequality violation can be obtained, the
maximum of the bump is correctly reproduced by the present
approximation, even if, once again, it is used outside the strict
domain of validity of our approximation. In practice, we have checked
that this is always the case (see also left panel of
\Fig{fig:checkBell}).
\section{The Large Squeezing Limit}
\label{sec:largesqueezing}
When restricted to its strict domain of validity, the approximation
scheme developed in \Sec{sec:appr:scheme} leads to even
simpler formulas. These formulas describe the large squeezing limit
approximation. In this section, we derive the behavior of the 
correlation functions in this regime.

\par

Let us start with $\langle \Psi_{2\, {\rm sq}}\vert
\hat{S}^{(1)}_z(\ell) \hat{S}^{(2)}_z(\ell) \vert \Psi_{2\, {\rm
    sq}}\rangle$. In the regime defined \Eqs{eq:cond:1}
and~(\ref{eq:cond:2}), $\gamma_1\ll 1$ and $\gamma_3 \ll 1$. Making
use of the formulas given in footnote~\ref{footnote:theta:asympt},
\Eq{eq:SzSz:appr} gives rise to
\begin{align}
\langle \Psi_{2\, {\rm sq}}\vert \hat{S}^{(1)}_z(\ell) \hat{S}^{(2)}_z(\ell)
\vert \Psi_{2\, {\rm sq}}\rangle
& \simeq 
-\frac{4}{\sqrt{\pi\gamma_2}\ell}
\vartheta_4\left(0,\ee^{-\gamma_2\ell^2/4}\right)
-{\rm erf}
\left(\sqrt{\gamma_2}\frac{\ell}{2}\right)
+2\sum_{q=0}^\infty (-1)^q\left(u_{q/2+1/2}+u_{q/2-1/2}-2u_{q/2}\right)
\, ,
\label{eq:SzSz:appr:largesqueezing}
\end{align}
an equation significantly simpler than \Eq{eq:SzSz:appr}.

\par

In fact, this is mainly for the correlation functions of
$\hat{S}_x(\ell)$ and $\hat{S}_y(\ell)$ that one really obtains
drastic improvement. Indeed using again the asymptotic behavior of the
Jacobi functions, one can write
\begin{align}
\sum_{n,m=-\infty}^\infty X_{n,m}^{(1)}\simeq 
\sum_{n,m=-\infty}^\infty X_{n,m}^{(3)}
&\simeq \frac{1}{2}\frac{\pi}{\sqrt{\gamma_1\gamma_2}}
\ee^{-\frac{\gamma_3^2}{\gamma_1}\ell^2}\left(\sum_{q=-\infty}^\infty
\mathcal{X}_{2q}^{(1)}+\sum_{q=-\infty}^\infty\mathcal{X}_{2q+1}^{(1)}\right)\, ,\\
\sum_{n,m=-\infty}^\infty X_{n,m}^{(2)}\simeq 
\sum_{n,m=-\infty}^\infty X_{n,m}^{(4)}
&\simeq 
\frac{1}{4}\frac{\pi}{\sqrt{\gamma_1\gamma_2}}
\ee^{-\left(\frac{\gamma_4^2}{\gamma_2}+\frac{\gamma_2}{4}\right)\ell^2}
\left(\sum_{q=-\infty}^\infty\mathcal{X}_{2q}^{(2)}+
\sum_{q=-\infty}^\infty\mathcal{X}_{2q+1}^{(2)}\right)\, ,
\label{eq:appr:largesqueezing:sumX2}
\end{align}
where the sums over $q$, making again use of the formulas given in
footnotes~\ref{footnote:theta:234}, can be expressed as
\begin{align}
\label{eq:appr:largesqueezing:sumX1q}
 \sum_{q=-\infty}^\infty \mathcal{X}_{2q}^{(1)}+
\sum_{q=-\infty}^\infty \mathcal{X}_{2q+1}^{(1)} & \simeq
2\sum_{q=-\infty}^\infty
(u_{q-1/2}+u_{q+1/2}-2u_{q})
-\frac{4}{\ell \sqrt{\pi \gamma_2}}
\vartheta_4\left(0,\ee^{-\gamma_2\ell^2/4}\right)
\, ,\\
 \sum_{q=-\infty}^\infty \mathcal{X}_{2q}^{(2)}+
\sum_{q=-\infty}^\infty \mathcal{X}_{2q+1}^{(2)} & \simeq
4\sum_{q=-\infty}^\infty
\Re\mathrm{e}\left(u_{q-1/2+i\gamma_4/\gamma_2}
+u_{q+1/2+i\gamma_4/\gamma_2}
-2u_{q+i\gamma_4/\gamma_2}
\right)
-\frac{8\ee^{\frac{\gamma_4^2}{\gamma_2}\ell^2}}
{\ell\sqrt{\pi\gamma_2}}\vartheta_4
\left(\frac{\gamma_4}{2}\ell^2,\ee^{-\gamma_2\ell^2}\right)
\, .
\label{eq:appr:largesqueezing:sumX2q}
\end{align}
Although \Eq{eq:appr:largesqueezing:sumX1q} cannot be further
simplified, this is not the case for
\Eq{eq:appr:largesqueezing:sumX2q}. Indeed, noticing that, in the
regime defined by \Eqs{eq:cond:1}-(\ref{eq:cond:2}), one also has
$\gamma_4/\gamma_2\ll -1$, the sum appearing in
\Eq{eq:appr:largesqueezing:sumX2q} can be rewritten in a more friendly
manner\footnote{Here, we make use of the relation $\erf(a+i
  b)\underset{b\rightarrow -\infty}{\longrightarrow}
  \frac{i}{\sqrt{\pi} b}\ee^{(b-ia)^2}$} and one obtains that
\begin{align}
\label{eq:sumucomplexeinter}
u_{q-1/2+i\gamma_4/\gamma_2}
+u_{q+1/2+i\gamma_4/\gamma_2}
-2u_{q+i\gamma_4/\gamma_2}\simeq
-\frac{1}{\sqrt{\pi\gamma_2}\ell}\left\lbrace
\ee^{\left[\frac{\gamma_4}{\gamma_2}-i\left(q-\frac{1}{2}\right)\right]^2\ell^2\gamma_2}
+\ee^{\left[\frac{\gamma_4}{\gamma_2}-i\left(q+\frac{1}{2}\right)\right]^2\ell^2\gamma_2}
-2\ee^{\left(\frac{\gamma_4}{\gamma_2}-iq\right)^2\ell^2\gamma_2}
\right\rbrace\nonumber \\ 
+\sqrt{\frac{\gamma_2}{\pi}}\frac{i}{\ell\gamma_4}\left\lbrace
\left(q-\frac{1}{2}\right)\ee^{\left[\frac{\gamma_4}{\gamma_2}-i\left(q-\frac{1}{2}\right)\right]^2\ell^2\gamma_2}
+\left(q+\frac{1}{2}\right)\ee^{\left[\frac{\gamma_4}{\gamma_2}-i\left(q+\frac{1}{2}\right)\right]^2\ell^2\gamma_2}
-2q\ee^{\left(\frac{\gamma_4}{\gamma_2}-iq\right)^2\ell^2\gamma_2}
\right\rbrace\, .
\end{align}
When summing over $q$, the terms of the first line in the above
equation~(\ref{eq:sumucomplexeinter}) give rise to an elliptic theta
function
that exactly cancels out with the second term of
\Eq{eq:appr:largesqueezing:sumX2q}. It follows that
\begin{align}
 \sum_{q=-\infty}^\infty \mathcal{X}_{2q}^{(2)}+
\sum_{q=-\infty}^\infty \mathcal{X}_{2q+1}^{(2)}&\simeq 
 \frac{4}{\ell\gamma_4}\sqrt{\frac{\gamma_2}{\pi}}
\sum_{q=-\infty}^\infty\Re\mathrm{e}
\biggl\lbrace i
\left(q-\frac{1}{2}\right)
\ee^{\left[\frac{\gamma_4}{\gamma_2}-i\left(q-\frac{1}{2}\right)\right]^2\ell^2\gamma_2}
+i\left(q+\frac{1}{2}\right)
\ee^{\left[\frac{\gamma_4}{\gamma_2}-i\left(q+\frac{1}{2}\right)\right]^2\ell^2\gamma_2}
\\ & 
-2iq\ee^{\left(\frac{\gamma_4}{\gamma_2}-iq\right)^2\ell^2\gamma_2}
\biggr\rbrace\, .
\end{align}
An important remark is that all the terms of this sum are absolutely
summable and, hence, the sum can be reordered. This leads to a simpler
expression in terms of a Jacobi function. Concretely, one
has\footnote{Here, by derivating the
  relations~\cite{Gradshteyn:1965aa}
\begin{align}
\sum_{p=-\infty}^\infty  \ee^{-a p^2}\cos(b p)  = \vartheta_3\left(\frac{b}{2},\ee^{-a}\right)\, ,\quad\quad
\sum_{p=-\infty}^\infty  \ee^{-a \left(p+\frac{1}{2}\right)^2}\cos\left[b \left(p+\frac{1}{2}\right)\right]  = \vartheta_2\left(\frac{b}{2},\ee^{-a}\right)
\end{align}
with respect to $b$, one obtains
\begin{align}
\sum_{p=-\infty}^\infty  p  \ee^{-a p^2}\sin(b p) & = - \frac{\partial}{\partial b} \vartheta_3\left(\frac{b}{2},\ee^{-a}\right)
\\  
\sum_{p=-\infty}^\infty \left(p+\frac{1}{2}\right) \ee^{-a \left(p+\frac{1}{2}\right)^2}\sin\left[b \left(p+\frac{1}{2}\right)\right] & = - \frac{\partial}{\partial b} \vartheta_2\left(\frac{b}{2},\ee^{-a}\right)
\end{align}
which we make use of.
}
\begin{align}
 \sum_{q=-\infty}^\infty \mathcal{X}_{q}^{(2)}+
\sum_{q=-\infty}^\infty \mathcal{X}_{2q+1}^{(2)}& \simeq 
 \frac{8}{\ell\gamma_4}\sqrt{\frac{\gamma_2}{\pi}}
\sum_{q=-\infty}^\infty\Re\mathrm{e}
\left[ i
\left(q-\frac{1}{2}\right)\ee^{-\ell^2\gamma_2
\left(q-\frac{1}{2}+i\frac{\gamma_4}{\gamma_2}\right)^2}
-iq\ee^{-\ell^2\gamma_2\left(q+i\frac{\gamma_4}{\gamma_2}\right)^2}
\right]\\& =
\frac{8}{\ell\gamma_4}\sqrt{\frac{\gamma_2}{\pi}}
\ee^{\frac{\gamma_4^2}{\gamma_2}\ell^2}\sum_{q=-\infty}^\infty\biggl\lbrace
\left(q-\frac{1}{2}\right)\ee^{-\gamma_2\ell^2
\left(q-\frac{1}{2}\right)^2}\sin
\left[2\ell^2\gamma_4\left(q-\frac{1}{2}\right)\right]
\nonumber \\ &
-q\ee^{-\gamma_2\ell^2q^2}\sin\left(2\ell^2\gamma_4 q\right)
\biggr\rbrace\\& =
\frac{8}{\ell\gamma_4}\sqrt{\frac{\gamma_2}{\pi}}
\ee^{\frac{\gamma_4^2}{\gamma_2}\ell^2}\left[\vartheta_3^\prime
\left(\gamma_4\ell^2,\ee^{-\gamma_2\ell^2}\right)-
\vartheta_2^\prime\left(\gamma_4\ell^2,\ee^{-\gamma_2\ell^2}\right)
\right]
= \frac{4}{\ell\gamma_4}\sqrt{\frac{\gamma_2}{\pi}}
\ee^{\frac{\gamma_4^2}{\gamma_2}\ell^2}\vartheta_4^\prime
\left(\frac{\gamma_4}{2}\ell^2,\ee^{-\frac{\gamma_2}{4}\ell^2}\right)
\, ,
\end{align}
where a prime denotes derivative with respect to the first argument of
the theta functions. Plugging this result into
\Eq{eq:appr:largesqueezing:sumX2}, one obtains
\begin{align}
\sum_{n,m=-\infty}^\infty X_{n,m}^{(2)}\simeq 
\sum_{n,m=-\infty}^\infty X_{n,m}^{(4)}
&\simeq 
\sqrt{\frac{\pi}{\gamma_1}}\frac{1}{\ell\gamma_4}\ee^{-\frac{\gamma_2}{4}\ell^2}
\vartheta_4^\prime\left(\frac{\gamma_4}{2}\ell^2,\ee^{-\frac{\gamma_2}{4}\ell^2}\right)
\, .
\label{eq:appr:largesqueezing:sumX2:last}
\end{align}
A last remark is that under the conditions defined by \Eqs{eq:cond:1}
and~(\ref{eq:cond:2}), $\gamma_4\ll -1$. In such a limit, one has
$\vartheta_4^\prime\left(\gamma_4\ell^2/2,\ee^{-\gamma_2\ell^2/4}\right)/(\ell\gamma_4)\rightarrow
0$. Since the $1/\sqrt{\gamma_1}$ prefactor in
\Eq{eq:appr:largesqueezing:sumX2:last} cancels out with the one of
\Eqs{eq:SxSx:exact} and~(\ref{eq:SySy:exact}), this means that one can
take
\begin{align}
\sum_{n,m=-\infty}^\infty X_{n,m}^{(2)}\simeq 
\sum_{n,m=-\infty}^\infty X_{n,m}^{(4)}
&\simeq  0 
\end{align}
in this limit. 

\par

Combining all these results, one obtains the large squeezing
approximation for the two-point correlation functions of
$\hat{S}_x(\ell)$ and $\hat{S}_y(\ell)$. It reads
\begin{align}
\langle \Psi_{2\, {\rm sq}}\vert \hat{S}^{(1)}_x(\ell) \hat{S}^{(2)}_x(\ell)
\vert \Psi_{2\, {\rm sq}}\rangle &
\simeq - \langle \Psi_{2\, {\rm sq}}\vert \hat{S}^{(1)}_y(\ell) \hat{S}^{(2)}_y(\ell)
\vert \Psi_{2\, {\rm sq}}\rangle\\ 
& \simeq
\ee^{-\frac{\gamma_3^2}{\gamma_1}\ell^2}
\left[\sum_{q=-\infty}^\infty
(u_{q-1/2}+u_{q+1/2}-2u_{q})
-\frac{2}{\ell \sqrt{\pi \gamma_2}}
\vartheta_4\left(0,\ee^{-\gamma_2\ell^2/4}\right)\right]\, .
\label{eq:SxSx:appr:largesqueezing}
\end{align}

\begin{figure}[t]
\begin{center}
\includegraphics[width=0.49\textwidth,clip=true]{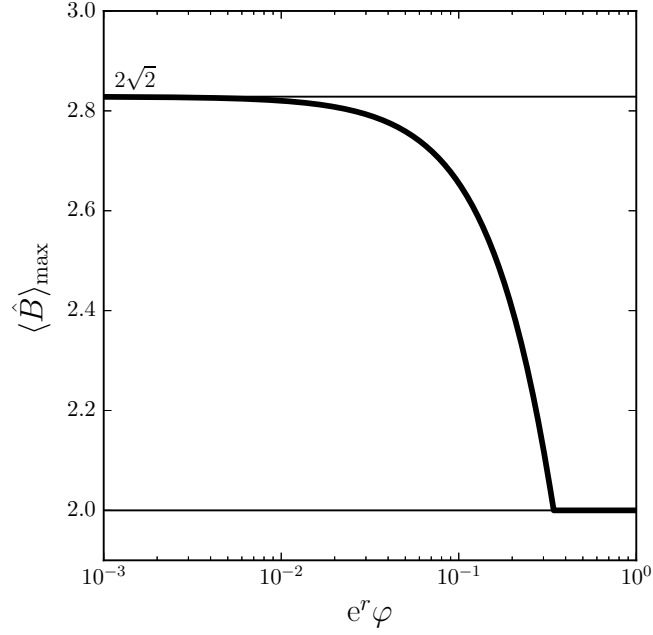}
\caption{Maximum Bell's operator expectation value $\langle B
  \rangle_\mathrm{max}$ (where the maximum of the Bell operator
  expectation value has been identified to the ``bump'', see the
  discussion in \Sec{sec:BellInequality} for an accurate
  definition of the bump) as a function of the combination of
  squeezing parameters $\ee^r\varphi$, in the large squeezing limit.}
\label{fig:largesqueezing}
\end{center}
\end{figure}
It is now interesting to notice that in the asymptotic formulas we
derived, \Eqs{eq:SzSz:appr:largesqueezing}
and~(\ref{eq:SxSx:appr:largesqueezing}), the squeezing parameters only
enter through the combinations $\gamma_2$ and
$\gamma_3^2/\gamma_1$. In the regime defined by \Eqs{eq:cond:1}
and~(\ref{eq:cond:2}), these are given by
\begin{align}
\frac{\gamma_3^2}{\gamma_1} &\simeq \frac{1}{\gamma_2}\simeq \frac{1}{2}\ee^{2r} \varphi^2\, .
\end{align}
Therefore, in the large squeezing limit, all correlators depend only
on a single combination of the squeezing parameters, namely
$\ee^r\varphi$. In \Fig{fig:largesqueezing}, the maximum Bell's operator
expectation value $\langle B \rangle_\mathrm{max}$ is displayed as a function
of $\ee^r\varphi$, in the large squeezing limit. One can see that in
this regime, Bell's inequalities violation can be realized if and only
if
\begin{align}
\label{eq:violationCriterion:largesqueezing}
\ee^r\varphi<0.34\, .
\end{align}
This is why in \Fig{fig:map}, the line $\varphi=0.34\, \ee^{-r}$ has
been displayed, and one can check that when $r\gtrsim 3$, this line
accurately delimits the region where Bell's inequality violation
occurs. Moreover, we also verify that all isocolor lines are indeed
aligned with it. Finally, let us note that the two asymptotic regimes
of \Fig{fig:largesqueezing} can be understood as follows. When
$\ee^r\varphi\gg 1$, $\gamma_3^2/\gamma_1\gg 1$ hence $\langle
\hat{S}_x\hat{S}_x\rangle\ll 1$ in \Eq{eq:SxSx:appr:largesqueezing}
[one can also show that $\langle \hat{S}_z\hat{S}_z\rangle\ll 1$ in
\Eq{eq:SzSz:appr:largesqueezing}] and no Bell's inequality violation
can be realized. In the opposite limit when $\ee^r\varphi\ll 1$, one
can take $\gamma_3^2/\gamma_1\simeq 0$ in
\Eq{eq:SxSx:appr:largesqueezing} and the limit $\gamma_2\gg 1$ can
also be performed. This leads to $u(x)\simeq\vert x\vert$. Then, in
the sum of \Eq{eq:SzSz:appr:largesqueezing}, only the term $q=0$ gives
a non vanishing contribution, and since $\vartheta_4(0,\ee^{-a})\simeq
1$ when $a\gg 1$, one obtains $\langle\hat{S}_z\hat{S}_z\rangle\simeq
1$. In the same manner, the only non vanishing term of the sum in
\Eq{eq:SxSx:appr:largesqueezing} is the one for which $q=0$, yielding
$\langle\hat{S}_x\hat{S}_x\rangle\simeq 1$. This is why Bell's
inequality is maximally violated in this limit.
\section{The Dual Case $\varphi\simeq \pi/2$}
\label{sec:varphi=pi/2}
The strict validity regime of the generic approximation scheme
developed in \Sec{sec:appr:scheme} requires that, in the large
squeezing limit, the squeezing angle $\varphi$ is not too close to
$\pi/2$, see \Eq{eq:cond:1}. This condition notably ensured that
$\gamma_1\ll 1$. In the opposite regime, where
\begin{align}
\label{eq:cond:3}
r\gg 1\quad\mathrm{and}\quad\cos\varphi\ll \ee^{-r}\, ,
\end{align}
one has $\gamma_1 \simeq 1/(\ee^{-2 r}+\ee^{2r}\cos^2\varphi)\gg 1$,
$\gamma_2\simeq 2\ee^{-2r} \ll 1$ and $\gamma_3\simeq 1/\gamma_4\simeq
-\tan\varphi\ll -1 $, and the approximation scheme of
\Sec{sec:appr:scheme} does not apply.

However, as noticed in \Sec{sec:spinPheno}, the two situations
are dual and connected through the
formulas~(\ref{eq:dual:corr:SzSz})-(\ref{eq:dual:corr:SySy}). Therefore,
for any configuration such that $\varphi\simeq \pi/2$, one can always
study the dual configuration for which $\varphi\simeq 0$ making use of
the approximation scheme of \Sec{sec:appr:scheme}, and then
use \Eqs{eq:dual:corr:SzSz}-(\ref{eq:dual:corr:SySy}) to obtain all
correlation functions [in particular, from \Eq{eq:B:optimizedAngles},
it is clear that the expectation value of the Bell operator is the same in
the two dual configurations].

Notwithstanding, in this section we develop an approximation scheme
specific to configurations such that $\varphi\simeq \pi/2$. The reason
is that, as one will see, such a scheme relies on a completely
different technique from the one used in \Sec{sec:spinPheno},
namely saddle-point approximations. Such methods may be preferred,
notably since they easily allow one to go to arbitrarily higher order
in the approximation, something which is not possible with the scheme
of \Sec{sec:spinPheno}. Therefore, if one wanted to compute
higher order corrections of the results of \Sec{sec:spinPheno}
in the case $\varphi\simeq 0$, one would simply have to study the dual
configuration when $\varphi\simeq \pi/2$ and make use of the
saddle-point techniques developed in this section. In this work, we
therefore provide two alternative approximation schemes, and for any
configuration, one can use one or the other, depending on convenience
or aimed accuracy.
In this section, the leading order of the approximation is derived
only (and it will be checked that it leads to maximal Bell's
inequality violation, in agreement with what was noticed at the end of
\Sec{sec:largesqueezing} for the dual configuration), but
generalization to higher order can directly be obtained.
\subsection{Correlation Function $\langle \Psi_{2\, {\rm sq}}\vert
  \hat{S}^{(1)}_z(\ell) \hat{S}^{(2)}_z(\ell) \vert \Psi_{2\, {\rm
      sq}}\rangle$}
\label{sec:phi=pi/2:z}
Let us first work out $\langle \Psi_{2\, {\rm sq}}\vert
\hat{S}^{(1)}_z(\ell) \hat{S}^{(2)}_z(\ell) \vert \Psi_{2\, {\rm
    sq}}\rangle$ in the regime of \Eq{eq:cond:3}. After performing the
change of integration variable $x=z+n+m$, \Eq{eq:Z1:def:simp} can be
rewritten as
\begin{align}
Z^{(1)}_{n,m} = \frac{\ell}{2}\sqrt{\frac{\pi}{\gamma_2}}
\int_p^{p+1}\dd x\ee^{\gamma_1 g(x)}f(x)\, ,
\label{eq:appr:phi/2:Z1:x}
\end{align} 
where
\begin{align}
f(x)= \erf\left[\sqrt{\gamma_2}\frac{\ell}{2}
\left(x-p-q\right)\right]+\erf\left[\sqrt{\gamma_2}
\frac{\ell}{2}\left(x-p+q\right)\right]\, , 
\quad\quad g(x)=-\frac{\ell^2}{4}x^2\, ,
\end{align}
and where one recalls that $p=n+m$ and $q=n-m$. Since $\gamma_1\gg 1$
appears in the exponential argument of \Eq{eq:appr:phi/2:Z1:x}, as
discussed before, the idea is to treat this expression with a
saddle-point approximation. Within the integration domain, the
function $g(x)$ is always maximal at one of the integral boundaries,
that we denote $x_0$. If $p \geq 0$, one has $x_0=p$, while if $p<0$,
then $x_0=p+1$. Let us now perform the change of integration variable
$x=x_0+y/\sqrt{\gamma_1}$. One obtains
\begin{align}
Z^{(1)}_{n,m} = \frac{\ell}{2}\sqrt{\frac{\pi}
{\gamma_2\gamma_1}}\int_{\sqrt{\gamma_1}(p-x_0)}^{\sqrt{\gamma_1}(p+1-x_0)}
\dd y\ee^{\gamma_1 g(y)}f(y)\, .
\end{align}
Since $\gamma_1\gg 1$, the integral is dominated by its contribution
close to $y\simeq 0$ and one can expand the function $f(y)$ at first
order in $1/\sqrt{\gamma_1}$,
\begin{align}
f(y) = & \erf\left[\sqrt{\gamma_2}\frac{\ell}{2}\left(x_0-p-q\right)\right]
+\erf\left[\sqrt{\gamma_2}\frac{\ell}{2}\left(x_0-p+q\right)\right]
+\sqrt{\frac{\gamma_2}{\gamma_1\pi}}\ell
\left[\ee^{-\frac{\gamma_2}{4}\ell^2\left(x_0-p-q\right)^2}
+\ee^{-\frac{\gamma_2}{4}\ell^2\left(x_0-p+q\right)^2}\right]y
+\mathcal{O}\left(\frac{1}{\gamma_1}\right)\, .
\label{eq:appr:phi-pi/2:f(y):expand}
\end{align}
The integral can then be performed exactly. If $p \geq 0$, $x_0=p$ and
the first term of the above expansion vanishes. One then has
\begin{align}
\left.f(y)\right\vert_{p\geq 0} = & 2\sqrt{\frac{\gamma_2}
{\gamma_1\pi}}\ell\ee^{-\frac{\gamma_2}{4}\ell^2q^2}y+\mathcal{O}
\left(\frac{1}{\gamma_1}\right)\, ,
\end{align} 
from which it follows that 
\begin{align}
\left. Z^{(1)}_{n,m} \right\vert_{p \geq 0} & = \frac{\ell^2}{\gamma_1}
\ee^{-\frac{\gamma_2}{4}\ell^2 q^2}\int_0^{\sqrt{\gamma_1}}
\ee^{-\frac{\gamma_1\ell^2}{4}\left(p+\frac{y}{\sqrt{\gamma_1}}\right)^2} y\dd y
 \simeq 
    \begin{cases}
    \displaystyle
      \frac{4}{\ell^2 p^2 \gamma_1^2}\ee^{-\left(\gamma_1 p^2 
+ \gamma_2 q^2\right)\frac{\ell^2}{4}}& \mathrm{if}\ p>0 \\
      \displaystyle
	  \frac{2}{\gamma_1}\ee^{-\gamma_2 q^2\frac{\ell^2}{4}}& \mathrm{if}\ p=0 
    \end{cases}
\, ,
\end{align}
where in the second line, we have expanded the result at leading order
in $1/\sqrt{\gamma_1}$. If $p<0$, $x_0=p+1$ and one has
\begin{align}
f(y)=\erf\left[\sqrt{\gamma_2}\frac{\ell}{2}\left(1-q\right)\right]+
\erf\left[\sqrt{\gamma_2}\frac{\ell}{2}\left(1+q\right)\right]
+\mathcal{O}\left(\frac{1}{\sqrt{\gamma_1}}\right)\, ,
\end{align}
from which one obtains
\begin{align}
\left. Z^{(1)}_{n,m} \right\vert_{p < 0} & = \frac{\ell}{2}
\sqrt{\frac{\pi}{\gamma_1\gamma_2}}\left\lbrace 
\erf\left[\sqrt{\gamma_2}\frac{\ell}{2}\left(1-q\right)\right]+
\erf\left[\sqrt{\gamma_2}\frac{\ell}{2}\left(1+q\right)\right] \right\rbrace 
\int_{-\sqrt{\gamma_1}}^0\dd y\ee^{-\gamma_1\frac{\ell^2}{4}
\left(p+1+\frac{y}{\sqrt{\gamma_1}}\right)^2}\\
& \simeq 
    \begin{cases}
    	\displaystyle
     -\frac{1}{\ell\gamma_1(p+1)}\sqrt{\frac{\pi}{\gamma_2}}
\left\lbrace \erf\left[\sqrt{\gamma_2}\frac{\ell}{2}\left(1-q\right)\right]+
\erf\left[\sqrt{\gamma_2}\frac{\ell}{2}\left(1+q\right)\right] 
\right\rbrace \ee^{-\gamma_1(p+1)^2}\frac{\ell^2}{4}  &  \mathrm{if}\ p<-1 \\
 	\displaystyle
	  \frac{\pi}{2\sqrt{\gamma_1\gamma_2}}\left\lbrace 
\erf\left[\sqrt{\gamma_2}\frac{\ell}{2}\left(1-q\right)\right]+
\erf\left[\sqrt{\gamma_2}\frac{\ell}{2}\left(1+q\right)\right] 
\right\rbrace & \mathrm{if}\ p=-1 
    \end{cases}
\, ,
\end{align}
where in the second line, we have again expanded the result at leading
order in $1/\sqrt{\gamma_1}$. From these expressions, it is clear that
the terms such that $p>0$ or $p<-1$ are exponentially suppressed in
the limit $\gamma_1\gg 1$. In some sense, the situation is similar to
the one of the large $\ell$ limit, see \Sec{sec:largeelllimit}, since
only the terms $p=0$ and $p=-1$ give a non-negligible
contribution. Recalling that $p$ and $q$ must have same parity, one
then has
\begin{align}
\sum_{n,m=-\infty}^\infty (-1)^{n+m} Z_{n,m}^{(1)} = \frac{2}{\gamma_1}
\sum_{q\,\mathrm{even}}\ee^{-\gamma_2\ell^2 q^2/4} -  \frac{\pi}{2\sqrt{\gamma_1\gamma_2}} \sum_{q\,\mathrm{odd}} \left\lbrace \erf\left[\sqrt{\gamma_2}
\frac{\ell}{2}(1- q)\right]+
\erf\left[\sqrt{\gamma_2}\frac{\ell}{2}\left(q+1\right)\right] 
\right\rbrace\, .
\end{align}
It is now time to make use of the fact that, in the regime under
consideration, $\gamma_2\ll 1$. In this limit, the second error
function can be expanded around the argument of the first one, and one
obtains
\begin{align}
\sum_{n,m=-\infty}^\infty (-1)^{n+m} Z_{n,m}^{(1)} \simeq 
\frac{2}{\gamma_1}\sum_{q=-\infty}^\infty\ee^{-\gamma_2\ell^2 q^2} 
-  \sqrt{\frac{\pi}{\gamma_1}}\ell \sum_{q=-\infty}^\infty\ee^{-\gamma_2\ell^2 q^2}\, .
\end{align}
It is interesting to notice that the first term, corresponding to
$p=0$, is subdominant at leading order in $1/\sqrt{\gamma_1}$, which
is consistent with the fact that the leading term in
\Eq{eq:appr:phi-pi/2:f(y):expand} vanishes for $p\geq 0$ and
$x_0=p$. One then obtains
\begin{align}
\sum_{n,m=-\infty}^\infty (-1)^{n+m} Z_{n,m}^{(1)} \simeq -  \sqrt{\frac{\pi}{\gamma_1}}\ell \vartheta_3\left(0,\ee^{-\gamma_2\ell^2}\right)
\simeq -\frac{\pi}{\sqrt{\gamma_1\gamma_2}}
\, ,
\end{align}
where we have expanded the final result in the $\gamma_2\ll 1$
limit. Making use of \Eq{eq:SzSz:exact}, this gives rise to
\begin{align}
\langle \Psi_{2\, {\rm sq}}\vert \hat{S}^{(1)}_z(\ell) \hat{S}^{(2)}_z(\ell)
\vert \Psi_{2\, {\rm sq}}\rangle = -1\, .
\end{align}
This is in agreement with \Eq{eq:dual:corr:SzSz} and the limit derived
at the very end of \Sec{sec:largesqueezing}.
\subsection{Correlation Functions $\langle \Psi_{2\, {\rm sq}}\vert
  \hat{S}^{(1)}_x(\ell) \hat{S}^{(2)}_x(\ell) \vert \Psi_{2\, {\rm
      sq}}\rangle$ and $\langle \Psi_{2\, {\rm sq}}\vert
  \hat{S}^{(1)}_y(\ell) \hat{S}^{(2)}_y(\ell) \vert \Psi_{2\, {\rm
      sq}}\rangle$}
Let us now make use of the same technique to calculate the correlation
function of the $x$ and $y$-components of the spin operators. Let us
start with the term $X_{n,m}^{(1)}$ given by \Eq{eq:X1:def}, our goal
being to put this expression under a form similar to
\Eq{eq:appr:phi/2:Z1:x}. Straightforward manipulations lead to
\begin{align}
X^{(1)}_{n,m} = \ee ^{-\gamma_1\ell^2/2}\ell \sqrt{\frac{\pi}{\gamma_2}}
\int_{2p}^{2p+1}\dd x\ee^{\gamma_1 g(x)}f(x)\, ,
\label{eq:appr:phi/2:X1:x}
\end{align} 
where, now, the functions $f(x)$ and $g(x)$ are defined by
\begin{align}
f(x)= \cos\left[\gamma_3\ell^2\left(x+1\right)\right]
\left\{\erf\left[\sqrt{\gamma_2}\frac{\ell}{2}
\left(x-2p-2q\right)\right]+\erf\left[\sqrt{\gamma_2}
\frac{\ell}{2}\left(x-2p+2q\right)\right]\right\}\, , 
\quad g(x)=-\frac{\ell^2}{4}x(x+2)\, .
\end{align}
Then, the calculation proceeds exactly in the same way as before. The
integral is dominated by contributions coming from the point $x_0$
with $x_0=2p$ when $p\ge 0$ and $x_0=2p+1$ when $p<-1$. One then write
$x=x_0+y/\sqrt{\gamma_1}$ and expands the results in inverse powers of
$\gamma_1$. Notice that, at leading order, the cosine function present
in $f(x)$ never contributes since the corresponding first correction
is quadratic in $x$. For $x_0=2p$, one obtains
\begin{align}
\left.X_{n,m}^{(1)}\right\vert_{p\ge 0}
\simeq \frac{8}{\gamma_1^2\ell^2(1+2p)^2}
\cos\left[\gamma_3\ell^2\left(2p+1\right)\right]
\ee^{-\gamma_2\ell^2q^2}
\ee^{-\gamma_1\ell^2(1+2p+2p^2)/2}.
\end{align}
Notice that this expression is perfectly valid when $p=0$ contrary to
the corresponding case for the $z$-component spin correlation function. In
the case $x_0=2p+1$, similar considerations lead to
\begin{align}
\left. X^{(1)}_{n,m} \right\vert_{p < -1} 
& \simeq 
     -\frac{\ee^{-\gamma_1\ell^2/2}}{\ell\gamma_1(p+1)}\sqrt{\frac{\pi}{\gamma_2}}
\cos\left[\gamma_3\ell^2\left(2p+2\right)\right]
\left\lbrace \erf\left[\sqrt{\gamma_2}\frac{\ell}{2}\left(1-q\right)\right]+
\erf\left[\sqrt{\gamma_2}\frac{\ell}{2}\left(1+q\right)\right] 
\right\rbrace \ee^{-\gamma_1\ell^2(1+2p)(3+2p)/4}\, ,  
\end{align}
while, if $p=-1$, one has
\begin{align}
\left. X^{(1)}_{n,m} \right\vert_{p =-1} \simeq 
2\ell \sqrt{\frac{\pi}{\gamma_1}}\ee^{-\gamma_1\ell^2/4-\gamma_2\ell^2q^2}	 
\, ,
\end{align}
It is clear that $p=-1$ gives the dominant contribution. The term
corresponding to $p=0$ is not exponentially killed but contains
additional power of $\gamma_1$ at the denominator. Therefore, it can
be discarded. Then, in order to calculate the first sum, we just have
to perform the sum over $q$ in the previous equation which can be
done, as usual, in terms of a Jacobi function. Using that $\gamma_2\ll
1$, one arrives at
\begin{align}
\sum_{n,m=-\infty}^\infty X_{n,m}^{(1)}\simeq \frac{\pi}{\sqrt{\gamma_1\gamma_2}}
\ee^{-\gamma_1\ell^2/4}.
\end{align}

The calculation of $X_{n,m}^{(2)}$ proceeds along the same lines, the
only (slightly) more complicated aspect being that one has now to deal
with complex error functions. Working out \Eq{eq:X2:def}, one arrives
at
\begin{align}
X^{(2)}_{n,m} = \ee ^{-(\gamma_1+\gamma_2)\ell^2/4-\gamma_4^2/\gamma_2\ell^2}
\frac{\ell}{2} \sqrt{\frac{\pi}{\gamma_2}}
\int_{2p}^{2p+1}\dd x\ee^{\gamma_1 g(x)}f(x)\, ,
\label{eq:appr:phi/2:X1:x}
\end{align} 
where the functions $f(x)$ and $g(x)$ can be expressed as
\begin{align}
f(x) &= 
\biggl\{\erf\left[\sqrt{\gamma_2}\frac{\ell}{2}
\left(x-2p-2q\right)+i\frac{\gamma_4}{\sqrt{\gamma_2}}\ell\right]
+\erf\left[\sqrt{\gamma_2}\frac{\ell}{2}
\left(x-2p-2q\right)-i\frac{\gamma_4}{\sqrt{\gamma_2}}\ell\right]
\nonumber \\ &
+\erf\left[\sqrt{\gamma_2}
\frac{\ell}{2}\left(x-2p+2q\right)+i\frac{\gamma_4}{\sqrt{\gamma_2}}
\right]
+\erf\left[\sqrt{\gamma_2}
\frac{\ell}{2}\left(x-2p+2q\right)-i\frac{\gamma_4}{\sqrt{\gamma_2}}
\right]
\biggr\}\, ,
\end{align} 
and $g(x)=-\ell^2x(x+2)/4$. As before, one must distinguish whether
one deals with the case $x_0=2p$ or $x_0=2p+1$ and it is easy to
convince oneself that the dominant contribution will be given by
$p=-1$ for which one has
\begin{align}
\left. X^{(2)}_{n,m} \right\vert_{p =-1} \simeq 
2\ell \sqrt{\frac{\pi}{\gamma_1}}\ee^{-\gamma_2\ell^2/4
-\gamma_4^2/\gamma_2\ell^2
-\gamma_2\ell^2q^2}	 
\, ,
\end{align}
Performing the sum over $q$ is now standard and one obtains
\begin{align}
\sum_{n,m=-\infty}^\infty X_{n,m}^{(2)}\simeq \frac{\pi}{\sqrt{\gamma_1\gamma_2}}
\ee^{-\gamma_2\ell^2/4-\gamma_4^2/\gamma_2\ell^2}.
\end{align}
Similar considerations can be made for the two remaining integrals. 

\par

Combining these results with \Eqs{eq:SxSx:exact}, one finally obtains
\begin{align}
\langle \Psi_{2\, {\rm sq}}\vert \hat{S}^{(1)}_x(\ell) \hat{S}^{(2)}_x(\ell)
\vert \Psi_{2\, {\rm sq}}\rangle \simeq
\ee^{-\gamma_1\ell^2/4}+\ee^{-\gamma_2\ell^2/4-\gamma_4^2/\gamma_2\ell^2}\simeq 1
\end{align}
where we have taken the limits $\gamma_1\gg 1$ and $\gamma_2,
\gamma_4^2/\gamma_2\ll 1$. In the same way, using \Eq{eq:SySy:exact},
one has
\begin{align}
\langle \Psi_{2\, {\rm sq}}\vert \hat{S}^{(1)}_y(\ell) \hat{S}^{(2)}_y(\ell)
\vert \Psi_{2\, {\rm sq}}\rangle \simeq -\ee^{-\gamma_1\ell^2/4}
+\ee^{-\gamma_2\ell^2/4-\gamma_4^2/\gamma_2\ell^2}\simeq 1
\, .
\end{align}
These results are in agreement with \Eqs{eq:dual:corr:SxSx}
and~(\ref{eq:dual:corr:SySy}) and the limit derived at the very end of
\Sec{sec:largesqueezing}. This also confirms that Bell's inequalities
are maximally violated in the large squeezing limit when
$\varphi=\pi/2$.

\section{Rotation in Phase-Space}
\label{sec:PhaseSpaceRotation}
In this section, we investigate the situation where the pseudo-spin
operators are defined with respect to a direction in phase space that
is different from the position $Q$ considered in the rest of this
paper. Let us therefore introduce the rotation in phase space
\begin{align}
\label{eq:rotation:Q}
\overline{\hat{Q}}_i &= \hat{Q}_i\cos \alpha - \hat{P}_i \sin \alpha ,\\
\label{eq:rotation:P}
\overline{\hat{P}}_i &= \hat{P}_i\cos \alpha +\hat{Q}_i \sin \alpha ,
\end{align}
where $\alpha$ is a real angle parameter and $i=1,2$. One can easily
check that
$[\hat{Q}_i,\hat{P}_i]=[\overline{\hat{Q}}_i,\overline{\hat{P}}_i]$
and this transformation is therefore canonical. As a consequence, it
can be represented by a unitary operator $U_i$ which, in the present
case, takes the following form~\cite{1994AnPhy.232..292A}
\begin{align}
U_i=\exp\left[\frac{i}{2}\ln (\cos \alpha)\left(\hat{P}_i\hat{Q}_i
+\hat{Q}_i\hat{P}_i\right)\right]
\exp\left[\frac{i}{2}\sin (2\alpha)\hat{Q}_i^2\right]
\exp\left(-\frac{i}{2}\tan \alpha \hat{P}_i^2\right).
\end{align}
One can indeed check that the transformation given by
\Eqs{eq:rotation:Q} and~(\ref{eq:rotation:P}) is realized by
$\overline{\hat{Q}}_i=U_i\hat{Q}_iU_i^{\dagger}$ and
$\overline{\hat{P}}_i=U_i\hat{P}_iU_i^{\dagger}$. The next step is to
study how the state~(\ref{eq:qstateposition}) transforms under this
canonical transformation~\cite{1971JMP....12.1772M}. For this purpose,
let us consider the eigenstates $\vert Q_i\rangle $ and $\vert
\overline{Q}_i\rangle $ of the operators $\hat{Q}_i$ and
$\overline{\hat{Q}}_i$, respectively. One can sandwich
\Eqs{eq:rotation:Q} and~(\ref{eq:rotation:P}) between $\langle
Q_i\vert $ and $\vert \overline{Q}_i\rangle $ and use the fact that
$\langle Q_i\vert \hat{P}_i\vert \overline{Q}_i\rangle =-i\partial
\langle Q_i\vert \overline{Q}_i\rangle /(\partial Q_i)$ and $\langle
Q_i\vert \overline{\hat{P}}_i\vert \overline{Q}_i\rangle =i\partial
\langle Q_i\vert \overline{Q}_i\rangle /(\partial
\overline{Q}_i)$. This results in two differential equations that
reads
\begin{align}
\frac{\partial }{\partial Q_i}
\langle Q_i\vert
\overline{Q}_i\rangle &= \frac{i}{\sin \alpha}
\left(\cos \alpha Q_i-\overline{Q}_i\right)\langle Q_i\vert
\overline{Q}_i\rangle, \\
\frac{\partial }{\partial \overline{Q}_i}
\langle Q_i\vert
\overline{Q}_i\rangle &= -\frac{i}{\sin \alpha}
\left(Q_i-\cos \alpha \overline{Q}_i\right)\langle Q_i\vert
\overline{Q}_i\rangle,
\end{align}
and leads to
\begin{equation}
\langle Q_i\vert
\overline{Q}_i\rangle =C\exp\left(-\frac{i}{\sin \alpha}Q_i\overline{Q}_i
+\frac{i}{2\tan \alpha}Q_i^2
+\frac{i}{2\tan \alpha}\overline{Q}_i^2\right),
\end{equation}
where $C$ is a constant. From this expression, one can now infer the 
wavefunction of the system after the canonical transformation. It is 
given by
\begin{equation}
\langle \Psi_{\rm 2\,
  sq} \vert \overline{Q}_1,\overline{Q}_2\rangle 
\equiv \Psi_{\rm 2\,
  sq}\left(\overline{Q}_1,\overline{Q}_2\right) 
=\int \langle \Psi \vert Q_1,Q_2\rangle 
\langle Q_1\vert
\overline{Q}_1\rangle
\langle Q_2\vert
\overline{Q}_2\rangle {\rm d}Q_1{\rm d}Q_2,
\end{equation}
where $\langle \Psi_{\rm 2\,
  sq} \vert Q_1,Q_2\rangle =\Psi_{\rm 2\,
  sq}\left(Q_1,Q_2\right) \propto \ee^{A\left(Q_1^2+Q_2^2\right)
  -BQ_1Q_2}$ is given by \Eq{eq:qstateposition}. This integral can
easily be performed since it is a Gaussian integral. Concretely, one 
has 
\begin{align}
\Psi_{\rm 2\,
  sq}\left(\overline{Q}_1,\overline{Q}_2\right) 
\propto e^{i(\overline{Q}_1^2+\overline{Q}_2^2)/(2 \tan \alpha)}
\int e^{-{\bf Q}^{\rm T}M{\bf Q}/2-{\bf J}^{\rm T}{\bf Q}}{\rm d}Q_1{\rm d}Q_2,
\end{align}
where 
\begin{align}
{\bf Q}=
\begin{pmatrix}
Q_1 \\ Q_2
\end{pmatrix},
\quad 
{\bf J}=\frac{i}{\sin \alpha}
\begin{pmatrix}
\overline{Q}_1 \\
\overline{Q}_2
\end{pmatrix},
\quad
M=
\begin{pmatrix}
-2A-\frac{i}{\tan \alpha} & B \\
B & -2A-\frac{i}{\tan \alpha}
\end{pmatrix},
\end{align}
and straightforward calculations leads to
\begin{align}
\Psi_{\rm 2\,
  sq}\left(\overline{Q}_1,\overline{Q}_2\right) 
\propto e^{i(\overline{Q}_1^2+\overline{Q}_2^2)/(2 \tan \alpha)}
e^{{\bf J}^{\rm T}M^{-1}{\bf J}/2}=\ee^{\overline{A}\left(\overline{Q}_1^2
+\overline{Q}_2^2\right)
  -\overline{B}\,\overline{Q}_1\overline{Q}_2},
\end{align}
where the new quantities $\overline{A}$ and $\overline{B}$ can be
expressed as
\begin{align}
\overline{A}&=\frac{i}{2\tan \alpha}+\frac{1}{2\sin^2\alpha \det M}
\left(2A+\frac{i}{\tan \alpha}\right)=A(r,\varphi+\alpha)\\
\overline{B}&= -\frac{B}{\sin ^2\alpha \det M}=B(r,\varphi+\alpha).
\end{align}
We conclude that, after the rotation, the wavefunction keeps its
shape unmodified, the squeezing parameter also remains unchanged but
the squeezing angle becomes $\varphi+\alpha$.

Interestingly enough, the above result can also be established
directly in phase space. A convenient tool to carry out this
calculation is the Wigner
function~\cite{1932PhRv...40..749W,Case:2008}
\begin{align}
W_{\rm 2\, sq}(Q_1,P_1,Q_2,P_2)=\frac{1}{\left(2\pi\right)^2}\int\dd x \dd y
\Psi_{\rm 2\, sq}^*\left(Q_1-\frac{x}{2},Q_2-\frac{y}{2}\right)\ee^{-i P_1 x -i P_2 y}
\Psi_{\rm 2\, sq}\left(Q_1+\frac{x}{2},Q_2+\frac{y}{2}\right),
\end{align}
which is a quasiprobability distribution in phase space that provides
an equivalent description of the quantum state than that of the
wavefunction or the density matrix. For the two-mode squeezed
state~(\ref{eq:qstateposition}), Gaussian integration leads
to~\cite{Case:2008, Martin:2015qta}
\begin{align}
W_{\rm 2\, sq}(Q_1,P_1,Q_2,P_2) =\frac{1}{\pi^2}\exp &\left[
-\cosh(2r)\left(Q_1^2+Q_2^2+P_1^2+P_2^2\right)
+2\sinh(2r)\sin(2\varphi)\left(Q_1P_2+Q_2P_1\right)
\right. \nonumber \\ & \left. 
+2\sinh(2r)\cos(2\varphi)\left(Q_1 Q_2 - P_1 P_2\right)
\right]\, .
\label{eq:Wigner:2msq}
\end{align}
If one replaces $Q_i$ and $P_i$ by their
expressions~(\ref{eq:rotation:Q}) and (\ref{eq:rotation:P}) in terms
of $\overline{Q}_i$ and $\overline{P}_i$ in this formula, one directly obtains
\begin{align}
W_{\rm 2\, sq}(\overline{Q}_1,\overline{P}_1,\overline{Q}_2,\overline{P}_2) =
\frac{1}{\pi^2}\exp &\left\lbrace
-\cosh(2r)\left(\overline{Q}_1^2+\overline{Q}_2^2
+\overline{P}_1^2+\overline{P}_2^2\right)
+2\sinh(2r)\sin\left[2\left(\varphi+\alpha\right)\right]
\left(\overline{Q}_1\overline{P}_2+\overline{Q}_2\overline{P}_1\right)
\right. \nonumber \\ & \left. 
+2\sinh(2r)\cos\left[2\left(\varphi+\alpha\right)\right]
\left(\overline{Q}_1 \overline{Q}_2 - \overline{P}_1 \overline{P}_2\right)
\right\rbrace\, .
\end{align}
Comparing with \Eq{eq:Wigner:2msq}, one notices that the same Wigner
function is obtained, except that the squeezing angle has been
redefined according to $\varphi\rightarrow \varphi+\alpha$. The
wavefunction in the $\overline{Q}$-position representation is therefore
given by the same expression as \Eq{eq:qstateposition}, if one
replaces $\varphi$ by $\varphi+\alpha$, which is exactly the result we
have obtained previously.

\par

As a consequence, if one defines the pseudo-spin operators with
respect to $\overline{Q}$ instead of $Q$, that is to say, if one
replaces $Q$ by $\overline{Q}$ everywhere in
\Eqs{eq:projectoru:def}-(\ref{eq:defsplus}), then one obtains the same
results as the ones derived above except that $\varphi$ must be
replaced by $\varphi+\alpha$. The analysis presented in this paper
therefore allows one to deal with any orientation between the squeezing
angle of the wavefunction and the pseudo-spin operators.

%
\twocolumngrid
\bibliography{bell}
\end{document}

%% file: newcommands.tex
\newcommand{\ie}{{i.e.~}}

\newcommand{\eg}{\textsl{e.g.~}}

\DeclareMathOperator{\erf}{erf}

\newcommand{\dd}{\mathrm{d}}
\newcommand{\ee}{e}

\newcommand{\umax}{\mathrm{max}}

\newcommand{\setR}{\mathbb{R}}

\newcommand{\beq}{\begin{equation}}
\newcommand{\eeq}{\end{equation}}
\newcommand{\bea}{\begin{eqnarray}}
\newcommand{\eea}{\end{eqnarray}}

\newlength{\wsingfig}
\newlength{\wdblefig}
\newlength{\wquadfig}
\newlength{\wtriplefig}
\newcommand{\Eq}[1]{Eq.~(\ref{#1})}
\newcommand{\Eqs}[1]{Eqs.~(\ref{#1})}
\newcommand{\Fig}[1]{Fig.~{\ref{#1}}}
\newcommand{\Figs}[1]{Figs.~{\ref{#1}}}
\newcommand{\Ref}[1]{Ref.~{\cite{#1}}}
\newcommand{\Refs}[1]{Refs.~{\cite{#1}}}
\newcommand{\Sec}[1]{Sec.~\ref{#1}}
\newcommand{\Secs}[1]{Secs.~\ref{#1}}